\documentclass[a4paper]{article}

\pdfoutput=1
\usepackage[T1]{fontenc}
\usepackage{ifthen,relsize,multicol,marvosym,float}
\usepackage{tabularx,tabulary,multirow,lastpage,lineno}
\usepackage[pdftex]{graphicx}
\usepackage{epic,eepic,epsfig}
\usepackage{siunitx}
\usepackage[polutonikogreek,french,german,english]{babel}

\usepackage{titling}
\usepackage[centertags]{amsmath}
\usepackage{pbox,fancyhdr,textcomp}
\usepackage{amssymb,amsfonts,oldgerm,mathrsfs,ipa}
\usepackage[text={6in,8.25in},centering]{geometry}

\usepackage{chicago}

\newcommand{\afourlengths}{\geometry{a4paper,text={15cm,22.6cm},centering}}

\newcommand{\hyperrefpdf}{\usepackage[colorlinks=true,citecolor=red,linkcolor=blue,urlcolor=blue,pdftex]{hyperref}}
\hyperrefpdf
\newcommand{\skipline}[1][1]{\vspace*{#1\baselineskip}}

\newcommand{\noitemsep}{\setlength{\itemsep}{0in}}

\newcommand{\underscore}[1][1]{\underline{\hspace*{#1ex}}}

\newcommand{\smidge}{\hspace*{.1em}}

\newcommand{\hem}[1][1]{\hspace*{#1em}}

\newcommand{\equalsdf}{=_{\mbox{\tiny df}}}

\newtheorem{theorem}{Theorem}[subsection]

\newtheorem{defn}[theorem]{Definition}

\newcommand{\french}[1]{\selectlanguage{french}#1\selectlanguage{english}}

\newcommand{\hto}{\texttt{H$\mathtt{_2}$O}}

\afourlengths

\predate{\skipline[-.75] \begin{center} \large}
\postdate{\end{center} \skipline[-2.75]}

\hyperrefpdf
\long\def\symbolfootnote[#1]#2{\begingroup%
\def\thefootnote{\fnsymbol{footnote}}\footnote[#1]{#2}\endgroup}

\title{Schematizing the Observer and the Epistemic \\ Content of
  Theories\thanks{This paper is forthcoming in \emph{Studies in
      History and Philosophy of Modern Physics}, 2020, with the
    following changes: what appears in this version as
    \S\ref{sec:what-meas-obsr} (``What Measurements and the Observer
    Are'') and \S\ref{sec:precis} (``Appendix: Pr\'ecis''), will not
    appear in the published version due to length constraints;
    otherwise the two versions are identical (except for trivialities
    noted in the text).  I thank Jeremy Butterfield for delightfully
    edifying discussion and for his usual detailed, extensive and
    penetrating comments on a draft.  I thank Adam Caulton, Neil
    Dewar, Robert DiSalle, Michael Friedman, Bill Harper, Sebastian
    Lutz, Lydia Patton, Chris Pincock, Chris Smeenk, and Jim
    Weatherall for many enjoyable and illuminating conversations
    covering all the topics I treat here.  I thank an anonymous
    referee for a superogatory report that helped me clarify a few
    important matters.  I am, as always, grateful to Howard Stein for
    many conversations over many years on these topics, and for having
    written papers that continue to inspire me.  This paper owes a
    clear and great debt especially to his ``Some Reflections on the
    Structure of Our Knowledge in Physics''.  I am not certain that
    Stein would endorse all my arguments and conclusions, but I have
    hopes he would have sympathy with the overall thrust.  Gef\"ordert
    durch die Deutsche Forschungsgemeinschaft (DFG) -- Projektnummer
    312032894, CU 338/1-1.  (Funded by the German Research Foundation
    (DFG) -- project number 312032894, CU 338/1-1.)}}

\author{Erik Curiel\thanks{\textbf{Author's address}: Munich Center
    for Mathematical Philosophy, Ludwig-Maximilians-Universit\"at;
    Black Hole Initiative, Harvard University; Smithsonian
    Astrophysical Observatory, Radio and Geoastronomy Division;
    \textbf{email}: \href{mailto:erik@strangebeautiful.com}
    {\texttt{erik@strangebeautiful.com}}}}

\date{}

\begin{document}
\thispagestyle{empty}
%\thispagestyle{plain}
%\thispagestyle{headings}
%\thispagestyle{myheadings}
%\thispagestyle{fancy}
%\english
\maketitle

%\skipline

% \begin{quote}
%   \begin{tabbing}
%     \hspace*{2em}\=When every year and month sends \=\kill
%     \>The \\
%     \\
%     \>\>Williams \\
%     \>\>\emph{Paterson} 
%   \end{tabbing}
% \end{quote}

\skipline

\noindent I dedicate this paper, with affection and admiration, to
Howard Stein, mentor and friend, one of the few people I have met in
my life to deserve the honorific `philosopher' in the deep Platonic
sense.

\skipline

\begin{quote}
  So I suggest that the principal difficulty is not that of how to
  leave the theory outside the laboratory door, but that of how to get
  the laboratory inside the theory.  \P \space\space Well, how
  \emph{do} we do it?  For of course we do put theory and experiment
  in relation to one another; otherwise it would be impossible to test
  theories, and impossible to apply them.  It would also, I should
  add, be impossible to \emph{understand} a theory, as anything but a
  purely mathematical structure---impossible, that is, to understand a
  theory \emph{as} a theory of physics---if we had no systematic way
  to put the theory into connection with observation (or experience).
  \begin{flushright}
    Howard Stein  \\
    ``Some Reflections on the Structure of Our Knowledge in Physics''
    \\
    (italics are Stein's)
  \end{flushright}
\end{quote}
\nocite{stein-struct-know}

\skipline

\begin{quote}
  \begin{center}
    \textbf{ABSTRACT}
  \end{center}  

  Following some observations of Howard Stein (1994), I argue that,
  contrary to the standard view, one cannot understand the structure
  and nature of our knowledge in physics without an analysis of the
  way that observers (and, more generally, measuring instruments and
  experimental arrangements) are modeled in theory.  One upshot is
  that standard pictures of what a scientific theory can be are
  grossly inadequate.  In particular, standard formulations assume,
  with no argument ever given, that it is possible to make a clean
  separation between, on the one hand, one part of the scientific
  knowledge a physical theory embodies, \emph{viz}., that encoded in
  the pure mathematical formalism and, on the other, the remainder of
  that knowledge.  The remainder includes at a minimum what is encoded
  in the practice of modeling particular systems, of performing
  experiments, of bringing the results of theory and experiment into
  mutually fruitful contact---in sum, real application of the theory
  in actual scientific practice.  This assumption comes out most
  clearly in the picture of semantics that naturally accompanies the
  standard view of theories: semantics is fixed by ontology's shining
  City on the Hill, and all epistemology and methodology and other
  practical issues and considerations are segregated to the ghetto of
  the theory's pragmatics.  We should not assume such a clean
  segregation is possible without an argument, and, indeed, I offer
  many arguments that such a segregation is not feasible.  It follows
  that an adequate semantics for theories cannot be founded on
  ontology, but rather on epistemology and methodology.
\end{quote}

\skipline

\tableofcontents

\section{The Complexity of the World and the Simplicity of Theory}
\label{sec:complex-simple}

For essentially every physical theory we have (Navier-Stokes theory,
general relativity, quantum field theory, \emph{etc}.\@), we have very
little detailed knowledge of the structure of generic solutions.
Usually, we know exact solutions only under conditions of perfect or
near-perfect symmetry or some other unrealistic assumption (two
bodies, no external perturbative influences, \ldots), and then argue
that we can apply such solutions to real physical systems, because the
approximation is adequate in appropriately controlled circumstances.
That is to say: we have no real idea \emph{at all}, in a
representational sense, ``what the world would be like if the theory
were true or largely true of it'', for any physical theory.

This becomes particularly clear when one stops to think, to really
think hard, about the mind-blowing complexity and richness of texture
of real physical systems in the world.  Take this glass of water
resting on the table, seemingly in equilibrium with its environment.
The water does not in fact have a constant, static density, pressure,
temperature, shear-stress, heat flux, fluid velocity and volume, as we
would naturally ascribe to it in a basic treatment using Navier-Stokes
theory, the classical theory of visco\"elastic, thermoconducive
fluids.  There are tiny temperature gradients from the edge of the
glass to the interior, and likewise tiny pressure gradients.  These
drive microscopic flows and eddies and vortices, which generate
fluctuations in the shear-stress.  There is evaporation of the water
at the surface, adsorption of a thin layer of particulate matter at
the surface, and thereby absorption of particulates from the air and
of air molecules themselves.  There is thus exchange of thermal energy
at the boundary, complicated by the little air bubbles and particulate
matter suspended there, and inhomogeneities in the glass at the
boundary.  The water itself, at a finer level of detail, consists of a
hyper-dense, stereometrically complex network of Hydrogen bonds with
rich characteristic symmetry patterns as web to the weft of instances
of dozens of species of ionic combinations of \texttt{H} and
\texttt{O} ($\hto$ being not even a majority, only a plurality).  The
water is pervaded and perturbed by the ambient electromagnetic field,
itself composed of endless linear superpositions of radiation from the
antenn{\ae} of radio stations playing Berg's ``Lulu'' and the BBC
news, mobile phones all over the Earth, magnetic induction from small
movements of power cords, the Earth's own magnetic field, infra-red
radiation and radio waves from Sagittarius A$^*$ (the supermassive
black hole at the center of the Milky Way), the cosmic microwave
background radiation, the jiggling of the water's own molecules, the
rotation of the iron core of some planet 30 light-years away, and the
gamma ray bursts from distant astronomical cataclysms.  The water is
inundated by cosmic rays.  In every atom of every molecule, continual
weak and strong nuclear reactions manifest themselves.  And on and on
and on.

I think we tend to forget in philosophy, we lose sight of, how complex
real physical systems are and what a miracle it is that our almost
naively, recklessly simple-minded theories, and the childishly
sketched models we construct in those theories, can still capture them
with astonishing accuracy, and do so in ways, moreover, that seem to
give us real understanding of the nature of the world in a broader
sense.  And correlatively, we forget how much distance there is
between those simple models we do know in any given theory and the
real physical systems they purport to represent, and so we forget how
many of the theory's other models would do the representational tasks
required and \emph{prima facie} do them better (by adding ever more
finely grained detail, for example), if one of them does at all---how
many of those models that we have no idea how to construct in any way
graspable by the human mind, or indeed even how to identify if someone
gave us one gift-wrapped.\footnote{To guard against misunderstanding,
  I want to emphasize the difference between, on the one hand,
  knowing---articulating, identifying, grasping---individual solutions
  to a theory's equations and, on the other, knowing propositions that
  state general properties of classes of solutions.  We know a great
  many such propositions for every theory in physics.  Although we
  know a vanishingly small number of exact solutions to the Einstein
  field equation in general relativity, we also have in hand a great
  number of deep, extensive and powerful theorems that characterize
  generic properties of large classes of solutions, none of which need
  be known in its peculiarity.  The singularity theorems of Penrose,
  Geroch and Hawking are excellent examples: any spacetime that
  satisfies a few generic conditions (positivity of energy, for
  example, and lack of causal pathology) possesses incomplete,
  inextendible causal geodesics.  (See, \emph{e}.\emph{g}.,
  \citeNP{curiel-sing} for a detailed exposition of the theorems,
  geared towards philosophers.)  Such propositions do not help the
  proponent of what I will call the standard view---sketched
  immediately below---for they do not allow one to grasp individual
  models one does not already know.}

The standard view today in much if not most of philosophy of science
in general, and philosophy of physics in particular, however,
especially since the seminal work of
Suppes~\citeyear{suppes-mngs-uses-mods,suppes-mods-data},
presupposes that a theory is characterized in a strong sense by its
family of ``models'', where a model is always meant to be something
that, at a minimum, captures the essence of a solution to the theory's
equations of motion or dynamical field equations.  Crucially,
moreover, it is presumed that those models have intrinsic physical
significance accruing to their formal structure independent of any
consideration of how the theory is applied in practice.  It presumes,
in other words, that we can cleanly separate one component of the
scientific knowledge a theory embodies, that encoded in the theory's
formal machinery, from all other components, including that involving
the practical application of the theory by experimental scientists.

This is true whether one hews to the semantic view of theories
\cite{suppe-srch-underst-sci-thry,fraassen-sci-image} or the
Best-Systems picture \cite{cohen-callender-better-best} or a
semantics based on possible worlds
\cite{lewis-def-theor-terms,butterfield-dual-equiv-phys-thrs}, or
one is a neo-Carnapian \cite{demopoulos-log-phil-leg}, or a
structuralist
\cite{stegmuller-struc-view-theors,costa-french-models-sci-reason},
or a neo-Kantian \cite{friedman-dyns-reason}, or one tries to
reconcile the syntactic and the semantic views by the use of category
theory \cite{halvorson-tsementzis-cats-sci-theor}, or one uses
category theory directly to embody the models of the theory
\cite{weatherall-cats-class-st-thrs}, or one champions a
sophisticated syntactic view---declared dead many times during and
after a long period of mordant vilification---on its own
\cite{lutz-right-synt-approach}, or one adopts any other of the
contemporary popular accounts of scientific theory.  The presumption
of the clean separation of different components of knowledge comes out
clearly in every one of these examples.  The semantic view treats its
models as having ``empirical content'' while explicitly denying that
those models need to represent actual experiments in order to have
such content.  The syntactic view introduces reduction sentences or
coordinating principles or correspondence rules to relate explicitly
demarcated languages, the theoretical and the observational.  Those
who have used category theory have treated the objects and morphisms
of the relevant categories as representing physical systems and their
behavior based solely on the postulation of a few minimal interpretive
principle that scrupulously avoid mention of experiment.  And so
on.\footnote{I find, moreover, that philosophers of physics who do not
  work directly on the semantics and structure of theories, but rather
  work on problems that either directly or implicitly require an
  account of a semantics, without hesitation, indeed usually without
  comment or (I suspect) thought, employ the standard view.  It is in
  this sense as well that I intend the claim that the view I sketch is
  ``standard'' today.  It may well be that very few today believe the
  standard view as I sketch it---but then I cannot understand the
  arguments they do make, \emph{e}.\emph{g}., in trying to articulate
  a criterion for theory equivalence based on formalism alone, with
  vague gestures at how it is all to make contact with empirical
  content as an afterthought.}

In light of my remarks about the paucity of our knowledge of actual
solutions to our theories' equations, therefore, to accept the
standard view is to accept one of two options:
\begin{enumerate}
  \noitemsep
    \item we stipulate by fiat that there is a family of actual (or
  possible, in some sense) physical systems, though we cannot identify
  most of them in practice, that are the ones represented by the
  solutions, and those are the models;
    \item we stipulate by fiat that the family of solutions themselves
  (\emph{e}.\emph{g}., vector fields on phase space), though we cannot
  represent them in anything like closed form or even identify almost
  any of them if they were to bite us on the ass, are the models.
\end{enumerate}
Call this `the standard view's dilemma'.  In both cases, we are saying
that the overwhelming majority of the content of the theory is
something that we do not know, are in fact nowhere near knowing, and
have good reason to think we will never know in anything resembling
thoroughness and detail---epistemologically speaking, either choice is
an act of faith.\footnote{I thank an anonymous referee for this
  marvelous phrase.}  In other words, it is---should be recognized to
be---utterly mysterious how the models are supposed to have intrinsic
physical significance.  Nonetheless, the story goes, there is a
preternatural relation of subsistence among the formalism, the models,
and the physical systems that unambiguously fixes them; and that in
turn gives rise to a magical relation of ``reference'' or
``representation'' that zaps out from the formalism to the world and
latches on to the salient physical systems in a physically significant
way.  We, in our supernal cognitive puissance, just \emph{know} this.

We are talking here of a Fichtean faculty of pure intellectual
intuition that grasps without cognitive, perceptual or practical
mediation the Kantian \emph{Ding an sich}.  Or more precisely, we are
talking of nothing at all.  A semantic relation that is not known and
cannot be known by humans in our current epistemic circumstance has no
possible use in actual science and has no possible bearing on a
fruitful and illuminating analysis of scientific knowledge.  

My rhetoric may make it sound as though only the realists are subject
to my wrath, but that is not so.  I intend my wrath to be directed as
well at all philosophers beholden in any way to something like the
standard view, whether of a realist or an anti-realist bent.  What is
this fairyland of formal models that are the semantic tools the
instrumentalist uses to make his predictions?  All those who hew to
something like the standard view, no matter how they think of the
character of the semantic relations between a theory and the world in
detail, have the same general problem: what is the family of formal
structures that ``characterizes'' the theory, what are the physical
systems the theory appropriately and adequately treats, what is the
relation between them, and how can we grasp all this in a
comprehensive way?\footnote{I characterize my technical notion of
  ``appropriate and adequate'' in \S\ref{sec:breakdown-regimes}
  below---for the time being, I use it as a term of art.  I use
  `propriety' as the nominal form of `appropriate' when I intend it in
  this (to be characterized) technical sense.}  That is to say, which
formal structures of the theory are even such as to be appropriate to
serve as adequate representations of parts of the world, and how can
we know this, given our woeful lack of knowledge about the details of
the formal structures and of the world?  For not all formal models in
the theory can designate or represent in a meaningful way all physical
systems we naively think they should be able to, though we have no way
of determining this in general.  \emph{What is the theory?}

We need a mechanism or device for bringing a precisely characterizable
situation in the world of experience into substantive, physically
significant contact with the mathematical structures of our theories.
This mechanism would be an intermediary that identifies the junctions
where meaningful connections can be made between the two and embodies
the possibility of the epistemic warrant we think we construct for our
theories from such contact and connection.  That mechanism, I shall
argue, is the schematic representation of the observer.  To quote
\citeN[pp.~649--650]{stein-struct-know}, who introduces the idea:
\begin{quote}
  In actual fact, the experimental physics is treated separately as a
  discipline in its own right, that is partly an art: an affair of
  both knowledge and manipulative and perceptual skill.  But the
  possibility of connecting this art with the theory is closely
  connected with a certain possibility within the mathematical
  structure that is the theoretical framework: \ldots\@ the
  possibility of representing experiments, and of representing the
  observer, ``schematically.''  \ldots\@ I want to speak \ldots\@ of
  ``schematizing the observer within the theory''; \ldots\@ the
  intention is \ldots\@ to secure empirical content---content within
  experience---for an abstract structure.
\end{quote}
\label{pg:stein-schema-quote}

The standard view gets right the fact that the family of physical
systems a theory appropriately and adequately treats is a central,
essential component of the epistemic content of a theory.  The
standard view goes wrong in thinking that it makes sense to
characterize this family in the abstract, without recourse to and
reliance on all the scientific knowledge we have pertaining to the
theory and its applications.  To identify that family of systems, we
need to demarcate the theory's regime of applicability.  To do that,
we need to fix error tolerances, acceptable levels of precision,
\emph{etc}., in our experimental practice.  To do that, we need real
models of actual experimental techniques, representing the actual
tools we have at our disposal for performing such measurements, for
the regime of applicability changes over time with advances in
practical knowledge---we need to schematize the
observer.\footnote{Although the recent wave of philosophical work on
  practice in science, exemplified by such thinkers as Margaret
  Morrison and Hasok Chang, has much in common with the views I
  propound here, I do not explicitly engage with the literature.  It
  is too vast, and often the motivations and aims of the work differ
  too much from those of this paper, for me to do so without expanding
  the scope of the paper beyond what it can profitably bear.  For the
  same reason, though it pains me, I do not directly engage with the
  literature on complexity in science, such as the marvelous work on
  generative entrenchment, levels of organization and causal thickets
  by William Wimsatt.  And again, I cannot engage with the interesting
  recent work being done on measurement by philosophers such as Eran
  Tal and Alistair Isaac.  As to why I am not situating this work in
  the current literature in a more standard way, but rather blowing it
  all off with brusque regret, I can say in my defense only two things
  to the anonymous referee who pushed me on it: first, this paper
  faces many more large issues taken at once than is standard in a
  philosophy of science paper these days; second, it most often does
  so in idiosyncracy.  It would take a book for me to work out how
  these ideas compare to---in their similarities and differences, in
  their virtues and demerits---the most influential contemporary
  accounts of all these issues.}

I give more detailed and extended arguments in
\S\ref{sec:need-schem-obsr}, and throughout the paper after, to defend
these claims.  I give here a simple, short, yet powerful one.  If
demarcating the regime of applicability did not depend on practical
knowledge of the kind that changes, and that cannot be encoded in the
pure formalism of the theory---the current reach of experimental
technique that determines fixed levels of precision and accuracy, the
availability of approximative theoretical techniques and heuristic
forms of argument to make calculations tractable, and so on---then
there would be a fixed spatial length, once and for all---say
3.14159265358{\ldots}\texttt{x}10$^{-3}$ \texttt{cm}---at scales
smaller than which the representational capacities of Navier-Stokes
theory breaks down.  But that is simply false.  It breaks down at
different scales for different types of fluids under different
conditions.  The increasing acuity and resolution of our measuring
instruments, moreover, in conjunction with our increasingly powerful
control of experimental error, and the increasing sophistication of
the models we can construct of our measuring devices and the fine
details of their interactions with different Navier-Stokes fluids
under different conditions, all mean that such numbers become ever
smaller.  They are not fixed \emph{sub specie
  {\ae}ternitatis}.\footnote{Although it is, lamentably, a theory that
  has received almost no philosophical attention, I will use
  Navier-Stokes theory as a recurring example to illustrate and
  substantiate my arguments throughout the paper.  I do this for
  several reasons, among the most important to my mind being that: it
  provides excellent models of complex, real phenomena we are all
  familiar with (water---\hto---under normal circumstances, for
  example); it is, with respect to physical content, a well understood
  theory; and it is a straightforward theory, in the sense that its
  physical content can be explicated without the use of any heavy
  technical machinery.  (The \emph{loci classici} for the general
  theoretical treatment of hydrodynamic phenomena are
  \citeNP{lamb-hydro} and \citeNP{landau-lifschitz-fluid}, and
  that for issues of stability and equilibrium in particular is
  \citeNP{chandrasekhar-hydro}; see, \emph{e}.\emph{g}.,
  \citeNP{pope-turb-flows} and \citeNP{foias-et-nse-turb} for a
  treatment of non-equilibrium fluid phenomena such as turbulence.)
  Nonetheless, the full richness and complexity of the theory, both
  mathematically and physically, lie well beyond us---we have only a
  poor grasp of the mathematical character of generic solutions to the
  Navier-Stokes equations, and many of the phenomena most
  characteristic of Navier-Stokes fluids, such as turbulence, remain
  among the most puzzling and mysterious that we are aware of.  As
  such, its philosophical contemplation readily throws up interesting
  questions and problems that philosophers' obsession with quantum
  theory and general relativity blind them to.}  (I explain the ideas
of a breakdown scale and the regime of applicability of a theory more
fully in \S\ref{sec:breakdown-regimes}.)

Thus, neither is the family of physical systems that a theory
appropriately and adequately treats fixed \emph{sub specie
  {\ae}ternitatis}.  I think this is right.  A scientific theory is
not a Platonic form.  It does not exist independently of our actual
epistemic state.\footnote{Even though I insist on the importance of
  attending to the complexity of the world in contraposition to the
  simplicity of our theories, I reject work such as
  \citeN{cartwright99}, which wants to do away with theories in favor
  of something like a patchwork of models.  Formal frameworks and
  theories \emph{unify} all those disparate models, providing a common
  structure they all instantiate, which can serve as the groundwork
  for both philosophical and physical investigation.  This is one
  important role for formalism.  My discussion, in
  \S\ref{sec:what-theory-is}, of the many possible models of the black
  hole SgrA$^*$ illustrates how this works: without general relativity
  serving as the common work-room in which all the models are
  constructed, examined and then put to further use, providing the
  common context that allows each to be compared in a meaningful way
  with all the others, it would be difficult if not impossible to
  understand them all as models of the same physical system or as
  making reference to the same physical quantities in different ways.}

This is a long and dense paper, and the overall argument has not leant
itself to a clean, simple, linear exposition.  I therefore sum up in
an appendix (\S\ref{sec:precis}) the main claims of the paper, along
with my main criticisms of what I call the standard view of
theories.\symbolfootnote[2]{This paragraph, and the appendix, do not
  appear in the published version of the paper, due to length
  constraints.}

\section{What Is a Theory?}
\label{sec:what-is-theory}

What is the theory of general relativity?  Here is one way to make
that question somewhat more precise: what is the more or less secure
fund of scientific knowledge that constitutes the theoretical and
empirical content of the theory---its \emph{epistemic
  content}?\footnote{If one is permitted to speak of explicating a
  question, one may consider this a first step on the way to a precise
  explication of the question ``What is a physical theory?''}  This
formulation of the question is certainly not a standard one in
philosophy of science.  On that standard view, the emphasis is on
constructing formal models corresponding to something like solutions
to the field equations or equations of motion of a theory, and perhaps
asking when two such families of models (or other formal structure)
are isomorphic in a (supposedly) relevant sense, with perhaps a
gesture at the possible ``representational capacities'' of the
formalism as an
afterthought.\footnote{\citeN{weatherall-cats-class-st-thrs} is a
  classic example.  I do not single this paper out because I think it
  is a poor one; to the contrary, I think it is a beautiful paper,
  rich with important insights.  It does, nonetheless, serve as a
  useful foil to my views, in part because of its clarity and depth.}

I do not like a formulation of the question that admits as an
appropriate answer such a formal structure, along with perhaps the
fixing of something like a Tarskian semantics or a sketch of
representational capacities.  A formulation of that kind assumes, with
no argument ever given, that it is possible to make a clean separation
between, on the one hand, one part of the scientific knowledge general
relativity embodies, \emph{viz}., that encoded in the pure formalism
and, on the other, the remainder of that knowledge.  The remainder
includes at a minimum what is encoded in the practice of modeling
particular systems, of performing experiments, of bringing the results
of theory and experiment into mutually fruitful contact---including
all the \emph{in}exact mathematical methods that cannot be formalized,
approximative and heuristic techniques motivated by loose physical
arguments and principles not part of the formalism or of any
interpretive postulates, and justified only by practical
success---including, in sum, real application of the theory in actual
scientific practice.\footnote{Because I shall often cite
  ``approximative techniques'' as an exemplar of scientific knowledge
  that cannot be encoded in a theory's formalism, I want to clarify
  what I mean.  There are often rigorous and well developed methods of
  approximation available that lend themselves to comprehensive
  formalization (\emph{e}.\emph{g}.,
  \citeNP{fillion-corless-back-err-anal-pertbn}).  One sometimes
  even has formalized methods for controlling the errors such
  approximative methods introduce (\emph{e}.\emph{g}.,
  \citeNP{fillion-corless-epist-mod-comput-err}).  Nonetheless, not
  all approximative and heuristic arguments can be formalized.  I
  discuss in \S\ref{sec:forms-aspects-know} an example, to wit,
  arguments to the effect that Navier-Stokes theory is essentially an
  equilibrium theory.}  This assumption comes out most clearly in the
picture of semantics that naturally and usually accompanies the
standard view of theories: semantics is fixed by ontology's shining
City on the Hill, and all epistemology and methodology and other
practical issues and considerations are segregated to the ghetto of
the theory's pragmatics.

We should not assume without argument that the epistemic content of a
theory can be cleanly and exhaustively divided into two parts, one
consisting of what is encoded in the pure formalism alone and the
other a catch-all for the rest, including the mess of real
application.  Much less should we assume without argument that two
such parts, no matter how characterized in detail, can be cleanly and
exhaustively segregated from each other in an adequate analysis of
that epistemic content.  That, however, is what the standard view
does.  Call this the `segregation problem'.

If one accepts my formulation of the question as fruitful and
interesting, then it becomes plausible that the theory of general
relativity does not---can not---in any interesting sense consist of
the theory of Lorentzian 4-manifolds along with a few interpretive
postulates.  Now, it may turn out that a compelling analysis
attempting to answer my formulation will show that, in the end, the
standard assumption of the clean segregation of the relevant domains
of knowledge---the ``theoretical'' and the ``practical'', for lack of
better terms---is a good and justified one.  I personally doubt it.

I will argue that the interplay between theory and practice
(experimentation, observation, modeling, heuristic and approximative
forms of argument, \ldots) is far more subtle, nuanced, complex and
rich than many if not most philosophers steeped in formal traditions
tend to appreciate, and in particular that the two are inextricably
intermingled in a strong sense.  The sum total of scientific knowledge
constituted by the theoretical and practical content of the
theory---its epistemic content---includes knowledge we have gained,
and more importantly could have gained only, through practical
modeling, experimentation and observation, heuristic and approximative
techniques of argument, and in particular the application of
representations and models of such experiments and observations in our
theory by use of heuristic and approximative techniques of argument.
It contains such knowledge, moreover, in a way that cannot at bottom
be cleanly segregated from any knowledge grounded in the theory's pure
formalism.  That is why I think my formulation of the question based
on the idea of knowledge is a fruitful sharpening and refinement, for
it is this sum of epistemic content that distinguishes general
relativity as a \emph{physical} theory from a merely mathematical
structure.  (It also, in the end, is what must serve as the decisive
basis for any criterion by which we judge general relativity to be
both a different and a better theory than, \emph{e}.\emph{g}.,
Newtonian gravitational theory.)  I emphasize from the start that my
conception of knowledge here is not limited to anything like
``justified true belief'', but includes more importantly comprehension
and understanding of the sort that often accompanies explanatory
practice and inspires novel discovery (on which, more below).

I will further argue that a crucial component of that epistemic
content consists in knowing how to schematize the observer in
theoretical representations of experiments, and cannot be grasped
without models of experiments in which the observer is schematized.
By ``schematize the observer'', as will become clearer, I mean
something like: in a model of an experiment, to provide a
representation of something like a measuring apparatus, even if only
of the simplest and most abstract form, that allows us to interpret
the model \emph{as} a model of an experiment or an
observation.\footnote{\citeN[ch.~12]{fraassen-sci-rep}, for example,
  an exemplar of what I am calling the standard view, explicitly
  argues for the opposite conclusion, even though he attempts to
  construct what he argues to be a pragmatist account of
  representation of physical systems by formal theory.  Indeed, in
  \citeN{fraassen-mod-meas-emp-grnd} he claims, rightly to my mind,
  that the empirical grounding of theoretical quantities as they
  appear in a theory's models must involve an account of how those
  quantities are ``related to measurement'' (p.~773).  He then argues
  that criteria for what counts as a measurement must display the way
  that that measurement procedures are theory-dependent.  He does not,
  however, explain how this can be done if the theory's models do not
  at least have the capacity to represent measurements.}  The force of
the segregability problem thus lies in the claim that one cannot
characterize and account for any of the epistemic content of a theory
without schematic representation of the observer.  As I discuss below,
even such ``trivial'' schematic representations as a line of sight for
telescopic observations in astronomy can allow one to determine and
illuminate important and substantial parts of the way that theory and
experiment come into contact, in this example by way of grounding an
analysis, \emph{e}.\emph{g}., of stellar aberration.  Such schematic
representation, moreover, is not possible in a physically significant
and cognitively substantive sense without recourse to all the other
components of what I am calling the practical part of scientific
knowledge, whose explication I give in \S\ref{sec:forms-aspects-know}
below.

\section{Why We Need to Schematize the Observer}
\label{sec:need-schem-obsr}

Because the meaning of scientific terms and propositions must rest on
the knowledge we have of the physical world, and most of all on the
knowledge we have gained through controlled observation and
measurement, that is, through experiment, epistemic content accrues to
a scientific theory in no small measure through the construction of
empirically successful representations of physical systems.  At
bottom, then, what secure epistemic content a scientific theory has
must rest in large part on the meanings expressed in the sound
articulation of experimental knowledge, for that is the final arbiter
of empirical success.  This requires at a minimum that we be able, at
least in principle, to construct appropriate and adequate
representations of actual experiments and observations in the
frameworks of our best scientific theories, that is, representations
of physical systems and experimental apparatus in relation to each
other as required by actual experiments, not just representations of
physical systems \emph{simpliciter}, in abstraction from experimental
practice.\footnote{Indeed, it is our incapacity to do this in a
  consistent way in the context of quantum theory that lies at the
  bottom of the Measurement Problem.  This alone shows the importance
  of the idea.}  It is characteristic of the ways that such models are
constructed and applied, moreover, that they require methods and
warrant---forms and aspects of knowledge---that go far beyond what can
be claimed to be captured by the theoretical formalism of a theory
alone, no matter how many representational capacities one ascribes to
it or interpretive postulates one affixes to it.

What is at issue here is not the relation of representation itself,
which can at most tell us ``what the world would be like if the theory
were true or largely true of it'', but rather the understanding and
comprehension we have of the world in so far as we are warranted in
thinking of our theories' models \emph{as} representations, which goes
far beyond a description of how the world would be under given
conditions (even if one generously admits ``nomological structure''
for inclusion in the description).  Such understanding and
comprehension come from arguments and investigations that cannot
always be ``represented'' by the formalism of ``the theory itself'',
but rather requires for their elucidation a rich and nebulous halo of
inexact mathematical techniques and heuristic physical principles
applied in ways that do not lend themselves to clean formalization.

There are, as I have intimated, several reasons why I claim that
schematizing the observer is required for an adequate philosophical
analysis of the structure and semantics of theories.  I discuss now a
few of them in more detail.  The first is a shallow but still
important one: sometimes the nature of the observational process
itself results in ``distortion'' of the magnitudes measured, and a
proper computation of the ``real'' values of the magnitudes of the
system's properties requires explicit modeling of the interaction
between the measuring instrument and the system itself to correct for
the effect.\footnote{For simplicity, construe this claim only in the
  context of non-quantum theories; the same reason applies in many
  cases in quantum mechanics, but there an appropriate analysis
  requires far more mathematical, physical and conceptual
  sophistication and subtlety than I have room for in this paper, and,
  more to the point, than I feel capable of giving a sound
  philosophical articulation of, given our current lamentable state of
  understanding of quantum theory.}  Call this `the problem of seeing
through a glass darkly'.\footnote{I thank Jeremy Butterfield for
  suggesting this excellent moniker.}  An example is stellar
aberration: when light from a star enters a telescope, the motion of
the telescope transverse to the path of the light while the light
traverses the telescope (because of the diurnal rotation of the Earth,
the orbital motion of the Earth around the Sun, and so on) makes the
star appear displaced from its actual position in the sky; in order to
correct for the effect, one must compute the actual motion of the
measuring device, which requires an explicit representation of it in
one's model of the observation.\footnote{See any good book on
  astrometry, such as \citeN{kovalevsky-seidelmann-astro}, for a full
  treatment of aberration and the details of computing corrections for
  it.}

The second reason is a middling deep one.  The quantitative results of
all measurements and observations, even the best ones, deviate from
those predicted by theory, even if only by a small amount; likewise,
there is an inevitable imprecision in the measured values.  Such
deviations largely accrue to measurements on account of systematic
errors arising from the idiosyncracy of the particular experimental
apparatuses used and the ways they are configured and deployed during
the measurement process; the imprecision is an inevitable artifact of
the limited acuity of any probe.\footnote{I ignore error arising from
  unknown sources, ``noise'', which raises its own fascinating set of
  problems, well beyond the scope of this paper.}  In order to compute
reasonable values for the expected errors and imprecision (so as, for
example, to be able to say when a measured result differs by an
inadmissible amount from a theoretically predicted result), one must
often take account of fine details of the measuring apparatus and the
particulars of its coupling to the system under study in one's model
of the experiment.  Call this `the expected error problem'.
Thermometry provides an excellent example: different sorts of
thermometers (bulbs of gas, pyrometers, inhomogeneous thermocouples,
\emph{et al}.\@) couple to systems in radically different ways, the
fine details of which must be handled on a case-by-case basis in order
to correct for the effects of such phenomena as convective currents in
fluids.  (I discuss the example of temperature at greater length at
the end of this section.)

The third reason is a deeper one.  Theory must be able to provide
guidance to experiment in the design of new types of tools for probing
novel sorts of phenomena, in the design of new types of tools for
probing known phenomena in novel ways, and in the design of the
experimental configuration itself, \emph{i}.\emph{e}., how the
instruments are used; conversely, experiment must be able to provide
guidance to theory in modeling practically constructed novel ways of
coupling to systems known and unknown so as to place constraints on
the possible soundness of theoretical description and prediction.  In
order for each to carry out these tasks, theory must be able to
represent the fine details of the apparatus as it is to be used in the
experiment.  This particular interplay between theory and experiment,
in theoretical guidance in the construction of instruments and the
design of experimental configuration on the one hand, and in
experimental constraint on the soundness of theory on the other, is
one of the most profound ways that theory and experiment are able to
make contact with each other; without it, it is difficult to see how
any epistemic content could accrue to theory in the first place.

The search by Hertz for ways to produce and detect free
electromagnetic waves as predicted by Maxwell's theory provides a
beautiful illustration of the delicate dialectic required between
theory and experiment, especially in the construction and modeling of
instruments in the attempt to produce and probe a phenomenon so poorly
understood.\footnote{See \citeN[\emph{passim}]{hertz-electric-waves},
  including the preface by Helmholtz, for an absorbing account.}
During most of the career of the investigation, Hertz had little idea
what sorts of arrangements of what sorts of physical system would
produce electromagnetic waves in the first place, and even less of
what sorts of instrument could reliably detect them.  His search
necessarily included the construction of detailed models both
experimental and theoretical, each guiding the other in turn, of
different proposed methods of coupling the electromagnetic field to
its environment and different instruments to try to realize those
couplings.\footnote{\citeN{buchwald-creat-sci-effs-hertz} provides a
  detailed and comprehensive exposition of Hertz's search for an
  instrumental design and a theoretical representation that jointly
  would do the job.  There is a close affinity between my own
  arguments and Buchwald's conclusion that, in the end, Hertz required
  only a highly abstract, ``schematized'' representation of the
  instrument---the dipole oscillator---for his purposes, and yet he
  still did need an explicit representation of it in his theoretical
  work, no matter how ``simple'' and abstract.  I take the lesson to
  be that Hertz needed that schematized representation in order for
  the theoretical and the experimental parts of his work to make
  substantive contact with each other.}  Of course, everyone at the
time knew that one could derive the wave equation for the
electromagnetic field from Maxwell's equations, and they knew the
plane-wave solutions and the principle of linear superposition for
electromagnetic fields, but I would say that, at that point in the
history of Maxwell theory, the theory did not include a
characterization of electromagnetic radiation with substantive
epistemic content.  In one sense, as Hertz famously remarked,
Maxwell's theory may be his equations, but the epistemic content,
including the practical knowledge, that characterizes those equations
\emph{as} a theory of \emph{electromagnetism} does not exist in a
meaningful way until we understand it well enough to use it to make
testable predictions---but a prediction is not testable if we don't
know in principle how to test it!\footnote{This all too brief
  discussion of Hertz bears profitable comparison with Harry Collins'
  idea of the experimenter's regress; see, \emph{e}.\emph{g}.,
  \citeN[ch.~4]{collins-chng-ordr-replic-induc}.  In contradistinction
  to Collins, however, I consider this relation between theory and
  experiment to be not only unproblematic, and in particular not
  circular in the sense of leading to a regress, but rather necessary
  for the progress of science.  I am thus more aligned in some ways
  with the views of \citeN{franklin-neglect-exp}, many of whose
  criticisms of Collins I endorse.}  This is not verificationism about
meaning.  It is rather the simple point that, if we do not know how to
bring a theoretical structure into contact with the world by
experimentation or observation, then we have little or no grounds for
trying to understand that structure as representing anything in the
world.\footnote{See also the philosophically rich essays by Maxwell
  \citeyear{maxwell-intro-lect-exp-phys,maxwell-sci-apparat}
  himself on this necessary sort of interplay between theory and
  experiment.}

The fourth reason is the deepest, and depends in part on the previous
three.  Without an understanding of the fine details of the way that
theories do and do not successfully model actual experiments on the
kinds of physical systems they purport to treat, we have no way to
demarcate the regime in which the theory does in fact appropriately
and adequately apply for the representation of those kinds of
systems---its \emph{regime of applicability}.  It is only by dint of
the practical forms and aspects of scientific knowledge, in other
words, that we can determine a theory's regime of
applicability---when, \emph{e}.\emph{g}., what we pre-theoretically
conceive of as a body of fluid, manifesting some dynamical behavior
(say, what looks like turbulence), is appropriately and adequately
treated by Navier-Stokes theory and when it is not.  Knowledge of that
regime, however, embodies perhaps the most important part of the
epistemic content of a theory---without it, we have no warrant at all
for trusting our use of the theory, much less for understanding it
\emph{as} a theory of a particular class of physical systems.  Nothing
in the theory's formalism can tell us this, without the input of
knowledge that can be had only by experimental practices that require
schematizing the observer in theoretical models.\footnote{Although it
  does not bear on the question of how the schematic representation of
  the observer informs and is required by work in physics itself, I
  feel it is important to note that deep philosophical investigation
  of the foundations of physical theories also at times relies on it.
  The work of \citeN{geroch-pred-gr},
  \citeN{malament-obser-indist-sts}, and
  \citeN{manchak-know-glob-struc-st} on predictability in and
  observational indistinguishability of relativistic spacetimes, and
  the work of \citeN{malament-rot-nogo,malament-rel-orbit-rot} on
  defining relative rotation in relativistic spacetimes, are exemplary
  in this regard: the analyses and arguments of each requires
  schematizing the observer as a timelike worldline.  The arguments
  could not be made without reference to the possible observations of
  such schematized observers.}  Much of the rest of the paper will be
dedicated to working this idea out.

\section{What a Theory Is}
\label{sec:what-theory-is}

What, then, is the theory of black holes?  It cannot be the subset of
Lorentzian 4-manifolds with non-trivial event horizon.\footnote{I put
  aside here the fascinating question of what physicists mean in the
  first place by the idea of a black hole, the manifest fact that
  different fields of physics, and often even different physicists in
  the same sub-field, have different, often mutually inconsistent, but
  still closely related definitions of `black hole'
  \cite{curiel-many-defns-bh}.  This phenomenon, by the way---the
  existence of different, even inconsistent definitions for the ``same
  type of thing''---is ubiquitous in physics.  To my mind, this
  strongly suggests already that a semantics for theories based on
  designation as the fundamental semantic relation cannot suffice for
  a representation of our knowledge in physics.}  That cannot tell us
that the Milky Way has at its center a black hole of about 4 million
Solar masses, about 88 million kilometers across, referred to by
astronomers as `Sagittarius A*' (abbreviated `SgrA$^*$', pronounced
`saj ay-star')---even putting aside the fact that it is likely that
the thing has nothing associated with it resembling an event horizon
in the standard, formal sense.  Even if we throw in interpretive
postulates or a Tarskian-like semantics, it cannot tell us how we know
this, why we think we are justified in believing it, the ways we may
be justified in using it as evidential warrant for further knowledge
claims, and so on---all of which counts as part of the epistemic
content of the theory.

To this observation, an advocate of the standard view may reply that a
particular prediction of the theory starting from particular initial
data is not ``part of the theory'' in any interesting sense---we must,
she will say, sharply distinguish between, on the one hand, the formal
structures of the theory that encode its form of nomic possibility
and, on the other, the possibly realized individual physical systems
characterized by contingent initial or boundary conditions.  Nonsense,
I say.  First, an \emph{ad hoc} argument: on the standard view,
SgrA$^*$ \emph{is} one of the Tarskian models, possible worlds, or
what have you, that characterizes the theory.  Thus scientific
knowledge that we have about SgrA$^*$ forms part of the theory even on
the standard view.  That leads to the deeper response, to point once
again to the question whether one can cogently articulate the
character and content of that knowledge in isolation from the other
forms or aspects of scientific knowledge, the more practical ones, we
have about SgrA$^*$: one cannot identify the putative Tarskian-like
model of the theory, or find the correct application of the
interpretive postulates required to identify a solution to the field
equations as representing SgrA$^*$, in isolation from the practical
knowledge used to characterize it as a physical system amenable to
representation by general relativity.  Standard interpretive
postulates such as ``freely falling particles traverse timelike
geodesics'' are inadequate for the job.

The problem for the standard view is even more severe than these
remarks suggest.  Criteria for what we are to count as part of
SgrA$^*$ and what not to count for the purpose of tallying the
theory's models are never specified by advocates of the standard view.
Should we include the accretion disk?  The mini-cluster of rapidly
orbiting stars immediately surrounding it?  The frozen magnetic field
surrounding the horizon and ionized plasma jets shooting along it?
Does adding or subtracting each of those yield a new model of the same
target system or just a model of a different target system?  Does
including inhomogeneities in the accretion disk or the magnetic field
or the plasma count as refining an existing model or as creating (or
identifying) a new model?  What about accounting for the tidal
deformation of the orbiting stars?  And so on.  And in each case, once
we have decided, then what type of model are we to use to represent
the phenomena in question?  One using magnetohydrodynamics to model
the accretion disk, or a statistical mechanics of charged plasma?
Both are encompassed in the framework of general relativity.

In our current state of knowledge---given the actual epistemic content
of general relativity as we know it---there is no single,
unambiguously correct answer to those questions.  Different ways of
modeling the phenomena yield different families of solutions to the
Einstein field equation in general relativity, none of which are
related to each other in any simple way: none of the families
``reduce'' to the others as one lets some parameter smoothly go to
zero, none of the families are subsets of the others, none of the
families non-trivially intersect all the others, and so on.  One can,
for example, depending on one's purposes, treat the accretion disk as
an uncharged dust or fluid or as a charged plasma, and, if the last,
one can either treat the plasma as a source or treat it as not a
source (``test matter'') for the ambient magnetic field.  This will be
justified, \emph{e}.\emph{g}., by a solid practical understanding of
the aspects, features and components of the system one's observational
instruments will couple to in the kinds of measurements one plans to
make.  For some wavelengths of electromagnetic radiation one aims to
detect, one can ignore the inhomogeneities and anisotropies in the
magnetic field arising from treating the plasma as a source, and for
others not, depending, \emph{inter alia}, on whether those
inhomogeneities and anisotropies are comparable in scale to the
wavelengths one plans to study.  Whether one treats it as charged is
irrelevant, for example, if one is making measurements in the
gamma-ray wavelength spectrum, as the electromagnetic radiation of
that type produced by the system is insensitive to the differences.
Similarly, when one treats the constituents of the accretion disk as
uncharged, then whether one models it as a dust or as a fluid is
irrelevant if one is making observations in the x-ray band, as the
system's generation of x-rays is insensitive to the
differences.\footnote{I struggled to find a few references to cite to
  substantiate these claims.  I failed.  This is not the sort of
  knowledge explicitly presented in research papers nor in textbooks
  (though it is often implicitly there, if one knows how to read
  between the lines).  I learned it, as all those in the field do, by
  asking questions at advanced research seminars and colloquiums and
  having detailed discussions with knowledgeable practitioners.  The
  discussion in \S\ref{sec:forms-aspects-know} of the relevance of how
  one learns different forms of knowledge will clarify the import of
  this remark.}

In all these cases, it may seem as though the one type of model is
just a simplification of the other, as (say) one allows some quantity
or coupling to go to zero.  This is not so.  One may think that the
models based on dust are just the fluid models in which the pressure
and viscosity are taken to zero.  The procedure just described,
however---``let pressure and viscosity go to zero''---is ambiguous, as
there are many ways to construct such a limiting family in general
relativity that each yields a different kind of spacetime as the limit
\cite{geroch-lim-sts}.  In any event, the ``obvious'' ways of doing
it give the wrong answer in some cases: it is a theorem of
\citeN{ellis-dyns-press-free-matt-gr} that if a pressureless dust is
shear-free, then either its twist or its expansion must be zero.  Many
fluid models of accretion disks, however, are both shear-free and have
non-vanishing twist and expansion, but the theorem tells us that there
are no dust models that correspond to such behavior.  (Recall that the
Einstein field equation is non-linear, so the behavior of the limits
of such procedures as ``take this parameter to 0'' cannot always be
simply ``read off'' the mathematics in a naive way.)  There is no
systematic relation between the dust and the fluid models of the
accretion disk.\footnote{Indeed, even just from a mathematical
  perspective, the situation is more complex than I have sketched, and
  more dire for the standard view.  What, \emph{e}.\emph{g}., is the
  level of differentiability we demand or require of solutions?
  Depending on the answer to the questions, we may or may not be able
  to represent such phenomena as shock waves.  It is still, in fact,
  an open theoretical and mathematical question whether generic shock
  waves in the standard sense \cite{landau-lifschitz-fluid} can be
  represented in general relativity outside the regime of exact
  spherical symmetry \cite{reintjes-temnple-shock-wv-gr-met-smooth}.
  So, are representations of shock waves in non-symmetrical systems
  part of the theory of general relativity itself?  Please do not say
  there is an answer to that question in the Platonic heaven of the
  space of formal solutions to the Einstein field equation.  That
  would not answer the question about the physics., for it would
  implicitly assume that there is such a thing as ``the way'' that
  models in general relativity represent systems in the real world,
  irrespective of how complex the models and the systems may be, and
  irrespective of the methods at our disposal for probing them.}

In sum, there is no single, canonical model that represents SgrA$^*$,
nor even a single naturally inter-related family of models.  We rather
have several different, not unambiguously related families of models,
each of which captures some aspect, feature or component of the system
not captured by the others, and each of which is appropriate and
adequate for the different investigative purposes one may have---each
gives what we believe to be excellent answers to kinds of questions
that the others answer terribly.\footnote{It is perhaps a commonplace
  now in philosophy of science that the same target system can have
  multiple models, each in a different theory or ``semi-autonomous''
  or ``phenomenological'' and all in manifest tension in some way with
  each other, as forcefully argued, \emph{e}.\emph{g}., in
  \citeN{morrison-one-phn-many-mods}.  My point is different, and, to
  my mind, much more striking: this happens within one and the same
  theory.}  The propriety and adequacy of the model for many given
purposes, moreover, is determined in large part by the kind of
knowledge that can be had only by comprehension of the practical part
of the theory---what sorts of instruments with what sorts of
sensitivities are appropriate and adequate for probing and studying
those different aspects, features and components of the system, in
which states and under what sorts of conditions, and the kinds of
arguments (almost never articulable strictly in the theoretical
formalism) that show this.

There is also the fact that we do not think general relativity is at
bottom correct as a theory of spatiotemporal structure and its
relation to matter; at some point, presumably, quantum effects must be
taken into account, and we have little to no idea how to do that.  In
that sense, \emph{no} model general relativity produces can possibly
be ``the right one'' for SgrA$^*$.  Physicists almost universally
ignore this fact in their work, and with good reason: even were we to
have a satisfactory theory of quantum gravity, it is almost certainly
the case that we would be unable to use it to construct models of
SgrA$^*$ to answer the astrophysical questions about it we are
interested in.

Indeed, a crucial part of the epistemic content of the theory is the
knowledge of the physical regime in which it is appropriate and
adequate for representing a given class of phenomena at all.  As
\citeN[pp.~6--7]{geroch-hyp-rel-diss-fl} concisely puts it, in
discussing attempts to formulate a relativistic version of
Navier-Stokes theory,
\begin{quote}
  \label{pg:geroch-regime-quote}The Navier-Stokes system [of
  equations], in other words, has a ``regime of applicability''---a
  limiting circumstance in which the effects included within that
  system remain prominent while the effects not included become
  vanishingly small.
\end{quote}
The physical quantities modeled by a theory's equations have a regime
in which they are simultaneously well-defined, satisfy the theory's
equations, and have values stable with respect to fluctuations and
other effects not representable in the theory (such as quantum effects
in classical general relativity, or molecular-scale fluctuations in a
theory of fluids).  That regime itself forms part of the core of the
theory, in so far as it determines what the scope and depth of the
theory's empirical content can be.  As I'll discuss below in
\S\S\ref{sec:breakdown-regimes}--\ref{sec:sems-epist-onto}, moreover,
until one has specified a regime of applicability for the theory, such
questions as I've raised about SgrA$^*$ cannot be answered, but to
specify such a regime requires of necessity knowledge that can in part
be had only by the construction of models of experiments that include
schematic representation of the observer.  One cannot determine the
regime of applicability by examination of the formal structures on the
one hand and a clearly articulated family of interpretive postulates
on the other.  The interpretive postulates on their own cannot do the
job, for they require the systems to which they apply to fall already
within the regime of applicability; they do not determine it.

That we feel secure in claiming that SgrA$^*$ is a black hole of that
mass and size---that we believe this to be \emph{scientific}
knowledge---depends in large part on the confidence we have in the
experiments and observations that delivered the data that allow us to
identify and characterize SgrA$^*$ as a black hole, and moreover as
\emph{that} black hole.  That confidence in turn depends on the
confidence we have in the methods we use to model those experiments
and observations and to manipulate and impose structure on the raw
data collected so as to put it into a form that can make fruitful,
physically significant contact with the abstract, formal structures of
pure theory.  And much of the knowledge that underlies the experiments
and observations and constitutes their results---the confidence we
have in those results, the way we use those results as evidence in
turn for other knowledge claims, the confidence we have in doing
that---both derives in large part from and consists in large part of
the knowledge that comes from constructing a model of the experiments
and observations themselves, including schematizing the
observer.\footnote{See \citeN{genzel-et-dark-mass-ctr-milky} and
  \citeN{ghez-et-acc-stars-orbit-bh} for accounts of the detailed
  infrared studies that nailed once and for all our confidence that
  SgrA$^*$ is indeed a black hole, and in particular the way that
  models of the experimental apparatus and the observations themselves
  were crucial to the arguments.
  \citeN{collmar-et-panel-proof-exist-bhs} and
  \citeN{eckart-et-superm-bh-good-case} provide illuminating
  discussions, almost all by the physicists directly involved in the
  work, of the fascinating methodological and epistemological problems
  associated with trying to ascertain that what we observe
  astronomically when we point our telescopes at SgrA$^*$ is indeed a
  black hole.}  This also necessarily involves epistemic content from
other theories---those, \emph{e}.\emph{g}., we use to model the
measuring instruments and the environment in which the experiment or
observation is taking place---and the knowledge we have of how the
epistemic content of those other theories interacts with that of the
theory at issue.  That itself forms a critical part of the epistemic
content of the theory we are considering, how it admits input from
other theories.

A theory in my sense thus includes such things as accounts of
experimental devices appropriate for studying the relevant kinds of
physical system, good practices for employing them, sound techniques
for the collection of raw data and for the statistical and other
analysis and organization of the same, reliable methods of
approximative and heuristic reasoning for constructing models and
solving equations, guidelines for determining whether a system of the
given kind is in a state and experiencing interactions with its
environment such as to be amenable to appropriate and adequate
representation by the theory (\emph{viz}., whether or not the system
falls into the theory's regime of applicability), and, more generally,
good principles of argumentation for regimenting all these and
bringing them to bear on each other.

When I speak of ``principles of good argumentation'' as forming part
of a framework, I do not mean to imply that I think that first-order
predicate logic thereby must be part of it.  I mean rather: forms of
argument \emph{peculiar to the domain of physical systems the
  framework purportedly treats}.  For some argumentative purposes,
some ``standard'' forms of mathematical and heuristic argumentation
will not be appropriate, even though they work well in other fields of
physics.  An example is how perturbations are treated, when understood
as leading terms in a (truncated) power expansion in some dynamically
relevant parameter.  That works well for systems that are well behaved
in an appropriate sense, but not for others, \emph{e}.\emph{g}.,
chaotic systems.  Thus, the argument in the 19th Century that the
Solar System is stable because it is so up to second order in a
standard expansion representing the perturbative effects of the
planets' gravitational forces on each other (according to the standard
methods of classical celestial mechanics as inaugurated by Laplace and
Lagrange at the end of the 18th Century), though unobjectionable in
and of itself as a standard form of perturbative reasoning, fails
because, as Poincar\'e showed in his epoch-making work \emph{Les
  M\'ethodes Nouvelles de la M\'ecanique C\'eleste}, the third-order
terms exhibit chaotic instabilities (as he explained, with an eye
towards the sort of issues I am concerned with here, in the popular
essay \citeNP{poincare-stab-sys-solar}).  The lesson is that one must
take care in applying perturbative analysis to any Newtonian
gravitational systems of more than two bodies.  The recognition and
respecting of the required care is part of the canon of ``good
argumentative forms'' for that theory.  Another good example is the
use of the linearized Einstein field equation as a truncated
perturbative expansion: one must take care to ascertain that the
phenomena at issue are in the appropriate ``weak field regime'' before
one is justified in using the linearized form in general relativity.
Even then, one must be careful of the sorts of arguments one will
employ the approximation for, for it will not be appropriate for many
sorts of claims one will want to investigate.  The exterior of
``large'' black holes can have curvature as close as one likes to
zero, and so will satisfy any criterion one reasonably lays down for a
``weak-field regime'', but solutions to the linearized Einstein field
equation on Minkowski spacetime can never replicate global features of
black holes, such as the presence of an event horizon.

Similarly, with regard to standards of what can count as good
evidence, only what is peculiar to the framework or theory forms a
part of it in my sense.  An example is stellar aberration.  Before
telecopic observational prowess, in conjunction with calculational
capacity, had developed to such a degree that the phenomenon could be
measured and controlled for, worrying about it did not form a part of
good evidentiary standards peculiar to the theory of astronomy as
formulated in the context, \emph{e}.\emph{g}., of Newtonian physics;
afterward, it did.

In sum, a theory must include all the practices and principles that
allow one to meaningfully bring formal theoretical structure and
practical experiment into contact with each other so that the former
may be used to interpret the latter, and the latter may be used to
constrain the former, \emph{i}.\emph{e}., so that data may be
structured in such a way as to be comparable with, and even identified
with, \emph{e}.\emph{g}., formal solutions to the equations of motion.
This will include, \emph{inter alia}:
\begin{itemize}
  \noitemsep
    \item mathematical structures, relations and formul{\ae} over and
  above the abstract equations of motion or field equations
  (\emph{e}.\emph{g}., in Newtonian mechanics that
  $\vec{v} \equalsdf \dot{\vec{x}}$, that mass is additive, that
  spacetime has a flat affine structure, and so on);
    \item standards of good argumentation (accepted approximative
  techniques for solving equations for the purpose of addressing
  particular kinds of questions, methods for stability analysis of
  perturbative results, sound heuristics for informal arguments, and
  so on);
    \item families of accepted experimental and observational
  practices for systems of different types;
    \item rules for connecting experimental outcomes with formal
  propositions (principles of representation, guidelines for the
  construction of concrete data models from raw observations, rules
  for reckoning expected experimental precision and error, and so on);
    \item rules of evidential warrant (what can be evidence, how to
  apply it, reckoning of error tolerance, and so on);
    \item and guidelines for judging the legitimacy of proposed
  modifications, extensions, and restrictions of all these.
\end{itemize}
(I make no pretense that this is an exhaustive list, only a sample of
characteristic components of a theory's epistemic content.)  Most of
this will be difficult if not impossible to articulate and record in
an exhaustive and precise way, so as to lend itself to use in formal
philosophical investigations.  We must trust that all such collateral
principles and practices are there, and can be, now and again, each
more or less precisely articulated as the occasion demands.  The same
holds true, however, for formal reconstructions of all forms of
reasoning in science (\emph{e}.\emph{g}., the `auxiliary hypotheses'
of the hypothetico-deductive method, which always hide an unruly mob
of philosophical sins)---which, I believe, those formal
reconstructions themselves lull us into forgetting.\footnote{See
  \citeN{curiel-fw-confirm-newt-abd} for a discussion of these kinds
  of informal practices and principles that necessarily attend real
  application of theoretical structure in a particular form of logical
  argument deployed in physics.}

On this view, the entirety of a theory is a dynamic entity, evolving
over time as new theoretical and experimental techniques and practices
are developed and accepted.  I believe this is the right way to think
about these matters for many if not most purposes in those parts of
philosophy of science studying scientific theories.  The contemporary
practice of treating theories as static, fixed entities, especially in
work of a more technical and formal character, can lead to serious
philosophical error.  An adequate semantics of a theory, for instance,
should reflect and accommodate its dynamic nature.

Before moving on, I must emphasize that I am not opposed to the use of
formal machinery in philosophy of science in general or in the study
of theories in particular, far from it.  I am opposed only to a
certain conception of how formalism can be fruitfully conceived of as
representing or capturing (part of) the structure of scientific
knowledge, \emph{viz}., as cleanly segregable from all practical
concerns.\footnote{As is so often the case,
  Suppes'~\citeyear{suppes-role-form-meths} discussion of this
  issue is insightful.  I am pleased to say my views have a great deal
  in common with his across the board, from the role of formal
  Tarskian-like models to the need for analysis of the pragmatics of
  experimental evidence.  I fancy many not directly familiar with his
  work may be surprised to hear this, as he is often caricatured as
  the formalist \emph{par excellence}.  He elaborates on his
  thoroughly pragmatist predilections and views, in most illuminating
  ways, in \citeN{suppes-pragm-phys} and
  \citeN{ferrario-schiaffonati-form-meths-emp-pract-suppes}.}

In the end, the standard view is a philosophical idealization.  Unlike
idealizations in physics, however, its use is not susceptible to
justification or defense by the use of a quantitative measure of the
expected error it introduces, and the subsequent judgement whether
that error falls within one's error tolerance.  The kinds of error the
standard view introduces are not controllable in that fashion.  One
must argue for the goodness of the idealization on a case by case
basis, in a way that respects the peculiar demands of the given
philosophical context.  There are cases in which it is manifestly a
good idealization.  A good example is
Earman's~\citeyear{earman-primer-determ} study of different
explications of the idea of ``determinism'' and whether or not
different physical theories respect them.  Relying only on the
mathematical formalism in conjunction with a few interpretive
postulates, Earman's constructions and arguments tell us something of
great conceptual and physical interest about those theories.  Specific
conclusions about what such explorations can teach us, however, must
be drawn with account taken of the fact that one ignores much if not
most of the epistemic content of a theory while exploring.  There are
cases in which it is manifestly not---just open any book on analytic
metaphysics.

\section[What Measurements and the Observer Are]{What Measurements and
  the Observer Are\symbolfootnote[2]{This section does not appear in
    the published version of the paper, due to length constraints.}}
\label{sec:what-meas-obsr}

The relevant notion of measurement and observation at play here is a
``non-primitive'' one, whose explication in the context of a
particular theory, or as an adjunct to a particular theory, is always
at least in part mediated by the theory.\footnote{It is, therefore,
  exactly the kind of measurement that does not pertain to the
  Measurement Problem in quantum mechanics.}  I mean something like:
``measurement as coupling between two individual systems, usually of
different types'', such that:
\begin{enumerate}
    \item one can distinguish the target system (the one being
  measured) from the measuring system, in the sense that there are
  some quantities of the target system and some quantities of the
  measuring system that dynamically evolve independently of each
  other;
    \item there is at least one fixed quantity (the \emph{measured
    quantity}) borne by the target system whose value will be
  determined by the measurement;
    \item there is a set of quantities (possibly a singleton, but not
  empty, the \emph{measuring quantities}) borne by the measuring
  system that jointly couple to the quantity being measured, in the
  sense that the measured quantity and possibly some of its
  derivatives are algebraic functions of the measuring
  quantities;\footnote{\emph{Cf}.\@ \citeN[pp.~3--4, arXiv
    version]{geroch-pdes-phys}:
    \begin{quote}
      Roughly speaking, two fields are part of the same physical
      system if their derivative-terms cannot be separated into
      individual equations; and one field is a background for another
      if the former appears algebraically in the derivative-terms of
      the latter.  The kinematical (algebraic) interactions are the
      more familiar couplings between physical systems.
    \end{quote}
    In this sense, some quantities borne by some types of physical
    system cannot be measured.  Entropy is an example: entropy
    mediates no physical interaction, does not couple with the
    quantities of any other type of physical system.  Its value can be
    determined only by measuring (in this sense) other quantities
    (\emph{e}.\emph{g}., temperature and free energy) and then
    calculating it from those values.  That is why there is no such
    thing as an entropometer.}
    \item the minimal temporal interval required for a measurement to
  take place is small compared to some relevant, characteristic
  time-scale associated with the dynamics of the state of the measured
  system during the time of measurement.
\end{enumerate}
Some simple examples will clarify what is meant.  For a gas-bulb
thermometer, the gas's volume (the measuring quantity) is an algebraic
function of the temperature of the system being measured; the
equilibration time, \emph{i}.\emph{e}., the time it takes for the
volume of the thermometer to achieve its final value directly
proportional to the temperature of the measured system, is small
compared to the characteristic time-scale in which the measured system
will change its temperature in response to any external influences
that may be present.  When the acceleration of a test-particle of
known mass and charge is used to measure the vectorial value of an
ambient, static electric field, the particle's acceleration is an
algebraic function of the value of the electric field; in this case,
the acceleration is an ``instantaneous'' response to the intensity and
orientation of the electric field.  Components of the Riemann tensor
are algebraic functions of the differential accelerations of a
collection of nearby freely falling particles, in accord with the
equation of geodesic deviation, and those differential accelerations
serve as measuring quantities for those components.

It is a deep and difficult question, to ask how one ``reads off'' the
results of a measurement from the values of the measuring quantities.
Naively, it seems as though the process must either continue in an
infinite regress or terminate in a measuring quantity whose values are
amenable to direct inspection by the ordinary human sensory apparatus.
But such a conclusion comes from too quickly accepting the raw
empiricist's naivet\'e---I think rather that the distinction drawn in
discussions of empiricism and realism, between ``theoretical'' terms
and ``observable'' terms, is a red herring.  Our sensory organs are
instruments, and we know---through theoretically grounded experimental
investigations---that they are often unreliable ones, and we know,
moreover, in quantifiable ways, experimentally determined under
precisely specifiable conditions, how they are unreliable: the ways
and the magnitudes of the errors.  We know the same thing about other
instruments of measurement we use in experiments, often in greater,
more precise and more quantitatively exact detail.  It is only
knowledge gained through such theoretically controlled means,
\emph{i}.\emph{e}., gained using instruments that we have some measure
of a theoretical grip on the inaccuracies of, that can contribute in
the strongest form to the epistemic content of a theory.  Thus, the
relevant distinction for the analysis of scientific knowledge and the
nature of scientific theories is between purely theoretical
propositions and experimental propositions based on measurements using
well understood instruments, whether those be sensory organs or
electron microscopes.  So far as \emph{scientific} knowledge is
concerned, the epistemological status of the different kinds of
instruments is the same, so long as their various forms of
inexactitudes and systematic errors are amenable to theoretical and
experimental characterization and testing---which I shall refer to as
``the epistemic primacy of experiment''.\footnote{My view has some
  affinity with that of \citeN[p.~12, from the English translation
  given in the bibliography reference]{helmholtz-geom-axioms}:
  \begin{quote}
    When we measure, we only effect with the best and most reliable
    aids known to us the same thing that we ordinarily ascertain
    through observation by visual estimation, judgment by touch, or
    pacing off.  In these cases, our own body with its organs is the
    measuring instrument that we carry about in space.  Our compasses
    are now the hand, now the limbs; or the eye, turning towards all
    directions, is our theodolite, with which we measure arc-lengths
    or surface-angles in the visual field.  \P \space\space Every
    comparison of magnitudes of spatial relations, whether by
    estimation or by measurement, thus proceeds from a presupposition
    about the physical behavior of certain natural bodies, whether our
    own body or the measuring instruments employed; a presupposition
    which, moreover, may possess the highest degree of probability,
    and may stand in the best agreement with all other physical
    relationships known to us, but which in any case reaches beyond
    the domain of pure spatial intuition.
  \end{quote}}

The results of perception are \emph{not} more epistemically
fundamental, primitive or secure \emph{with respect to} the
constitution of an evidentiary basis for a scientific claim.
\emph{That} is a true dogma of empiricism, that perception is the most
epistemically secure source of knowledge.  Both the results of
perception and those of experimentation must themselves be
secured---their reliability, veracity, accuracy, precision, scope,
fallibility, \emph{etc}.---by scientific investigation, each every bit
as much as the other.

To my mind, the theoretical in science, as supported and substantiated
by controlled experimentation, is on a more sound epistemic footing
than the perceptual.  To be clear, I mean that the theoretical both
has more epistemic warrant and plays a more fundamental role in the
construction of epistemic warrant than the perceptual \emph{with
  regard to their respective roles in scientific enterprises},
precisely because of the substantive contact of the theoretical with
the experimental, in the relation between which we have many powerful
methods of measuring, tracking, manipulating, controlling for, and
justifying essentially all the kinds of approximation, idealization
and error that such investigations must suffer, none of which we have
for perception.  And, in any event, the perceptual is almost entirely
irrelevant for the philosophical comprehension and understanding of
the nature of science and of scientific knowledge, as is shown by the
fact that, outside physiology, no science has the theoretical capacity
to include sense organs in its models.  This, I think, has been one of
the main and most pernicious sources of error in much of Twentieth
Century philosophy of science, inherited from Kant as worked through
the mangle of the Logical Empiricists: that the most relevant
epistemic distinction in the analysis of science is the theoretical
versus the observable.  It's not.  The most important and fundamental
distinction is between the theoretical and the experimental, and the
most important correlative question is, how does it happen that these
two dramatically different representations of the physical world not
only come into fruitful contact with each other, but indeed need and
rely on each other in such intimate ways in order to be able to come
into contact with the physical world in the first place.  (My
one-sentence sketch is, of course, a caricature of both Kant and the
Logical Empiricists, but it's one that, astonishingly, continues to
guide much contemporary philosophical work.)

Science itself shows us how unreliable our perceptual faculties are,
with great precision and rigor.  If we accept those findings---and I
think we must, if we are to accept anything in science at all, as
those findings are grounded in and derive from the same sorts of
experimentation, reasoning, \emph{etc}., as all our most deeply held
scientific results---then it follows \emph{a fortiori} that perception
and the deliverances thereof are less epistemically secure than
experimental knowledge as represented in our theoretical formalism,
for we rely on experiment to characterize the epistemic reach,
precision and accuracy of perception.

In any event, how a real human comes to know the result of a
measurement is an independent question from that of what
\emph{constitutes} a measurement and what constitutes an epistemically
warranted judgement of a sound and accurate \emph{outcome} of an
experiment.  That we see a pointer on an instrument indicating the
value measured as a result of an experiment to be 3 is a fact of
little to no relevant epistemic import.  How and that we have good
reason to believe that the result we expect is 3 and that the reading
`3' is the result of the correct functioning of the experiment as a
whole is what matters, and that comes entirely from our reasons for
believing that our experiment is latching on to the germane parts or
aspects of the world in the right way and that our theoretical
representations of the experiment appropriately and adequately
captures that latching on to.\footnote{Compare the remarks of
  \citeN[part~\textsc{ii}, ch.~5]{popper-log-sci-disc} to the effect
  that an observation sentence is not an autobiographical report, but
  a conventional statement of socially accepted scientific fact.  Also
  compare the remarks of \citeN[p.~237]{suppes-pragm-phys} concerning
  the fact that
  \begin{quote}
    experimentalists are not in any sense searching for epistemic
    certainty in designing their experiments or reporting their
    observations\ldots.  [T]he fantasies of philosophers about
    certainty of observations is not at all the right model of how to
    think about what experimental physicists are doing.
  \end{quote}
}

We can teach a person to record the temperature of soup with a given
thermometer, even if that person has no knowledge of temperature,
thermometers or even soup: ``stick this rod in the container of this
liquidy stuff, in such a way that this bulb at the end is completely
submerged; hold everything still for one minute, then write down the
number that appears at the other end of the rod''.\footnote{Compare
  the function of the illiterate ancient and medieval scribes who
  scrupulously duplicated manuscripts one letter at a time, having no
  knowledge at all of what each squiggle meant, perhaps not even aware
  that each squiggle was a conveyor of semantic content, nor indeed of
  how to syntactically individuate squiggles; which example, moreover,
  shows how important such workers, who have no knowledge of what
  they're doing, can nonetheless be in any human enterprise, even the
  intellectual ones.  We would have no works of Plato, none of
  Archimedes, Aristotle, Aeschylus, of any of those whose thought
  currently forms one of the cornerstones of the knowledge and culture
  of the present age, without the toil and travail of those now
  nameless and forever hapless laborers.}  Such a person, however, can
reach no epistemically relevant judgment on any scientific question we
may want to raise about the experiment: was the result not valid
because, \emph{e}.\emph{g}., of the presence of a confounding external
factor that disrupted the proper working of the experiment?  We can be
as thorough and detailed as we like in giving instructions to the
ignorant soul we entrust with the job, but there is always ambiguity,
always something we will not mention explicitly that can bear on
whether the recorded outcome is trustworthy, and which the amanuensis
will be in no position to recognize as relevant, in her ignorance.
Did we forget to mention not to hold the thermometer in place with
heated tongs?  Most likely.  Only someone with practical knowledge of
how the instruments work, informed by the right sort of theoretical
knowledge about both the instruments and the target system, can judge
such matters with epistemic warrant---and they can do so even if their
perceptual faculties are grievously impaired.

For all these reasons, I want to speak of experimentation, measurement
and observation as practices and processes that can be characterized
independent of the direct experience of humans, though certainly not
independent of the current epistemic state and capacity of humans.  By
``schematize the observer'', therefore, I mean something like: in a
model of an experiment, to provide a representation of something like
a measuring apparatus, even if only of the simplest and most abstract
form, that allows us to interpret the model \emph{as} a model of an
experiment or an observation.  Such schematic representation,
moreover, is not possible in a physically significant and cognitively
substantive sense without recourse to all the other components of what
I will call the practical part of scientific knowledge, as spelled out
in the next section.

\section{The Different Forms and Aspects of Scientific Knowledge}
\label{sec:forms-aspects-know}

\skipline[.5]

\begin{quote}
  \begin{tabbing}
    \hem[3]\=When every year and month sends\hem[8.5]\=\kill    
    \>In theory, there's no difference between theory and practice. \\
    \>In practice, there is.
    \\
    \>\>-- Yogi Berra
  \end{tabbing}
\end{quote}  

\skipline

An adequate semantics should be able to support the articulation of,
to represent the meaning of, all knowledge claims a theory embodies,
for all the forms and aspects of knowledge thus embodied.  As the
discussion of SgrA$^*$ in \S\ref{sec:what-theory-is} highlights, my
formulation of the question ``what is a theory?''  brings into focus
what I consider another shortcoming of the standard view---that,
according to it, all scientific knowledge is essentially of one
undifferentiated kind: approximately true (or at least epistemically
justified) propositions about states of the world, and perhaps about
the nomic structure of the world independent of the state the world is
actually in.  This comes out most clearly in the picture of a theory's
semantics that naturally accompanies the standard view of theories,
that it is fixed by ontology.

This is not an accurate picture of scientific knowledge.  Scientific
knowledge has many forms and aspects that the picture cannot well
capture.  An account of scientific knowledge that characterizes a
theory as a \emph{physical} theory of the types of physical systems it
purports to treat must include \emph{all} forms and aspects of
knowledge the theory has endowed us with, including that gathered in
the context of experimental practice and expressible only by using
resources that go beyond the theory's formalism.  A semantics that
captures only ``potential knowledge'' that it is impossible for humans
to know, such as that embodied in the ``family of all models or
possible worlds'' of the theory, is useless as a semantics on which to
ground an analysis of the nature of scientific knowledge and its
epistemic warrant, because science is a human enterprise.  If the
semantics captures only what it is impossible for humans to know, then
what good does it do?  And why should we care about it?  What problems
will it solve, or even illuminate?  Ones of a purely formal character,
no doubt.

A useful classification of what I have been calling the theoretical
and the practical forms of scientific knowledge is suggested by the
following observation of \citeN[p.~637, emphases
his]{stein-struct-know}:
\begin{quote}
  It is hard, but possible, to learn theory by self-study from books;
  it is surely much harder to learn experimental techniques without a
  teacher to help one acquire skills; but what I suspect to be
  impossible is to learn the principles of experiment without
  \emph{actual} experience with the relevant \emph{instruments}.
\end{quote}
How one is able to learn different parts of the epistemic content of a
theory may seem at first glance irrelevant to an analysis of the
character of the knowledge to be learned.  To the contrary, I think
Stein has put his finger on something deep about epistemic content:
how it is possible to come to learn something shapes and constrains
the kinds of epistemic warrant we can have for it, and thus constrains
how we may bring it to bear in possible evidential relationships to
other pieces of potential knowledge.  The nature of the object of
knowledge determines how one may learn about it, and so that
constraint itself can teach us something about its nature.  Thus, if
one wants to understand the structure and character of our knowledge,
in all its forms, one of the deepest and most powerful clues we have
is what we do and can learn, and how we do and can learn it, for how
knowledge can be learned tells us how it can (and cannot) be
represented---and thus whether it can be accommodated by a
representational regime consisting only of formalism in conjunction
with ``coordinating principles'' or ``representational capacities''.
One of the most important distinctions here is that between a subject
matter's being \emph{teachable} and its being
\emph{learnable}.\footnote{It is instructive in this regard to compare
  Stein's remarks about how one can and cannot teach and learn
  different forms of knowledge in physics with Plato's discussion, in
  \emph{Protagoras}, of how one can and cannot teach and learn virtue:
  Plato concludes (as I read the dialogue) that virtue is learnable
  but not teachable (at least not based solely on the output and
  intake of discursive language use); one must practice it in order to
  understand and master it.}

Theoretical knowledge of the sort Stein describes in the first part of
the quote has the peculiar and remarkable property of being
representable by a well arranged sequence of propositions (a well
written textbook, say).  If it did not have that property, we could
not learn it from such a sequence.  Epistemic warrant for knowledge
acquired in that way has its strengths and weaknesses.  It relies on a
particular form of trust in the scientific community,\footnote{Of
  course, not only the learning of theoretical knowledge requires such
  trust: ``When scientists present the results of an experiment they
  take responsibility for those results by attaching to them the most
  precious coin of the scientific realm: the individual scientist's
  pledge to speak the truth''
  \cite[p.~2]{lockman-remote-obs-good-obs}.}  and on a belief that
one's own idiolect shares enough in common with that of the author's
(or, at least, that one can learn enough of how the author uses
language) to be certain that one has understood the epistemic content
the author is trying to convey.  It also, however, admits of being
made precise and clear in a way that conduces to verification and
defense, and to application in further argument.

Practical knowledge cannot be effectively represented in this
way.\footnote{Practical knowledge in Stein's sense, then, and as I
  will further develop the idea in this section, is close kin{} to the
  notion of ``knowing how'' much discussed in standard epistemology
  since its introduction by Ryle, and even closer to its more
  specialized form, ``tacit knowledge'', as introduced by Polanyi in
  the context of studying the epistemology of engineering and
  technology.  The literature on those notions is too vast to engage
  with here.}  An anecdote will perhaps show this better than an
argument.  During the 2017 run of the Event Horizon Telescope
(\url{https://eventhorizontelescope.org/}), when a worldwide team of
astronomers were co-ordinating the use of several observatories around
the world to simultaneously record observations of SgrA$^*$ (for later
collation and structuring) in the attempt to record the first direct
image of a black hole, the head of the organization, Shep Doeleman,
masterminded the proceedings in a control center two doors down from
my office at the Black Hole Initiative.  Naturally, I spent some time
in the room watching them, as this was scientific history in the
making.  One evening, while the team was deciding whether or not to
make observations that night, I witnessed the following exchange
between Shep and an instrumentalist at ALMA, the Chilean observatory
(5000 meters above sea level), who was having trouble with one of the
specially designed instruments---it seemed to be malfunctioning in a
particular way, but the source of the problem was not immediately
clear.  Shep said, ``Oh, yeah, I've see this before.  Here's what you
do.  Remove the oscillators and take'em down to 1000 meters [above sea
level], and leave'em for a few hours; they'll settle down, and then
when you take them back up, they'll work.''  Chilean instrumentalist:
``Why does that work?''  Shep: ``I have no idea, but it does.''  This
is a part of the practical knowledge associated with observational
relativistic astronomy that I doubt anyone will ever find written in
any book---that sort of malfunction is likely the result of the quartz
crystal oscillators suffering from pressure-related issues associated
with high elevation.  Moreover, even if written, it is the sort of
thing one could not \emph{learn} from just having read the book.  Of
course, one could learn that particular trick---but to learn how to
\emph{play} with the instruments so as to acquire the feel for them
necessary to figure out such tricks for identifying and solving
problems---that is something that cannot be learned from a book, as
Stein so beautifully points out in the passage I
quoted.\footnote{Before making the anecdote public, I emailed Doeleman
  to ask his permission, asking him to read as well the remarks I made
  surrounding it.  He wrote me back an interesting elaboration on the
  anecdote, explaining the process by which he acquired the knowledge
  about the oscillator, which illuminates the theme even more:
  \begin{quote}
    [The anecdote you're relaying concerns] a quartz crystal
    oscillator we had running at Chile for a while that then went on
    the fritz.  After we brought it back to the SAO{} [Smithsonian
    Astrophysical Observatory] it seemed to return to normal, and we
    attributed that to the high elevation and low pressure (up at
    16,500 ft).  That's not to say that the standard operating
    procedure should be to send such crystals to a low elevation `spa'
    for rejuvenation every so often.  It was more just me speculating
    about what was going on.  However the general point you are making
    is very valid.  As with any profession, we develop intuition about
    the instruments and systems we work with, just as any craftsman
    understands the feel of good tools working good materials.  The
    lesson in the anecdote you quote is that the quartz oscillator in
    question was brought up to high altitude at the ALMA site as a
    reference to check the performance of the hydrogen maser, which is
    critical for VLBI.  When the crystal-maser comparison went South,
    there was some concern that it might have been the maser that was
    the problem, but after working with these atomic clocks for so
    long, I was sanguine.  There was a greater probability that it was
    the crystal (with a power supply not designed specifically for
    long-term operation at high altitude) that was at fault.  When I
    brought a second crystal to Chile to check it out, that was indeed
    the case.
  \end{quote}
  Doeleman then recommended that I read
  \citeN{lockman-remote-obs-good-obs}, which examines why going to
  the telescope and being there with the instruments in person can
  have great benefits, both epistemic and practical, over remote
  observing.  I did read it, and now I recommend that you do as well.
  Here is a gem from it (p.~3):
  \begin{quote}
    Someone who actively plays with the equipment, tries out various
    combinations of things, and constantly iterates on technique, not
    only gets data and a sense of its correctness, but also develops
    skills which can make the next data better.  The way to become a
    skilled observer is to participate in the observations as
    completely as possibly, to seek active control or understanding of
    every phase of the process; to try to recognize the difference
    between the basic limitations of an instrument and those
    limitations which are rooted in style or tradition.
  \end{quote}}

I think this lesson generalizes: how it is possible to learn different
pieces or areas of scientific knowledge is one important criterion for
determining the kind of knowledge one is dealing with.  This practical
form of knowledge, an inextricable part of the epistemic content of
the theory one considers, can not be formalized or even just well
captured in book form; most theoretical forms of knowledge can.  The
latter is, to speak crudely, propositional in a way the former is not.
That does not, however, \emph{ipso facto}, imply that it is more
important for a proper understanding of the character and content of
the sum of knowledge that a theory embodies, that characterizes it as
a \emph{physical} theory of \emph{that} sort of family of physical
systems.

The proponent of the standard view may well reply that it's an amusing
anecdote, but tells us nothing of the epistemic content of general
relativity.  I disagree.  It is knowledge of this sort, and the trust
we have in experimentalists and instrumentalists of the calibre of
Doeleman and his collaborators, that gives us the epistemic warrant we
have for believing that our astronomical observations provide us
understanding and comprehension of the physical systems that general
relativity provides us models of---that justifies our belief that
general relativity is a good theory of those systems, and that we know
what the mathematical formalism represents, and know in particular
which parts of the formalism represent which parts of the world.  More
importantly, this shows how the two types of knowledge are not cleanly
segregable---it is only practical knowledge of the kind had by the
experimentalists commingled with the formal theoretical apparatus that
allows us to understand the formalism as part of a physical
theory.\footnote{I do not want to leave the reader with the idea that
  I think all theoretical knowledge is articulable in propositions and
  learnable from books, nor that all practical knowledge can be had
  only by experience.  The situation is far more nuanced than that,
  and a complete discussion would take us too far afield.  To get an
  idea of how the simplistic picture I have sketched falls short,
  consider all the inexact techniques theoreticians learn and master
  in order to solve their problems, and how, moreover, they can in
  general learn to do so only by practice, by trial and error, not
  from textbooks---for example, the ``feel'' for when a given
  perturbative or approximative technique is the right one, how it
  should be applied, when it is legitimate, when illuminating, where
  it breaks down, \emph{etc}.  I have not found a good discussion of
  this with particular regard to theoretical physics, so I invite the
  reader to look at the great geometer William Thurston's remarkable
  discussion of what mathematics is and what it does
  \cite{thurston-proof-prog-math}, particularly his account, in \S4
  of that paper, of how he came to learn what a ``proof'' in
  mathematics really is and how it is he thinks that mathematicians
  produce and check them.}

To make this argument a little more precise, I again follow
\citeN[p.~635, italics his]{stein-struct-know}, now in distinguishing
among different aspects of scientific knowledge:
\begin{quote}
  I am construing the word knowledge in a wide and ambiguous sense.
  The reflections I am proposing have as their object (a) our
  knowledge in physics as an \emph{achieved result}: knowledge as
  \emph{the knowledge we have of X}; (b) our knowledge as susceptible
  of \emph{justification} or \emph{defense}---that is, as involving a
  structure of ``evidence'' for its asserted contents; and (c)
  knowledge---science---as (to appropriate a word of Isaac Levi's) an
  \emph{enterprise}: an activity aimed at increasing our knowledge in
  sense (a), by means appropriate to the constraints of (b).
\end{quote}
I would add to this list a fourth, closely related to, but distinct
from, Stein's (b): knowledge as ground for epistemic warrant for other
knowledge claims and for furthering the enterprise of science,
\emph{i}.\emph{e}., knowledge that is suited to playing the role of
evidence for other assertions.  Call this `(d)'.\footnote{Not all
  knowledge is suited to playing this role, and its being suitable may
  depend on facts about the wider scientific context, beyond what
  pertains only to the narrow investigation at hand.  It would take us
  too far afield to go into this now.  I thank Chris Smeenk for
  helping me to see how subtle this business can be.}

The standard view, I claim, captures at most one of these aspects,
knowledge as achieved result, and that at best only in part.  The
standard view's picture of the knowledge content of a theory, entirely
captured by indicative propositions about the state and nomological
structure of the world, does not allow one even to ask how such claims
are justified, how their justification depends on the knowledge we
have in virtue of other theories, the limitations on the scope of such
claims, and so on, for the practical knowledge I have been discussing
informs and provides much of the body of the theory's knowledge in all
these aspects, and the standard view does not allow this knowledge to
be represented.  Thus, it cannot explain the epistemic warrant we have
for trusting and using the theory, and thinking that the understanding
and comprehension it seems to give us of the world \emph{is about the
  world}.\footnote{\emph{N}.\emph{b}.: nothing I am saying commits me
  to any form of realism about theories---I can believe that our best
  theories tell us a lot about the world, and even that those theories
  provide us with deep understanding of some aspects of the world,
  without also believing that those theories are exactly or even
  approximately true of the world in any deep or strong
  representational sense.}

The standard view, however, is deficient, I claim, even with regard to
knowledge of aspect (a) (knowledge as achieved result), in so far as
it precludes fundamental understanding and comprehension we gain about
parts of the world that can be had only by heuristic, informal
arguments that cannot be formalized.  A good example is the way that
one of the fixed relations among quantities in Navier-Stokes theory,
\emph{viz}., that heat flux is always independent of the pressure
gradient \cite[ch.~\textsc{v}, \S49]{landau-lifschitz-fluid}, partly
encodes the fact that Navier-Stokes theory is valid only for fluids
quite close to hydrodynamical and thermodynamical equilibrium of a
certain kind.  (This relation is not one of the equations of motion,
the so-called Navier-Stokes equations; it forms part of what I call
the kinematical constraints of a theory, which I describe in
\S\ref{sec:breakdown-regimes} below.)  The only arguments I know of
for the fact that Navier-Stokes is essentially an equilibrium theory
are not only heuristic and approximative, but indeed are, strictly
speaking, incorrect in parts, incoherent as a whole, and even
inconsistent in several ways (all with respect to the formalism of the
theory).\footnote{I do not know of any published arguments for the
  proposition.  It is a fact widely known among physicists practicing
  in the relevant areas, part of the common lore of the field, passed
  down from teacher to student in lectures and laboratory
  apprenticeships and such.}  Here is a sample of some of the
propositions such arguments rely on:
\begin{itemize}
  \noitemsep
    \item that a ``particle'' of fluid could ``switch flow lines''
  only if the line it traversed crossed another (whereas in fact this
  can happen also if two lines merely osculate, or, in technical
  terms, at a point at which a Jacobi field of the flow vanishes);
    \item that flow lines can ``cross'' in a sense, even in the
  absence of turbulence, by converging to a single, isolated point, a
  ``singularity'' (as happens, more or less, when water runs down a
  narrow drain);
    \item that, strictly speaking, however, flow lines can never
  cross, whether in turbulent flow or smooth, as flow lines lose their
  definition the moment they converge.
\end{itemize}
The propositions wear their individual incorrectness and incoherence,
and their mutual inconsistency, on their sleeves.  Nonetheless, they
serve their purpose adequately in practice, allowing the formulation
of an argument that sketches a picture of a fundamental part of the
epistemic content of a theory of a genus of physical phenomena
(Navier-Stokes fluids), a picture that in essence conveys the
important physical features of the phenomena at issue.  In particular,
such arguments convey the kind of understanding and comprehension
Navier-Stokes theory yields of the systems it treats.  The nature of
the arguments, moreover, are such that they cannot be made sense of,
much less concluded successfully, within the confines of the formal
apparatus of the theory alone and any epistemic content that could
accrue to it from a pristine semantics based on ontology.  In any
event, ``close to hydrodynamical and thermodynamical equilibrium'' is
not a concept that lends itself to precise formalization once and for
all, independent of context.  It is, however, of the most fundamental
importance in coming to understanding the epistemic content of
Navier-Stokes theory that it is essentially an equilibrium theory.  No
model in the theory represents this fact.  No formal, rigorous
proposition can capture this idea once and for all.

Both theoreticians and experimentalists employ that sort of loose,
heuristic argument as a matter of course in their workaday activity,
as one of the most important tools in their respective tool-boxes.  To
my mind, the most striking and salient feature of such arguments lies
in the fact that, strictly speaking, one cannot prove them either
correct or incorrect, in so far as, strictly speaking, several of the
terms and relations used in such arguments lose their definition at
various points.  (The sentence ``$A$ bears the property $\phi$'' can
be neither true nor false if neither `$A$' nor `$\phi$' nor `bears the
property' has an unambiguous definition.)  One cannot analyze the
meaning of the propositions of such arguments by analyzing their
possible respective truth-conditions in isolation from those of the
others, by judging relations of reference to pure ontological models
for example.  And yet those propositions do have meaning, albeit not
precisely determined, and can be more or less appropriate and
adequate.  One must base the meaning of the propositions on the import
of the argument as a whole, and in particular on what we know to be
the sort of ``mild'' incoherencies and inaccuracies we can allow
ourselves in contexts like this, as determined and verified by the
experimental probing and practical fixing of the theory's regime of
applicability.  We are back in the neighborhood of needing to know how
to schematize the observer in order to grasp and articulate large and
important parts of the theory's epistemic content.\footnote{None of
  this has to do with meaning holism in anything like Quine's sense,
  as expressed, \emph{e}.\emph{g}., in the famous (and glib) aphorism,
  ``our statements about the external world face the tribunal of sense
  experience not individually but only as a corporate body''
  \cite[p.~41]{quine-2dogmas}.  The meaning of the---strictly
  speaking---incoherent and inconsistent propositions in the kinds of
  arguments I discuss indeed cannot be fixed without reliance on a
  body of other propositions.  Here, however, the required
  propositions concern either the propriety and adequacy of such
  descriptions in experimental models of fluids, or the way such
  propositions can be treated as approximations, as it were, of
  rigorous and precise theoretical propositions.  There is no
  implication that \emph{all} the propositions of theory face the
  tribunal of experience as a corporate body.}

I have argued that the standard view gives too little as an answer to
the question of the scope of a theory's epistemic content.  In another
sense, however, the standard view gives too much---for it says that
every question one can pose to the theory has a determinate answer.
In a formal sense, this may be correct.  Navier-Stokes theory is a
continuum theory, so in principle it answers questions about how
fluids behave at the scale of 100$^{-100}$ \texttt{cm}.  But if one
wants to know whether the answer given is \emph{relevant to or even
  just possibly meaningful in the real world}---whether the answer
counts as possible scientific knowledge in the senses, for instance,
of Stein's (b) and (c), and my (d), and indeed even for the factual
kind in Stein's (a)---then one has to know, among other things, the
scope of the theory's regime of applicability, \emph{i}.\emph{e}., the
spatial, temporal and energetic scales (\emph{e}.\emph{g}.\@) in which
it is a good theory and those where it breaks down.  For that, one
must have recourse to knowledge that one can come by only through
experimentation and the application of knowledge embodied by other
theories, and thus, more importantly, knowledge that is not and cannot
be encoded in the formalism of the theory itself.  Our knowledge of
quantum theory tells us that the answer to any question about how
fluids behave at the scale of 100$^{-100}$ \texttt{cm}, well below the
Planck scale, is not so much false as \emph{meaningless}---we have
good reasons to think that ordinary spatial concepts such as
``fraction of a \texttt{cm}'' are not meaningful past a certain point.
According to the standard view, however, all the answers Navier-Stokes
theory gives us about the behavior of fluid at such scales are simply
false, albeit meaningful.

In so far as knowledge of the regime constitutes knowledge in Stein's
aspect (b), moreover, as being susceptible to justification by
involvement in a network of evidentiary relations, it cannot be given
as a physical interpretive postulate in any standard sense
(\emph{e}.\emph{g}., a Reichenbachian coordinating principle, a
Carnapian reduction sentence, or contemporary gestures at
representational capacities), \emph{i}.\emph{e}., one that encodes (in
part) or at least makes contact with knowledge cleanly segregable from
any knowledge represented only by the theory's
formalism.\footnote{This, I think, is what is right about the claims
  of \citeN{friedman-dyns-reason} to the effect that, in so far as
  the relativized \emph{a priori} provides the grounds for endowing
  theoretical formalism with empirical content, it itself is not
  susceptible to justification by experimental means.}  Our knowledge
of where and how Navier-Stokes theory, \emph{e}.\emph{g}., breaks down
is in large part based on and justified by experimental results as
represented in and put into meaningful contact with the mathematical
formalism.  A physical interpretive postulate, therefore, that encoded
such knowledge would have to make inevitable reference to such facts
that are known only by experimentation, and thus require knowledge of
how the observer and measuring devices are represented in the theory.
Whether a given question, therefore, that seems in the abstract to be
one the theory can address is in fact one the theory can give a
physically relevant answer to---whether it is about a system that the
theory appropriately and adequately treats---can be determined only by
recourse to knowledge that goes well beyond that which can be encoded
only in the formalism or that can be expressed by propositions making
no reference to detailed experimental knowledge.  Nonetheless, such
knowledge ineliminably involves as well theoretical terms whose
definitions rely in part on some of the formal structure of the
theory.  It is knowledge, in other words, that cannot be cleanly split
into a formal part and a practical part.  The standard view cannot
apply.

Much of this discussion has implicitly relied on a picture of
knowledge that goes far beyond what can be captured in propositions
expressing true, justified beliefs.  The most important component of
the epistemic content of a theory, on my view, is more akin to what
the (notoriously problematic) concepts of ``comprehension'' and
``understanding'' try to get at.  I shall not attempt to give a
systematic account of these ideas, but I can say a few things to try
to convey my sense of them.  ``Understanding'' is something like ``the
capacity to operate successfully in the scientific enterprise, in all
its aspects and parts'': to use our representations and instruments as
the basis for the fruitful continuation of the enterprise, as part of
evidential warrant in testing, as basis for predictions and
characterizations, as inspiration for potentially fruitful new
investigations, as the grounds for conceptual clarification and
innovation in foundational work; and perhaps most of all to grasp how
our representations and our instruments relate to, inform, and
contribute to the constitution of each other, and to grasp that in
such a way that it grounds the work succesfully continuing the
enterprise.

Before moving on, I want to address a question that my discussion has
raised.  One may well object that the kind of knowledge shown by, say,
observational astronomers in good manipulation of their instruments
may be part of the epistemic content of \emph{some} theory, but surely
not that of general relativity.  That may well be right in some sense,
although I suspect that one can never impose a clean, sharp
demarcation, once and for all, between a given theory and auxiliary
theories needed by experimentalists in the study and use of it.  What
then can we say, if anything, about criteria for whether a particular
piece of knowledge has been delivered by a particular theory in the
relevant sense, so as to count as part of \emph{its} epistemic
content, rather than, say, as part of the content of an auxiliary
theory used to model an experimental apparatus?  I don't know.  At a
minimum, it should be a piece of knowledge that we could not, even in
principle, have without using the theory at issue, at least as
determined by the context of our current state of achieved knowledge
and our understanding of the web of evidentiary relations that support
it.  In principle, one should always be able to perform experiments of
different kinds, with instruments modeled by different theories, so as
to determine whether or not the theory at issue makes an ineliminable
contribution to the epistemic content of the knowledge, and so to
attempt to isolate that contribution.  A theory, however, at bottom,
is a nebulous congeries.  I doubt that there can be a comprehensive
criterion by which one could judge all attempts to answer any question
about whether this or that bit of knowledge exclusively belongs to the
epistemic content of this or that theory, an answer relevant and
appropriate for all foreseeable philosophical and scientific purposes,
much less for all those we cannot yet imagine.

In the end, all the problems I discussed in this section arise from
the same deficiency in the standard view, to wit, its inability to
allow schematization of the observer in a way relevant to the
articulation of the epistemic content of a theory: all the problems
devolve upon the need to bring theoretical and practical knowledge
into fruitful contact, and it is the capacity to schematize the
observer that allows that to happen.

\section{How Theory and Experiment Make Contact with and Inform Each
  Other}
\label{sec:theor-exper-contact}

That we must have the capacity in our theories to construct explicit
models of complete experimental situations including instruments and
the actual methods of their deployment in order to represent actual
observations, however, raises a serious problem, one that
\citeN[p.~290]{stein-carnap-not-wrong} trenchantly poses:
\begin{quote}
  \ldots\@ we have no language at all in which there are well-defined
  logical relations between a theoretical part that incorporates
  fundamental physics and any observational part at all---no framework
  for physics that includes observational terms, whether theory-laden
  or not\ldots.  I cannot think of any case in which one can honestly
  deduce what might honestly be called an observation.  What can be
  done, rather, is to represent \ldots\@ ``schematically,'' within the
  mathematical structure of a theoretically characterized situation,
  the position of a ``schematic observer,'' and infer something about
  the observations such an observer would have.
\end{quote}
In other words, we do not have a formal account of the epistemic
content of the theories of physics even minimally adequate for any
account of their actual empirical application; this is not to say that
such applications in real scientific practice have no foundation or
are unjustified, only that we have no adequate comprehension of the
process.  Forget how we get the theory into or out of the
laboratory---how do we get the laboratory into the theory?  This, I
think, is the fundamental issue one must address in trying to give an
account of the epistemic content of scientific theories and how its
components bear on each other, and in particular in answering the
question whether they are mutually extricable in the way demanded by
the standard story told by philosophers about the character of
scientific theories, and relied upon by them in their work.  (See
again the long passage from \citeNP{stein-struct-know} that I quote
on page~\pageref{pg:stein-schema-quote}.)

In order to address this problem, in this section I explain with a
little more precision what I mean by `theory', with the goal of
drawing a---schematic!---picture of how experimental results are
placed in physically significant contact with formal theoretical
structures, and what the nature of that contact is.  For my purposes,
it suffices to provide only a rough characterization sketched in broad
strokes.\footnote{I discuss here only one way that theory and
  experiment make contact, when one has already in hand a developed
  and articulated theory.  There are other ways, other roles for each
  of theory and experiment in bearing on the other, that I do not
  discuss because they are not relevant for my purposes, such as the
  use of experiment in guiding the development of new theory, and in
  expanding understanding.  I think excellent case studies for
  exploring those kinds of roles would be Newton's investigations on
  light and color (which have not received the philosophical attention
  they deserve---see
  \citeNP{stein-meta-meth-newton,stein-furth-consid-newt-meth} for a
  discussion of those investigations with particular regard to these
  issues, and \citeNP{stein-enterprise} for an analysis of how they
  illuminate questions of understanding in both physics and
  philosophy) and the role of Fizeau's experiments in the development
  of special relativity (for which, see
  \citeNP{patton-reconsid-exps}).  \citeN{patton-exp-thry-build}
  provides an insightful discussion of the problem more generally.}

A theory, as implied by the discussion up to this point, is a system
that allows one, among other things, to formulate propositions and
affirm them in principled ways based on evidence gathered according to
good principles, to apply them in turn as evidence for other
propositions, and to use them as the inspiration and basis for new
investigations.\footnote{My conception of a theory is in many ways
  similar to, and indeed in part inspired by,
  Carnap's~\citeyear{carnap-eso} conception of a linguistic
  framework, particularly in the way that a theory in my sense serves
  to define a fixed sense of physical possibility relevant to the
  kinds of system the theory treats, just as a Carnapian framework
  does for the types of entity whose existence is analytic in the
  framework.  Carnap's conception is too broad and vague, however, to
  do the work I require of it.  \citeN{stein-carnap-not-wrong}
  provides an insightful and illimunating, albeit brief, discussion of
  the differences between a Carnapian framework in the original sense
  and a theory in the sense of a formal structure in theoretical
  physics related to, but more restricted in scope than, the type I am
  sketching here.  \citeN{lakatos-fals-meth-sci-rsrch} has some
  affinity with the gist of this view, in particular his notion of the
  ``hard core'' of ``research programs'', though, again, the
  differences in detail outweigh the similarities.  There is perhaps
  more affinity with the ``research traditions'' of
  \citeN{laudan-prog-probs}, in so far as different ones can share and
  swap important methodological and theoretical principles, as can
  happen with theories in my sense.}  A theory in this sense is a
system that allows for the unified representation and modeling of a
particular kind of physical system so as to render that kind amenable
to investigation by scientific reasoning and practices of all forms.
A particular kind of physical system is one such that all individuals
falling under the kind bear the same physical quantities whose
properties are characterized by and whose behavior is governed by the
same set of equations of motion and collateral mathematical relations.
A theory in my sense also includes such things as accounts of
experimental devices appropriate for the probing of the relevant kind
of physical system, good practices for employing them, sound
techniques for the collection of raw data and statistical and other
analysis and organization of the same, reliable methods of
approximative and heuristic reasoning for constructing models and
solving equations, and guidelines for determining whether a system of
the given kind is in a state and experiencing interactions with its
environment such as to be jointly amenable to appropriate and adequate
representation by the theory (\emph{viz}., whether or not the system
falls into the theory's regime of applicability), and so on.

Before diving into the details of how I think that all these
components hang together so as to make it possible for the theoretical
and the experimental parts of a theory to make fruitful contact with
each other, I shall need further distinctions with regard to ``level''
of theoretical representation; it will be useful to draw them by
laying down terminology.  I shall not attempt to give formal, precise
definitions.  These ideas are for my purposes adequately clarified by
examples.

Theories are often formulated in the context of a framework. Newtonian
mechanics, the heart of which is embodied in the definitions and the
three laws Newton lays out in his \emph{Principia}, is perhaps the
exemplar \emph{nonpareil} of a framework.  Newton's Second Law in the
abstract is not the equation of motion of any particular kind of
physical system.  It is rather the template that any equation of
motion for any particular kind of system treated by the framework must
instantiate.  Newton's theory of gravity is a theory formulated in the
framework of Newtonian mechanics.  It treats that kind of physical
system characterized, \emph{inter alia}, by the possession of inertial
mass, of gravitational mass (equal in magnitude to the inertial mass),
of spatial position, and of a velocity expressible as the temporal
derivative of spatial position, such that its dynamical evolution is
governed by Newton's gravitational force law.  Navier-Stokes theory is
another theory formulated in the framework of Newtonian mechanics.  It
treats that kind of physical system characterized, \emph{inter alia},
by the possession of inertial mass, shear viscosity, bulk viscosity,
thermal conductivity, fluid velocity, heat flux, pressure, and
shear-stress, all satisfying among themselves fixed relations of
constraint (\emph{e}.\emph{g}., that heat flux is always independent
of the pressure gradient), and whose dynamical evolution is governed
by the Navier-Stokes equations.

The template in a framework for equations of motion and other
mathematical relations I call \emph{abstract}.  Canonical examples are
Newton's Second Law, the Schr\"odinger equation in non-relativistic
quantum mechanics, and so on.  Structure and entities at the highest
level of a theory formulated in a given framework I call
\emph{generic}.  In particular, generic structure has no definite
values for those quantities that appear as constants in the theory's
equations of motion and other mathematical relations.  The symbol
`$k$' appearing in the generic equation of motion of an elastic spring
modeled as a simple harmonic oscillator,
$\ddot{x} = - \frac{k} {m} x$, denotes Hooke's constant (the
coefficient of proportionality of a force applied to the spring and
the resulting diplacement from its equilibrium position), but
possesses no fixed value, and the same for the mass $m$.  It is
important to keep in mind, however, that all these formal
representations of physical quantities at the generic level in a
theory do have determinate physical dimensions, for Hooke's constant,
\emph{e}.\emph{g}.,
$\displaystyle \frac{\mathtt{m}}{\mathtt{t}^\mathtt{2}}$
(\emph{i}.\emph{e}., ``mass over time-squared'').  Otherwise, one
could not say that this is the generic equation of motion for, say, a
spring rather than a pendulum or oscillating string or electric
circuit or any other type of system whose dynamical evolution is
governed by the equation of motion for a simple harmonic oscillator.

One can, in the same way, write down generic solutions to the generic
equations of motion; these are formal representations of the
dynamically possible evolutions such systems can manifest.  One
generic solution to the equation of motion of an elastic spring is
$x(t) = \cos (\sqrt{\frac{k} {m}} \smidge t)$, where one may think of
`$k$' and `$m$' as dummy variables, not determinate real numbers.
Generic structure defines a \emph{genus} of physical system, all those
types of physical system the theory appropriately and adequately
treats.

One obtains \emph{specific} structure by fixing the values of all such
constants in generic structure, say $m = 1$ and $k = 5$ (in some
system of units) for the elastic spring.  This defines a
\emph{species} of physical system of that genus, all springs with
those values for mass and Hooke's constant.\footnote{One can as well
  consider mixed systems, with, say, a fixed value for mass but
  indeterminate value for Hooke's constant.  These raise interesting
  questions, but they are beside the point here.}  One now has a
determinate space of states for systems of that species, and a
determinate family of dynamically possible evolutions, \emph{viz}.,
the solutions to the specific equations of motion, represented by a
distinguished family of paths on the space of states.  A path in this
family is an \emph{individual model} (or \emph{individual solution})
of the specific equations of motion; one common way to fix an
individual model is to stipulate definite initial conditions for the
specific equations of motion.  An individual model, as the name
suggests, represents a unique physical system of the species, that
whose dynamical quantities satisfy the initial
conditions.\footnote{Everything I say so far applies only to
  deterministic theories that represent the dynamical behavior of its
  target systems with something like partial-differential equations.
  The discussion is easily modified to accommodate,
  \emph{e}.\emph{g}., stochastic theories, and theories such as
  classical thermodynamics that have no paths on a space of states
  representing physically possible processes.  It is not clear to me
  what the limits of this picture are in this regard---what its regime
  of applicability is, if you will.  It would be an interesting
  exercise to find an example of a physical theory that does not
  conform to it, and see whether one could emend the picture to as to
  account for the theory or not.  Either way, one would learn
  something of interest.}  In the case of the spring, that may mean
the unique one whose position and momentum at a given time have the
values given by the initial conditions.\footnote{I have deliberately
  taken my terminology from biological taxonomy, inspired by the
  remark of \citeN[p.~605]{peirce-doct-chnc}:
  \begin{quote}
    Now, the naturalists are the great builders of conceptions; there
    is no other branch of science where so much of this work is done
    as in theirs; and we must, to great measure, take them for our
    teachers in this important part of logic.
  \end{quote}
  There is much of insightful relevance in the lead-up to this remark,
  about how one individuates and characterizes genera and species of
  physical systems in my sense, which it would be illuminating to
  discuss, but it would take us too far afield.  I am tempted to
  describe structure at the level of a framework as \emph{phylar}, and
  to call the family of all physical systems treated by a framework a
  \emph{phylum}---and so all physical systems would fall under the
  kingdom of physics---but I suspect it would just be distracting to
  the reader.}

Finally, I call a \emph{concrete model} a collection of experimentally
or observationally gathered results structured and interpreted in such
a way as to allow identification with an individual
model.\footnote{This idea bears obvious and interesting comparison
  with the distinction between data and phenomena as drawn by
  \citeN{bogen-woodward-sav-pha}.  If I understand their distinction,
  their idea of data more or less corresponds with my idea of
  ``experimentally or observationally gathered results'', but their
  notion of phenomena, in so far as it seems to try to capture
  something like general patterns in the world, does not square with
  my conception of a concrete model, which is the result of
  structuring the results of a single experiment (or family of related
  experiments).  A complete discussion is beyond the scope of this
  paper.}  (I shall sometimes also refer to a concrete model as
\emph{structured data}.)  If I continually measure the position and
momentum of a spring with known values for $m$ and $k$, oscillating in
one dimension for an interval of time equal to its period of
oscillation, and graph the results ``in a natural way'',
\emph{e}.\emph{g}., as a curve parametrized by time on a Cartesian
plane whose $x$-axis represents position and $y$-axis represents
momentum, then I shall produce a curve that can be, ``in a natural
way'', identified with exactly one dynamically possible evolution on
the space of states used to represent that species of spring.
Kepler's organization and structuring of the planetary ephemerides
into parametrized ellipses satisfying the Area Law and the Harmonic
Law is a marvelous example from the actual history of physics of this
procedure.\footnote{I am making a serious over-simplification here.
  Not every concrete model contains data for a sufficient number of
  dynamic quantities to enable the identification of the concrete
  model with a unique individual model.  A set of data may contain,
  \emph{e}.\emph{g}., data only for the pressure of a Navier-Stokes
  fluid, which allows one to identify the concrete model only with a
  family of individual models, all those with those values for
  pressure, but having any values for all the other dynamic quantities
  Navier-Stokes attributes to fluids.  I also ignore the fact that, in
  light of the inevitability of expected error and experimental
  imprecision, one can \emph{always} identify a given concrete model
  with a family of individual models, \emph{viz}., all those whose
  values collectively lie within the error and imprecision intervals
  associated with the data.  I shall not discuss these issues, as they
  raise too many problems orthogonal to my purposes here, though the
  problems are of great interest in their own right.  I assume from
  hereon that a concrete model is such as to allow identification with
  a unique individual model.  In particular, I assume it contains
  values for enough dynamic quantities to determine individual
  states---at least as many as the ``dimension'' of the state
  space--and that determinate, individual values have been fixed for
  all data, not just intervals defined by the range of expected error
  and imprecisions.}

It is this identification of individual and concrete models that
embodies the substantive contact between theory and experiment
required to comprehend the full epistemic content of a theory; as we
will see, moreover, in general this cannot be done without
schematizing the observer.\footnote{This classification of levels of
  theory and data and the concomitant idea of identifying individual
  and concrete models have obvious similarities with the proposals of
  Suppes~\citeyear{suppes-mngs-uses-mods,suppes-mods-data}, and
  indeed much of my own thought has been inspired by my reactions to
  Suppes' extensive work on these questions (not just those two
  papers).}  There are many fascinating and deep problems associated
with characterizing the ways that experimental and observational
results are collected and appropriately transformed into structured
data, but I put them aside.\footnote{As is, again, so often the case,
  \citeN[p.~297]{suppes-mngs-uses-mods} sums the matter up with
  elegance and concision:
  \begin{quote}
    The maddeningly diverse and complex experience which constitutes
    an experiment is not the entity which is directly compared with a
    model of a theory.  Drastic assumptions of all sorts are made in
    reducing the experimental experience, as I shall term it, to a
    simple entity ready for comparison with a model of the theory.
  \end{quote}} Here, I shall give only a (schematic!)  outline of how
I envisage the procedure of bringing theory and experiment into
contact, based on the concepts I just articulated.  I give a more
finely grained analysis of the procedure in the next section.

The identification---I think essentially always---comes down in the
end to the brute comparison, one by one, of the values of
theoretically calculated quantities on the one side with the values of
the analogues of those quantities in the structured data on the other,
with the choice being made each time whether the values are close
enough, given one's understanding of the errors and imprecisions in
measurement, the introduction of skew in statistical reconstruction,
one's pragmatically chosen error tolerance, and other such factors.
An excellent example is provided by the comparison of the observed to
the calculated values for the planetary orbital periods and the
semi-major axes of the orbits in Book~\textsc{iii} of the
\emph{Principia}, in Newton's claim that the orbits of the planets and
of Jupiter's and Saturn's satellites all respectively satisfy Kepler's
Harmonic Law.  His explicit discussion of the errors involved in
constructing the concrete models, primarily due to limitations in the
observational instruments, and the errors involved in constructing the
individual models, primarily due to idealizations, and why they can
all safely be ignored when comparing the data to the theoretically
calculated results, exemplifies what I am trying to get at.  (See
especially Newton's discussions of Phenomenon~\textsc{i}, \textsc{ii},
and \textsc{iv}, and \citeNP{harper-newtons-sci-meth} for a
comprehensive and detailed discussion of Newton's analysis of the
data.)

Newton's discussion also exemplifies perhaps the most important
feature of such identification, at least for my purposes, \emph{viz}.,
the thoroughly and fundamentally pragmatic character of the whole
affair.  One does not evaluate the truth value of the proposition
``all the values are in close enough agreement''.  One rather chooses
to accept (or not) the theoretical structure as an appropriate and
adequate representation of the structured data, and one does this for
any of a number of reasons, one of which may be that the differences
in values one focuses on fall within one's pragmatically chosen error
tolerance.  The pragmatic nature of the acceptance is shown most
clearly by the fact that, even when all the values do \emph{not} fall
within one's error tolerance, there may be other principled grounds on
which one can reasonably choose to accept the identification of
theoretical structure with concrete model, \emph{e}.\emph{g}., to
serve as the ground of further, more refined investigation.  This is
what Newton did in his studies of the moon's librations and other
irregularities in its motion---he knew his initial, simple theoretical
models (2-body interaction ignoring the Sun, circular orbit) were
inadequate (in the sense of predictive accuracy) by any reasonable
measure, but he relied on them nonetheless in attempting to construct
better ones, taking into account ever more finely grained details of
the system (\emph{Principia}, Book~\textsc{i},
Proposition~\textsc{lxvi}, and Book~\textsc{iii},
Propositions~\textsc{xvii}, \textsc{xxii}, and
\textsc{xxv}--\textsc{xxxv}; see also
\citeNP{smith-newt-prob-moon-mot,smith-mot-lunar-apsis})---and he
had damn good reason to do so.  In such a case, there can be no
question of acceptance based on the evaluation of the truth value of a
proposition characterizing the accuracy of the data.

With regard to the exact relation between theory and experiment in
making such identifications, it is worth noting that one can
``Keplerize'' the planetary ephemerides without Newton's Laws or any
of the other most characteristic theoretical concepts of his system,
such as inertial mass and force.  One needs only the pre-theoretical
ideas of ``ellipse'', ``area swept out in given time'', ``distance to
sun'', ``orbital period'', and so on.  One constructs the structured
data using these pre-theoretical concepts, and then \emph{shows} that
it can be \emph{interpreted} using the concepts of the Newtonian
framework in such a way as to support identification of the concrete
model with a particular individual model.  And indeed that is what
Newton did.\footnote{My use of ``pre-theoretical'' here bears fruitful
  comparison with that of \citeN{hempel-stand-conc-sci-thrs}.}

To try to reassure the reader that I am not cherry-picking
particularly apt examples from the Greatest Hits in the History of
Science, I will also briefly discuss the recent detection of
gravitational waves by LIGO \cite{abbott-et-obs-gw-bin-bh}.  This
example also shows in a clear light where theory enters into the
design of the experiment and the construction of structured data, and
where it does not.  I give only a wildly simplified, schematic
description of the way observations are transformed into structured
data and compared with theory in the experiment.\footnote{I ignore
  many subtleties, such as the fact that radically different kinds of
  theoretical models and calculational methods are used to model
  different parts of the single process of in-spiral, coalescence and
  ring-down, and also the fact that the stress-strain is determined by
  the use of Michelson interferometers, and so optical theories are
  required as well as mechanical theories in the design and
  performance of the experiment and the structuring of the data.
  There are many fascinating philosophical problems of methodology and
  epistemology here.  See, for instance, the project being run by
  Lydia Patton on the problems associated with parameter estimation in
  LIGO
  (\url{https://www.researchgate.net/project/LIGO-and-parameter-estimation}),
  and her recent paper \citeN{patton-expand-thry-test-gr}.  See also
  the current doctoral work of Jamee Elder
  \cite{elder-phd-epist-gw-astro}, whose PhD thesis explores the
  whole area with great insight.}
\begin{enumerate}
  \noitemsep
    \item Design the instruments and their relative configuration so
  as to be susceptible kinematically to coupling with the kinds of
  physical phenomena one is interested in, according to the models of
  one's theories---in this case, differential stress-strains caused by
  ``distortion'' of spacetime as gravitational waves pass, in the
  persona of the relative acceleration of contiguous portions of
  continuous bodies (the instruments).
    \item Arrange the instruments in the desired configuration, fire
  them up, and begin collecting raw data.
    \item In the first pass through raw data, discriminate between
  ``noise'' and ``signal'', throwing out ``bad measurements'',
  winnowing signal pollution from other sources, ignoring mistakes and
  instrument malfunctions, and so on.  (This will have to be done at
  several points later in the process as well.)
    \item Match and collate different numerical ``data streams'' of
  raw stress-strain (in this case, using time-stamps of the measured
  stress-strain coming from instruments in geographically separated
  locations).
    \item Perform numerical interpolation, extrapolation, and
  manipulation (best-fit statistical analysis, \emph{etc}.\@) on the
  raw data so as to produce a continuous curve of aggregated
  stress-strain plotted against time; it is now structured
  data.\footnote{This description is too simplistic, even at the
    broad-brush level I am working at: theoretically generated models
    are partly used already here to separate signal from noise.  This,
    however, is controlled for by residuals testing,
    \emph{i}.\emph{e}., subtracting the best-fit waveform from data,
    and then comparing the residue to normal detector noise.  It is
    also controlled for by testing for phenomenological deviation, by
    comparing data with templates generated by analytically deforming
    templates away from general relativity.  In this way, one
    effectively removes dependence on general relativity from this
    stage of the process.  See \citeN{abbott-et-tests-gr-gw150914}
    and \citeN{elder-phd-epist-gw-astro} for more thorough
    discussion.}
    \item Using the theory of general relativity, construct models of
  the production of gravitational waves in different phases of the
  process (in-spiral of a binary black hole system, coalescence,
  ring-down of the resultant black hole to equilibrium); calculate the
  waveforms as they would appear when passing through the Earth.
    \item Based on those waveforms, construct continuous graphs
  representing the theoretically predicted responses of the
  instruments to the stress-strain resulting from them.
    \item Compare the structured-data graph with the theoretically
  constructed graphs.
    \item Decide, based on calculation of expected errors, statistical
  analysis of the rate of expected false positives, and so on, whether
  the structured-data graph matches one of the theoretical graphs to
  within one's error tolerance.
    \item If the match is good enough, use the physical principles of
  general relativity to interpret the measured stress-strain in the
  instruments as the result of passing gravitational waves produced in
  the coalescence of a binary black hole system.
    \item Identify the structured data with an individual model (or
  family of models) of a coalescing binary black hole system.
\end{enumerate}
One of the most interesting points for our purposes is where the
dynamics of general relativity, \emph{viz}., the Einstein field
equation itself, enters: only in steps 6 and 7.\footnote{It may also
  come into play in step 9, in the determination of expected errors,
  but that discussion is far beyond the scope of this paper.}  Step 1
may appear already to depend on the Einstein field equation, but the
appearance is illusory, arising from the description I used.  The
instrument needs only to be sensitive to stress-strain induced by
``gravitational forces'', irrespective of how one interprets,
conceives or represents such a thing.  Of course, one must know how to
interpret the relevant notion of ``gravitational force'' in the
conceptual framework of general relativity, and how to represent
stress-strain, for all of which the theory's kinematics suffices.  In
particular, one needs to know only how to measure the relative
acceleration of contiguous parts of continuous bodies (induced
stress-strain in the instrument), and to ensure that the stress-strain
does not arise from any source other than ``gravitational''
(electromagnetic forces, \emph{e}.\emph{g}., or mechanical
disturbances arising from vibrations in the Earth's surface caused by
passing semi trucks).  None of that depends on the assumption (or lack
thereof) that the Einstein field equation holds for the matter fields
one is working with.

The way that the rest of general relativity comes into play, moreover,
in steps 1, 10 and 11, does not make the experiment ``theory-laden''
in any interesting sense, because it is only the pre-theoretical
concepts of stress-strain, relative acceleration, and so on, that
ground the construction of the instruments and the structuring and
analysis of the data used to test the predictions.\footnote{These
  concepts are pre-theoretical with respect to general relativity, not
  with respect to the theories (\emph{e}.\emph{g}., of the continuum
  mechanics of solid bodies) that are used to construct detailed
  models of the instruments themselves and their reactions to imposed
  forces.}  The Einstein field equation comes into play only in
producing predictions to compare to observed results, and then the
rest of the conceptual framework of general relativity comes into play
in interpreting the results by the identification of a structured data
model with an individual model (or family of models) in the
theory.\footnote{At this point, the experiment has not ruled out the
  possibility that other theories of gravity besides general
  relativity might also produce predictions appropriately and
  adequately matching the structured data.  This fact highlights the
  important point that different theories---\emph{e}.\emph{g}.,
  general relativity and many alternative theories of gravity---can
  share a substantial amount of epistemic content, \emph{viz}., that
  embodied in various pre-theoretical concepts and general principles
  they obey, and that encoded in the structured data formed from
  observations that can be modeled by and can be used to test
  theoretical propositions in the different theories.  It should thus
  be clear that I reject the ``incommensurability'' of theories in any
  deep or interesting sense.}

I hope it to be clear how all 4 aspects of both types of knowledge (as
I sketched them in \S\ref{sec:forms-aspects-know}) come into play in
inextricable ways throughout the entire process.  The construction of
the graphs representing theoretical predictions of instrument
responses, for example---a necessary part of building the evidentiary
web of relations that constitutes the confirmation of general
relativity by the experiment---requires input from both experimental
and theoretical sources.  One cannot, however, point to the
``experimental part'' and the ``theoretical part'' of the graphs.  Nor
can one understand the graphs' empirical import using the formalism of
the theory in conjunction with minimal interpretive principles---among
other reasons, to do so can give no reason to believe that the graphs
have the capacity to represent systems in the theory's regime of
applicability.

One may well ask, however: if the only theoretical input to the design
of the instruments came from pre-theoretical notions, how can I claim
(as I did in \S\ref{sec:need-schem-obsr}) that the guidance of
instrument design is one of the most profound ways that \emph{theory}
and experiment make contact with each other?  The answer is twofold.
First, even though ``stress-strain'' as used here is pre-theoretical
with respect to general relativity in the sense that one need not use
general relativity to characterize it, it is still an essential part
of the conceptual framework of general relativity.  One must be able
to ensure that the way one characterizes stress-strain in general
relativity is consistent with, and indeed identifiable with, the
pre-theoretic notion one explicitly used in designing the instrument.
In that way, the conceptual structure of general relativity, its
kinematics, did indeed guide the design of the instrument.  Second,
the design of the configuration of the experiment, how the instruments
are deployed and used, must respect both the theory's kinematics and
its dynamics.\footnote{I explicate ``kinematics'' and ``dynamics'',
  and further discuss this point, in \S\ref{sec:breakdown-regimes}.}

\citeN[p.~639, emphases his]{stein-struct-know} speaks of ``\emph{a
  mathematical structure [of a theory] discernible in the world of
  phenomena, of observations, of experience}.''  I endorse this idea
by the requirement that structured data be identifiable with formal
theoretical structures.  I mean, however, something quite weaker than
what many contemporary philosophers would assume on reading Stein's
phrase in the absence of any context.  I mean only that the theory
facilitates the successful identification of individual and concrete
models in such a way as to make it possible for its entire formal and
conceptual apparatus to come into play and form the basis for further
successful reasoning about the systems at issue.  In particular, I
make no claims that structures in the individual models and in the
concrete models, much less ``in the world itself'', are ``isomorphic''
or ``homomorphic'' or ``similar'' in any way above and beyond the fact
that they are relevantly identifiable with each other.\footnote{See
  the discussion in the final section of
  \citeN{curiel-cm-lag-not-ham} for examples of how one can identify
  the same concrete model with individual models in different theories
  (the Lagrangian and Hamiltonian formulations, respectively, of the
  theory of an oscillating spring) without any physically significant
  morphism between any of the models at issue.}  (I do not even know
what it means to speak of an isomorphism or homomorphism or similarity
between a mathematical structure and ``the world
itself''.\footnote{\label{fn:isomorph}Jeremy Butterfield countered
  this claim in conversation by adverting to work on the
  correspondence theory of truth, such as
  \citeN{davidson-true-facts}, in which there seems to be a
  straightforward sense of isomorphism between a formal semantic
  structure and parts of the world.  I can accept the cogency of such
  work but still deny that the isomorphism at issue consists of
  anything more than the kind of bare, thin agreement between formal
  semantic structures and parts of the world captured by the
  identification of concrete models and individual models in my sense.
  For the claim that there is a cogent sense of isomorphism between a
  mathematical structure and the world to be true in a substantive and
  interesting sense, I would want to see an example like the
  following: there is an isomorphism between the (standard) topology
  of $\mathbb{R}^4$, as used in flat FLRW cosmological models in
  general relativity, and the topology of the real world.  I do not
  know, however, what ``the topology of the real world'' means.  I can
  certainly envisage ways to operationalize its meaning, but different
  operationalizations, each with manifest physical content, will yield
  different topologies.  One obvious way will yield the standard
  manifold topology (at least locally).  Another, however, based on
  time functions, will yield a different topology
  \cite{geroch-dom-dep}.  So which one is ``the'' topology of ``the
  world itself''?})

Indeed, the sense in which there may be any structure in the concrete
model isomorphic with structure in the individual model is etiolate at
best.  In the case of LIGO, the extrapolated continuous curve of
stress-strain data is isomorphic with the analogous curve
theoretically constructed on the basis of the individual model general
relativity yields, but that is only the minimal consistency condition
required in order to justify the identification of the two.  (Note,
moreover, that the theoretical curve ``isomorphic'' to the data curve
is, strictly speaking, not even part of the individual model in the
sense of the standard view!)  The additional structures in general
relativity one needs in order not only to justify the identification
but further to give empirical content to the individual
model---additional structure such as the relation of the relative
acceleration of neighboring non-geodetic curves to the Riemann tensor,
\emph{e}.\emph{g}.---have no correlate, isomorphic or otherwise, in
the concrete model.  The concrete model has no affine structure.

In any event, the identification is, strictly speaking, not a relation
between ``different structures'' at all, in the sense of being
representable by a proposition whose truth value one can evaluate
based only on consideration of theory and data alone---it is, as I
have already said, a pragmatic choice: does one accept that the two
can be identified with each other or not?  It is a pragmatic choice in
the same way as use of a framework in the sense of \citeN{carnap-eso}
is decided by pragmatic choice.  This is why pragmatics and
epistemological concerns, not ontology, lie at the ground of
semantics, as I will argue in \S\ref{sec:sems-epist-onto} below.
Nonetheless, I shall continue to speak of this identification as a
``representation relation'' for convenience, though I emphasize again
that by ``representation'' I do not mean a relation whose content can
be exhausted by reference to the theory and data alone, nor even by
reference to theory, data and theory-users (\emph{\`a la} van
Fraassen)---it is a relation that relies on schematizing the observer.

One can agree with this analysis of how theory and experiment make
contact with each other while remaining agnostic about all issues
pertaining to realism and anti-realism.  Indeed, none of this has
anything to do with any issue pertaining to structural realism in
particular, nor to debates about realism and anti-realism in general.
There is no claim made or needed that the structure manifest in the
structured data is ``really'' part of the furniture of the world, in
some deep metaphysical sense.  That it is manifest in the structured
data suffices for the soundness of the representational relations at
issue.  Those relations are agnostic about realism, as any good
epistemic relation should be.

I conclude this section with a summation of where in all this the
schematic representation of the observer enters.  In the case of
Newton's investigations, they appear explicitly in his defense of the
way he identifies his theory's models with the Keplerized orbits, by
explaining the way, \emph{e}.\emph{g}., one expects variation in the
observed size of Jupiter's apparent diameter, and thus in the distance
its satellites appear to be from its center, when using telescopes of
different sizes and resolving powers (Book~\textsc{iii},
Phenomenon~\textsc{i}).  In the case of LIGO, they appear in several
places, perhaps most notably for our purposes in the calculation of
the response of the instrument to applied stress-strain from passing
gravitational waves.  It is true that often, if not indeed almost
always, explicit representations of the observer and the instruments
will not appear in the concrete models---they get washed out in the
structuring of the data.  They are still there implicitly in the
background, however, and the eventual identification of individual and
concrete model, including the interpretation of the pre-theoretical
terms in the structured data using the concepts of the full theory,
can happen only because of them.

At a deeper level of analysis, schematic representation of the
observer is what grounds the circumscription of the regime of
applicability of a theory, because different kinds of instruments,
naively modeled and deployed, give different outcomes for ``measuring
the same thing'' near the boundaries of the regime.  Circumscribing
the regime, however, grounds all other aspects of what one may want to
call the semantics of a theory in particular and its epistemic content
more generally, so it is to this problem that I now turn.

\section{The Breakdown Scale and the Regimes of a Theory}
\label{sec:breakdown-regimes}

\skipline[.5]

\begin{quote}
  \begin{tabbing}
    \hspace*{8em}\=When every year and month stress-energy\=nds\kill
    \>This shaking keeps me steady. I should know. \\
    \>What falls away is always. And is near. \\
    \>I wake to sleep, and take my waking slow. \\
    \>I learn by going where I have to go. \\
    \>\>-- Theodore Roethke \\
    \>\>``The Waking''
  \end{tabbing}
\end{quote}

\skipline

It is characteristic of an appropriately unified kind of physical
system, one treated by a theory in my sense, that there exist a set of
scales at each of which all quantities the theory attributes to the
kind of system simultaneously lose definition.  In other words, every
theory, in so afar as it treats an appropriately unified kind of
physical system, not only has a regime of applicability, but it has a
\emph{single, unified} one, bounded on all sides by scales
characterized by the values of different combinations of its physical
quantities.  For classical fluids, for example, the definitions of
their pressure, fluid flow, viscosity, and all the rest break down at
spatial and temporal scales a few orders of magnitude greater than
those of the mean free-path of the fluid's constituent particles.  All
those quantities also lose definition when the fluid enters a strong
enough state of turbulence, which can be characterized by (\emph{inter
  alia}) a ratio of the fluid's kinetic energy to a measure of its
viscous damping---a scale independent of the small spatial and
temporal ones characterized by the mean free-path.  There is no
\emph{a priori} reason why the definitions of all the different
physical quantities represented by the theory should fail at the same
set of characteristic scales, even though, in fact, those of all known
examples do, not only for theories of classical fluids but for all
physical theories we have.\footnote{This is a fact that deserves
  philosophical investigation.  I discuss it in more detail in
  \citeN{curiel-prpy-basis-sems}, but with no pretense of even
  touching on all the interesting questions this fact raises.}  This
seems, indeed, to be one of the markers of a physical theory, the
existence of a set of characteristic scales for its physical
quantities, at each of which all the theory's physical quantities
simultaneously lose definition---``places'' where all the
kinematically and dynamically relevant structures of the theory break
down all at once, in the sense that the theory becomes inadequate for
an appropriate treatment of any system beyond the determined
boundaries.\footnote{
  \begin{quote}
    \skipline[-.5]
    \begin{tabbing}
      \=When every year and month \hspace*{2em}\=\kill
      \>Und wir:  Zuschauer, immer, \"uberall, \\
      \>dem allen zugewandt und nie hinaus! \\
      \>Uns \"uberf\"ullts.  Wir ordnens.  Es zerf\"allt. \\
      \>Wir ordnens wieder und zerfallen selbst. \\
      \skipline[-.3]
      \>\>-- Rainer Maria Rilke \\
      \>\>``Eighth Duino Elegy''
  \end{tabbing}
\end{quote}} Because all the quantities a theory attributes to the
systems it treats lose definition at the same time when the system
passes beyond a breakdown scale, the idea is unambiguous.

Although we perhaps naively tend to think of scales determined by
spatial, temporal and energetic quantities when considering how and
where theories break down in their capacity to provide sound
representations of phenomena, any quantity in any theory can provide
such a measure.  Velocity provides a breakdown scale for Newtonian
mechanics, and acceleration and scalar curvature provide different
breakdown scales for various theories of gravity, such as general
relativity.  No breakdown scale, moreover, can be a single number
holding for all systems the theory treats.  Navier-Stokes theory, for
instance, becomes inadequate for different fluids at different
energies and spatial and temporal scales.  Often it is not a bound on
a single quantity, such as a value of energy, a value of spatial
length, \emph{etc}.: classical Maxwell theory, \emph{e}.\emph{g}.,
breaks down when the ratio of the field's amplitude to its frequency
approaches $\hbar$.  Nor is it ever the case that there is a single
characteristic scale for each theory.  Navier-Stokes theory breaks
down when various measures of flow complexity indicate the fluid is
approaching turbulence, when the fluid is too viscous, when one tries
to use the theory to specify behavior at time scales comparable to
equilibration time after a sharp disturbance, when temperatures become
large enough that heat loss due to emission of blackbody radiation
becomes non-neglible, when the ambient electromagnetic field in the
environment becomes strong enough that the constituent molecules of
the fluid begin to denature and ionize, and on and on---which shows,
moreover, that sometimes a breakdown scale is determined by physical
quantities not even representable in the theory (such as the
electromagnetic field for Navier-Stokes).  A breakdown scale, then, is
something like the following: a measure of or function of or relation
among quantities, such that, when the joint state of the system and
its environment imply that the values of some of the system's
quantities do not satisfy the measure, function or relation, then the
theory can no longer provide good models of the system.\footnote{I
  gloss over a subtlety here: sometimes approximations used to
  construct models of particular phenomena, such as surface waves in
  fluid dynamics, have characteristic breakdown scales different from
  those of the material in which the phenomena manifest
  \cite[ch.~\textsc{ix}]{lamb-hydro}.  The theory can provide
  appropriate and adequate models of the systems in the relevant
  states, only not in the way the approximations definitive of the
  phenomena at issue require.}  (I provide a more precise definition
below.)

Breakdown scales can never be determined by analysis of the formalism
and theoretical machinery of the theory alone, without input from
knowledge acquired by experimentation in particular and empirical
investigation in general.  They are rather fixed by knowledge that one
can gather only from investigations grounded in that part of the
epistemic content of the theory not captured by the formalism by
itself.  As such, they change with the increasing scope and depth of
our experimental reach.\footnote{This is not to say that one can never
  find relations among quantities derived from purely theoretical
  considerations that determine \emph{some part} of the breakdown
  regime.  \citeN[pp.~2--4]{geroch-hyp-rel-diss-fl} points out, for
  example, that any relativistic theory of dissipative fluids---such
  as Navier-Stokes theory posed in the relativistic context---must
  fail on every combined time-distance scale $\tau$-$d$ such that
  $d^2 < \eta \tau$ and $ d > c \tau$, where $\eta$ is a value of a
  typical Navier-Stokes dissipation coefficient and $c$ is the speed
  of light.  Instead of a characteristic break-down scale, this
  requirement defines a characteristic break-down \emph{area} in the
  $\tau$-$d$-plane.  Note that the complement of this region in the
  plane, that is, the region in which the system remains valid (at
  least so far as these conditions are concerned), includes
  arbitrarily small $d$-values and arbitrarily small $\tau$-values
  (though not both at the same time).  This all follows from
  considerations of relativistic constraints on causal propagation
  alone.  Such relations, however, never exhaust the boundary of the
  complete breakdown scale, and in general form only a small part it.
  The determination of whether any given system in any given state
  satisfies such relations, moreover, is still sensitive to the state
  of experimental prowess---we can always improve the accuracy and
  precision of our measurements of $\tau$, $d$, $c$ and $\eta$.}
Every theory has many breakdown scales, putting constraints on the
values of all the quantities the theory treats.  For simplicity,
however, I shall speak as though all such scales are spatial or
temporal.

What does it mean to say that the theory cannot provide good models of
systems outside its breakdown scales?  One of the most important
markers of this is that the quantities the theory attributes to the
system lose unambiguous definition.  For a Navier-Stokes fluid, for
example, different sorts of thermometers that allow spatial
discrimination on scales only a couple of orders of magnitude greater
than the mean free path of the fluid's molecules will record markedly
different ``temperatures'' depending on characteristics of the joint
system that one can ignore at larger scales---the fine details of the
fluid's convective flow in relation to the geometry of the
thermometric system, for instance, and even the transparency of each
thermometer to the fluid's particles.\footnote{See,
  \emph{e}.\emph{g}., \citeN{benedict-funds-temp-press} for detailed
  exposition of the complex interplay among theory, model and
  experiment one must take account of in attempting to define a
  physical quantity such as temperature based on the behavior of real
  measuring devices, that is, to render epistemic content to any
  representation of it, and to deploy it meaningfully in real
  applications of theory.  It is not the most up-to-date reference
  with regard to the international agreement on defining the standard
  methods for the determination of temperature (for which see,
  \emph{e}.\emph{g}., \cite{crc-handbook-chem-phys-95}), but I have
  found no better guide to the nuts and bolts of thermometry when it
  comes to developing in a physically illuminating way how the
  mathematical models and the experimental techniques bear on each
  other.}  As the fluid approaches turbulence, to take another
example, the values of all its quantities begin to vary rapidly in
time and eventually cannot be measured by any conventional means---the
quantities are no longer well defined.  In practice, all that can be
measured are averages (usually time averages), leading to the
application of the idea of ensemble averages and the development of
statistical generalizations of the Navier-Stokes equations to treat
the fluid flow \cite{foias-et-nse-turb}.

This discussion shows how one cannot even define physical
quantities---\emph{e}.\emph{g}., temperature---without explicit
schematic representation of the observer, much less have understanding
of how to employ their representations in scientific reasoning in ways
that respect the regime of applicability.  Without such a definition
and such understanding, however, any theoretical proposition referring
to temperature can have no epistemic content.  The ideal of
segregability enshrined in the standard view is illusory.

In order to draw out the full import of the existence of breakdown
scales for the characterization of the epistemic content of a theory,
however, a few more definitions and distinctions must be
made.\footnote{See \citeN{curiel-kins-dyns-struc-theors} for a more
  comprehensive and detailed discussion of all the issues I treat
  here.}  The most basic distinction is between what I call the
kinematical and the dynamical parts of a theory.  The kinematical
parts, roughly speaking, are all those features of a theory's
treatment of a system that remain the same irrespective of the state
the system is in and irrespective of the kinds of interaction it has
with its environment.  The kinematic quantities are, therefore, those
which are assumed to be constant.  In relativity theory, this includes
the speed of light \emph{in vacuo}.  In Newtonian particle mechanics,
the masses of the particles are assumed to be fixed once and for all.
In Navier-Stokes theory, the fluid's shear and bulk viscosities and
its coefficient of thermoconductivity are assumed to be constant.  The
dynamic quantities are those that are allowed to vary as the system
evolves, such as position, velocity, momentum, shear-stress, and so
on.

All theories, moreover, attribute not only constant quantities to its
systems, but also fixed mathematical relations among both its
kinematic and dynamic quantities whose form is always the same, what I
call its kinematical constraints.  In Newtonian mechanics, the
velocity is always the time-derivative of position.  (This is not so
trivial a requirement as it may sound, as there are theories in which
it fails, \emph{e}.\emph{g}., in some formulations of relativistic
fluid mechanics; see \citeNP{landau-lifschitz-fluid} and
\citeNP{earman-stat-thermo-rel}.)  Navier-Stokes theory requires
that the shear-stress tensor be symmetric, and that heat flux be
independent of the pressure gradient.  The attribution of definite
values for the kinematic quantities to a species of system also is a
kinematical constraint, \emph{e}.\emph{g}., that water under
``normal'' conditions have the value (approximately) 1000
\texttt{\textmu Pa-s} (micro Pascal-seconds) for its shear viscosity.

Every theory postulates as well dynamical relations among its
quantities, the more familiar equations of motion or field equations.
These have the characteristic property that their specific form
depends on the interactions the system has with its environment.
($\mathbf{F}$ always equals $m \mathbf{a}$ in Newtonian mechanics, but
the functional form of $\mathbf{F}$ may be anything one wants,
depending on what forces the system at issue is subject to.)

Now, to be able to bring theory and experiment into physically
significant contact by having the capacity to identify individual and
concrete models, one must demonstrate that the structures intrinsic to
the theory are appropriate and adequate for representing and reasoning
about the genus of physical systems the theory purports to treat.  I
am now in a position to give (somewhat) precise definitions for these
terms.  The theory is \emph{appropriate} (or has \emph{propriety in
  representation}) for treating a given system if all the quantities
the theory attributes to a system are well defined and they jointly
satisfy all the theory's kinematical constraints, given the state of
the system in conjunction with the interactions it has with its
environment.  Thus, a theory can be used with propriety to treat a
type of physical system it putatively represents if and only if the
system's environment and its own state jointly permit the
determination, within the fineness and ranges allowed by their nature,
of the system's quantities over the variously relevant scales
appropriate for the representation of the relations among the
quantities manifested in the phenomena at issue.

Propriety is a property accruing to the representation of individual
states, not to the course of a dynamical evolution: one may not be
able to identify an entire concrete model with an entire individual
model, but, if one is to apply the theory in a meaningful way at all
to the treatment of a system, at a minimum one must be able to
identify those substructures of each that refer only to single states.
Since the kinematical constraints are the same for all states, it
makes sense to ask whether they are satisfied so long as the
quantities themselves are well defined.  Indeed, I now impose a
stronger criterion for the presence of a breakdown scale, in the form
of a definition.
\begin{defn}
  \label{def:breakdown}
  A \emph{breakdown scale} of a theory is a bound on or function of or
  relation among its quantities such that the kinematical constraints
  are satisfied to the required degree of accuracy given the
  experimental techniques used for probing them if and only if the
  joint state of the system and its environment imply that the bound
  or function or relation is satisfied.
\end{defn}
If the kinematical constraints are not satisfied, one has no reason to
think that the system at issue is one the theory can treat at all.

It is a brute fact about the actual physical theories we use in real
science that they become predictively inaccurate in regimes well
separated from all its breakdown scales, \emph{i}.\emph{e}., well
before the theory's kinematical constraints are no longer satisfied.
The theory is \emph{adequate} if its dynamical relations are satisfied
to some prescribed degree of accuracy, given the experimental
techniques used for measuring it.  Thus, a theory is adequate if and
only if entire concrete models can be identified with entire
individual models.  If the theory does not have propriety in
representation, it cannot be adequate.  It is only now, when we have
reached the point when it makes sense to inquire after a theory's
adequacy, that the issue of the accuracy of the theory's
\emph{predictions}---their ``truth''---becomes meaningful.

This discussion suggests the following.
\begin{defn}
  The \emph{regime of propriety} of a theory is the family of physical
  systems whose states, in conjunction with their interactions with
  their environments, are bounded on all sides, in all ways, by all
  the breakdown scales of the theory.
\end{defn}
In this regime, the theory's representational resources are
appropriate for modeling the kinds of system at issue, even if they
are not predictively accurate.  In particular, all quantities are well
defined, and all kinematical constraints are satisfied.  A system is
in the theory's regime of propriety if and only if one can identify
those parts of an individual model representing a single state with
those parts of a concrete model representing the same.
\begin{defn}
  The \emph{regime of applicability} of a theory is a subset of the
  regime of propriety, in which the theory's dynamical equations are
  predictively accurate in their modeling of the behavior of the
  family of systems at issue---the theory model's are adequate for
  ordinary scientific usage.
\end{defn}
A system is in the regime of applicability---it is adequate---if and
only if one can identify an entire individual model with an entire
concrete model.\footnote{Identifying the regimes in this way with
  ``the family of all systems such that\ldots\@'' does not fall prey
  to the problems the standard view faces when it attempts to do
  something similar: humans will never be able to identify the
  entirety of the regime of propriety of any theory in practice, but
  we can determine whether any given system belongs to it or not.
  That is what the standard view fails to do, blinkered by its own
  methodological restrictions.}

A theory may appropriately treat a family of phenomena even when it
does not model the dynamical behavior of all members of the family to
any prescribed degree of accuracy, \emph{i}.\emph{e}., even when the
equations of motion are not satisfied in any reasonable sense.  A
theory, that is to say, can and does tell us much about the character
and nature of physical systems for which it does not give accurate
representations so long as they are in its regime of
propriety---systems, in other words, it cannot soundly represent in
totality, cannot be true of, and so systems that, according to all the
standard contemporary accounts of theory structure and semantics, the
theory should have nothing to say about at all.  I shall now discuss
four roles that propriety, even in the absence of adequacy, plays in
informing and contributing to the epistemic content of a theory.

Fisrt, theories do not predict kinematical constraints; they demand
them.  I take a prediction to be something that a theory, while
appropriately modeling a system, can still get wrong.  Newtonian
mechanics does not predict that the velocity of a body equal the
temporal rate of change of its position; rather it requires it as a
precondition for its own applicability.  It can't ``get it wrong''.
If the kinematical constraints demanded by a theory do not hold for a
family of phenomena, that theory cannot treat it, for the system is of
a type beyond the theory's scope.  By contrast, if the equations of
motion are not satisfied, that may tell one only that one has not
taken all ambient forces on the system (couplings with its
environment) into account; it need not imply that one is dealing with
a different type of system.  Even in principle, one can never
decisively rule out the possibility that the equations of motion are
inaccurate only because there is a force one does not know how to
account for, not because the system is not appropriately treated by
those equations of motion.  This can never happen with a kinematical
constraint.  It is either satisfied, to the appropriate and required
level of accuracy given the measuring techniques available and the
state of the system and its environment, or it is not.  Thus, the
propriety of a theory constitutes its necessary preconditions of
applicability.\footnote{In this sense, the kinematical constraints
  play a role analogous to that proposed by
  \citeN{reichenbach-rel-apriori-know} and
  \citeN{friedman-dyns-reason} for what they call the relativized
  \emph{a priori} of a theory.  Kinematical constraints, however,
  differ in this respect: they are part of the theory itself, not
  supra-theoretical principles, as are the relativized \emph{a priori}
  of the neo-Kantians.  Also, satisfaction of the kinematical
  constraints can, indeed must, be \emph{experimentally} verified in
  order for one to ascertain that the theory represents the system
  with propriety.  To the contrary, satisfaction of the relativized
  \emph{a priori}, in the neo-Kantian sense, does not admit of
  empirical verification, but rather grounds the possibility of
  empirical investigation in a stronger sense.}

Indeed, satisfaction of kinematical constraints is required in general
for the equations of motion of a theory to be well posed or even just
consistent (the second role for kinematical constraints).  The
initial-value formulation of the Navier-Stokes equations, for example,
is well set (in the sense of Hadamard) only if the shear-stress tensor
is symmetric, and it is thermodynamically consistent only if the heat
flux is independent of the pressure gradient \cite[ch.~\textsc{v},
\S49]{landau-lifschitz-fluid}.  One cannot even formulate Newton's
Second Law if velocity is not the first temporal derivative of
position.  More generally, in a sense one can make precise
\cite{curiel-cm-lag-not-ham}, if the kinematical constraints of
Lagrangian mechanics are not satisfied
($\mathbf{v} = \mathbf{\dot{q}}$), then one cannot formulate the
Euler-Lagrange equation; and similarly, if the kinematical constraints
of Hamiltonian mechanics are not satisfied (the $\mathbf{p}$s and
$\mathbf{q}$s do not satisfy the canonical Poisson-bracket relations),
then one cannot formulate Hamilton's equation
\cite{curiel-cm-lag-not-ham}.  Thus satisfaction of the kinematical
constraints is required as a precondition for the appropriate
application of a theory in modeling a kind of system.

Third, it is essential that all the kinematical constraints be
satisfied for the theory to be able to represent any aspects of the
system at all, for it is those kinematical constraints, not the
equations of motion, that characterize the genus of system at issue.
Many systems have, \emph{e}.\emph{g}., shear-stress (Navier-Stokes
fluids, elastic continua, Maxwell fields, charged plasmas,
\emph{etc}.); what makes a shear-stress the shear-stress of a
Navier-Stokes fluid as opposed to that of a Maxwell field is its
satisfaction of the Navier-Stokes kinematical constraints,
\emph{e}.\emph{g}., that the shear-stress be, in a sense one can make
precise, transverse to the fluid flow.  By contrast, you can't throw a
rock without hitting a system whose equations of motion are that of a
simple harmonic oscillator: if I know only the equations of motion, I
cannot in general tell you what kind of system I am dealing
with.\footnote{Thus, kinematical constraints are constitutive of the
  systems the theory treats, also in a way analogous to the
  relativized \emph{a priori} of the neo-Kantians.}

Finally, it is the kinematical constraints, not the equations of
motion, that guide the experimentalist in the design of instruments
for probing and measuring the quantities the theory attributes to the
systems it treats.  (Recall the general discussion of this issue in
\S\ref{sec:need-schem-obsr}, and the analysis of this issue in the
particular case of LIGO in \S\ref{sec:theor-exper-contact}.)  An
instrument that is to measure Newtonian velocity, for instance, must
be sensitive to differences in spatial location at ever smaller
measured temporal intervals, even if only indirectly, in accord with
the kinematical constraint $\dot{\mathbf{x}} = \mathbf{v}$.  Such an
instrument, if well designed, does not care about how the system
accelerates, \emph{i}.\emph{e}., about its dynamics.  Similarly, an
instrument that would measure shear-stress of a Navier-Stokes fluid
must conform to the equality of pressure and reversed sense of shear
across imaginary surfaces in fluid that is represented by the symmetry
of the shear-stress tensor.  Again, the instrument need not care at
all about the dynamics of the fluid for it to measure the
instantaneous value of the quantity.  In this way, kinematical
constraints provide the foundation for the operationalization of the
meaning of theoretical terms.

There is one more subtlety in the idea of the regimes of a theory that
must be addressed.  It is in fact never the case that an individual
model of a given physical system represents the system appropriately
and adequately \emph{in toto}.  There is always a sense in which some
``parts'' of the model represent nothing physical at all.  Consider
again individual solutions to the Navier-Stokes equations.  Because
Navier-Stokes is a continuum theory, the solutions to its equations
allow one to make, indeed they \emph{necessarily encode}, what seem to
be physical predictions at arbitrarily small spatial and temporal
scales.  We know, however, that the theory is not appropriate for that
task, and we know we can ignore those ``predictions'' of the theory in
assessing its adequacy and soundness, for real fluids are not
continua.  General relativity is scaleless---one cannot formally
distinguish a model of a Schwarzschild black hole of radius
$44\mathtt{x}10^{-10^{39876}}$ \texttt{km} from one of radius 44
million \texttt{km}.  And yet we also know that the theory is
appropriate for treating the latter, not the former.  Indeed, even in
the case of a Schwarzschild black hole of radius 44 million
\texttt{km}, which does find appropriate and adequate representation
in the theory, not all ``parts'' of the individual model represent.
One finds again the same circumstance as in Navier-Stokes theory: we
have good reason not to trust ``predictions'' general relativity makes
in such models about the geometry of spacetime at spatiotemporal
scales approaching the Planck scale.  Propositions about geometry at
such scales formulated in the model have no empirical content.  That
in no way detracts from the sound epistemic content the model accrues
by the propositions it allows one to formulate about geometry at much
larger scales.

One comes to recognize and understand these limitations in the
representational capacities of theories only by knowledge of the
breakdown scales, and correlatively of the regime of propriety.  Thus,
the knowledge of how the mathematics is to be applied---how it
represents and what it represents---is not segregable from the
practical knowledge that grounds the determinations of the scales and
regimes.  This is not a matter of ``segregating part of the
mathematical structure as representational fluff (`gauge')''---for
there just is the solution itself.  The fact that one cannot use it to
model phenomena happening on small spatiotemporal scales does not mean
that ``part'' of the solution---``that part representing small
stuff''---is gauge, for there is no such part cleanly segregable from
the rest.  It is not like the case of the Faraday tensor and one of
its 4-dimensional gauge potentials, two different mathematical
entities with a fixed relation between them.  And it is not that there
is a simple rule, \emph{e}.\emph{g}., ``ignore everything in cubes
whose edge is smaller than the mean free-path of the fluid's
constituents'', because such rules will in general be incorrect.
Sometimes what happens on such scales is physically relevant (a
discontinuity in a boundary condition, say, or wild small-scale
fluctuations that render coarse-grained statistical averages
meaningless).  One has to use one's judgment, on a case by case basis,
to determine how the formalism encodes potential knowledge, how it
successfully represents parts of the world, how it can be used as part
of sound, legitimate reasoning.\footnote{This problem underscores the
  shallowness of much of the debate about realism in science, in
  particular the almost ubiquitous yet deeply problematic implicit
  assumption that ``realism'' is a holistic attitude---a mathematical
  structure \emph{in toto} ``really represents'' or it does not.
  That, however, is too coarse a conception.  Some aspects of a
  mathematical structure may represent, and others not,
  \emph{e}.\emph{g}., the integrated phase differences in a quantum
  wave-function (Berry phase) may be ``real'' whereas the phase itself
  may not, or those ``parts'' of a Navier-Stokes model representing
  behavior on scales greater than 10$^{-4} \mathtt{cm}$ may ``really
  designate'' whereas those parts representing smaller scale behavior
  may not.  \citeN{halvorson-real-qt} provides a splendid discussion
  of related matters.}

To hark back to the discussion, in \S\ref{sec:forms-aspects-know}, of
the different forms and aspects of scientific knowledge, I argued
there that that the standard view of theories gives one too little in
answer to the question of a theory's epistemic content.  Here, by
contrast, is a case in which the standard view tries to answer too
many questions, tries to embody too much knowledge in Stein's sense
(a)---simple, factual knowledge---in so far as it allows one to
formulate propositions that can have no empirical content \emph{even
  about systems falling in the theory's regime of applicability}.

This characterization of the regime of propriety and the regime of
applicability makes explicit the ways that schematizing the observer is
required for a theoretical structures to make substantive contact with
experimentation, and so for a theory as a whole to have non-trivial
epistemic content.

\section{Semantics Is Epistemology, Not Ontology}
\label{sec:sems-epist-onto}

The conclusions of my discussion of breakdown scales and the regimes
of a theory have a deep consequence for how meaning and truth may be
related in an adequate semantics for theories.  According to the kind
of semantics that naturally accompanies the standard view of theories,
all of a theory's propositions must be (at least approximately) true
of a system in order for the theory to represent it.  By my arguments,
however, the meanings of a theory's theoretical terms must be fixed by
the epistemic content associated with the regime of propriety, not the
regime of applicability, \emph{i}.\emph{e}., in part by situations in
which \emph{not} all of a theory's propositions about the system are
true---all those treating the dynamics of the system, all the
assertions standardly called ``predictions'', are false.  The meanings
of the theory's propositions, therefore, cannot be fixed solely by
their truth-conditions in any standard sense---the seductive intuition
grounding essentially all contemporary thought on the semantics of
scientific theories, as \citeN[ch.~B, \S7,
p.~22]{carnap-intro-semantics} concisely expresses it: ``\ldots\@ to
understand a sentence, to know what is asserted by it, is the same as
to know under what conditions it would be true.''  One cannot even
begin to investigate what the truth conditions of some of the
sentences in the theory may be, however---in particular, its dynamical
predictions---until one already knows enough about the meaning of its
terms to ascertain the truth of some of its other sentences,
\emph{viz}., the kinematical constraints, that guarantee that the
system at issue falls within the regime of propriety.

One can think of propriety in representation, in part, therefore, as
what a theory requires for it to have the capacity to produce
propositions whose truth-value can be investigated---not a fixing of
truth conditions, but rather the securing of the possibility to
investigate whether or how truth-conditions for a given proposition
can be determined in the first place.  A theory does not possess even
the capacity to be accurate or inaccurate in its treatment of a family
of phenomena if it does not represent the phenomena with propriety.
It follows that one can not even entertain questions about the truth
of many sorts of propositions---most of all those depending on the
identification of individual models with concrete models---until one
has determined that the theory has the resources, both practical and
formal, to represent the system at issue with propriety,
\emph{i}.\emph{e}., until meaning already has accrued to the formal
structures of the theory.  The fact that the regime of propriety is
strictly larger than the regime of applicability, therefore, shows
that the fundamental idea of semantics should rather be: to understand
a sentence, to know what is asserted by it, is the same as to know
under what conditions its constituent terms can be assigned meaning
and so allow one to investigate the possibility of ascertaining the
truth conditions of the sentence.

Indeed, I believe that all the problems I have discussed with the
standard view---in particular, its lack of accounting for all kinds of
knowledge, and its incapacity to handle the breakdown of models---boil
down to its implicit reliance on something like a truth-conditional
semantics, in conjunction with designation as the fundamental semantic
relation: semantics is fixed by ontology.  This is what the assumed
clean segregation of formal from practical spheres of knowledge
amounts to in current philosophical work, and, conversely, that clean
segregation is implied by such a semantics.\footnote{My reliance on
  the formally characterized distinction between kinematical and
  dynamical structures (\S\ref{sec:breakdown-regimes}) is not
  inconsistent with my contention that a good semantics allows no
  clean segregation of one part of epistemic content captured by the
  formalism from another part that is not.  It may still be the
  case---and in fact is the case---that different parts of the
  formalism play different roles in the analysis of the epistemic
  content, indeed in characterizing the integrity of epistemic content
  itself, not cleanly separable into formal and non-formal parts.}

More precisely, a view about the structure and semantics of physical
theory based ultimately on ontology, grounded in the assumption of a
clean split between the parts of the epistemic content of a theory
captured by its formalism and all other parts, is inadequate for (at
least) two reasons.  First, it does not allow us, within the scope of
the theory itself, to understand why models of systems in the regime
of propriety but not in the regime of applicability are not accurate
even though all the quantities the theory attributes to the system are
well defined and the values of those quantities jointly satisfy all
kinematical constraints the theory requires.  Second, we miss
something fundamental about the meaning of various theoretical terms
by rejecting such models out of hand merely on the grounds of their
inaccuracy.  It is part of the semantics of the term `hydrostatic
pressure', \emph{e}.\emph{g}., that its definition as a physical
quantity treated by classical fluid mechanics breaks down when the
fluid approaches turbulence closely enough; because, however, the
theory's equations of motion stop being accurate long before, in a
precise sense, the quantity loses definition in the theory and long
before the kinematical constraints of the theory stop being satisfied,
any account of the structure of theories and their semantics that
rejects the inaccurate models in which the term still is well defined
will not be able to account for that part of the term's meaning.
Thus, an adequate account of physical theory must be grounded on
notions derived from relations in some sense prior to the theory's
representations of the dynamical behavior of the physical systems it
treats, relations that govern the propriety of the theory's
representational resources for modeling the system at issue.  These
are the theory's kinematical constraints.

One may think that this discussion based on an analysis of how, where
and when theories break down more properly belongs to pragmatics (in
the sense of semiotic theory) than to semantics.  That is not so.  A
system of formal semantics that would ground itself in the family of
possible physical systems for which it provides sound models cannot
even get started until that family is demarcated.  But that is to
require an investigation of the boundary of the theory's regime of
propriety, which is thus logically and conceptually prior to any such
system of semantics.\footnote{\label{fn:intension}Jeremy Butterfield
  has suggested to me that an intensional semantics of the sort,
  \emph{e}.\emph{g}., that \citeN{lewis-gen-sems} propounds---which
  manifestly lends itself to the articulation of a semantics for
  theories in conformity with the standard view---can handle these
  pragmatic issues as part of the formal semantics itself, in so far
  as such pragmatic issues are encoded in the intensions.  I do not
  think that is right.  Such a semantics would evaluate a proposition
  in Navier-Stokes theory purporting to describe the behavior of a
  fluid at spatial scales of 10$^{-100}$ \texttt{cm} as false.  Such
  propositions are not false, however.  They are meaningless.  None of
  the quantities Navier-Stokes theory attributes to systems are well
  defined at such scales, so no ``proposition'' purportedly referring
  to them can have meaning.  It is folly to require that every well
  formed formula constructible in the terms of the formalism of a
  physical theory and derivable from its formal principles must have a
  determinate truth value, much more a meaning.  If one could re-tool
  such a scheme of intensional semantics so that its domain is not
  ``all possible worlds'' (or ``all possible models'' in some other
  appropriate sense), however, but rather a pragmatically
  characterized regime of propriety, then it might in fact be a useful
  tool for trying to define a semantics grounded on propriety in
  representation.  Even if one were to do so, however, one should
  still recognize that all the heavy lifting is performed by what are
  commonly considered to be the pragmatic elements of the intension,
  not by the determined relation of designation.  Indeed, those
  pragmatic elements must be investigated and fixed \emph{before} one
  can define designation: they are logically and conceptually prior to
  designation.  So one ought also ask: what work does the designation
  do?  What is its cash value (to echo William James)?  For the
  meaning of the terms is, by my lights, already in place once one has
  succeeded in determining the pragmatic elements that govern proper
  use---\emph{useful} use---and so, \emph{a fortiori}, govern the
  understanding that the proper use of the language can foster.
  Nonetheless, I do endorse the teamwork picture of
  \citeN{lewis-langs-lang}---that the 2 approaches to semantics
  (Gricean pragmatic versus formal language) are complementary, not
  contradictory; the problem today is to convince people on each side
  of the divide to reach out and work with those on the other side,
  and not to focus exclusively on one's own approach in studies and
  applications.  I do not, however, fully endorse Lewis's conclusion
  (p.~35) that \emph{both} ingredients are essential.  The
  Gricean/pragmatic approach seems to me essential; the formal
  approach seems to me useful.  The real problem with philosophers
  like Lewis, and the way they adhere to the standard view of
  theories, is that they believe more in their non-empirical theory of
  semantics then they believe in (\emph{e}.\emph{g}.\@) quantum
  mechanics.  (See, for instance, the dismissive remarks in the
  preface to \citeNP{lewis-phil-papers-2} about the reliability of
  quantum mechanics as a guide for trying to understand the world.)
  The referential relation of word to world required by Lewis's
  picture and others like his (in order to avoid, \emph{inter alia},
  Putnam's~\citeyearNP{putnam-real-reas-apa} permutation argument) is
  not grounded in any empirical fact, cannot be exposed or even only
  studied by any empirical investigation, and yet Lewis believes more
  firmly in \emph{that} than in the deliverances of physics.  I say
  (to echo Anscombe): I cannot understand such a man.}

According to a semantics that requires predictive accuracy, such as a
Tarskian one and in general almost every one conforming to the
standard view, the idea of the regime of propriety is meaningless.
Such a semantics cannot explain or even accommodate this fact about
our theories, for \emph{predictively inaccurate models cannot be
  Tarskian models or possible worlds or objects in the category of
  solutions to the equations of motion of the theory}.  Semantics
founded on the standard view, therefore, does not exhaust the
representational capacity of the theory, and the theory gains
non-trivial semantic content from everything it can significantly
represent, whether in all accuracy or not.  The set of possible worlds
picked out by satisfaction of the equations of motion is not a rich
enough family of worlds to express or encode all the information the
theory can give us about the possibilities of the actual physical
world.  The theory tells us more about physical quantities like
pressure in the actual world than there would be to learn about it in
a world the theory would be true of, in the standard sense.

Thus, the regime of propriety must be included as part of the theory's
semantics---at least so far as a real semantics of a real physical
theory goes, not just a formal semantics of a formal theory: if I
don't know the family of actual and physically (not mathematically)
possible systems the theory applies to, I don't, by the lights of the
standard view itself, know the semantics; but if I don't know the
regime of propriety, I don't know that family; and nothing in the
formalism of the theory itself can tell me the regime---I cannot fix
the ``ontology'' by reasoning grounded on a clear segregation of the
theory's epistemic content into one part captured by the formalism and
``interpretive postulates'' and another part corresponding only to
practice.

A semantics grounded on pure ontology assumes a kind of Fichtean
direct intellectual grasp of the world by our representational
systems: the posited relation between our symbolic systems and the
``objects in the world'' (irrespective of whether one is a realist
about such things or not) is unmediated by the actual state of our
knowledge and by the practices and techniques we have for probing,
experimenting on, and more generally investigating the world and
evaluating the results of those investigations.\footnote{These
  criticims do not apply to the use of any Tarskian-like semantics in
  logic and mathematics, where one cannot cleanly and unambiguously
  segregate our symbolic systems from the objects they purport to
  represent; or, at least, the kind of access we may have there to
  such objects is mediated only by the symbolic systems, not by
  experimental practice.}  Have we learned nothing from Kant?  What
sense is there in trying to articulate ``truth conditions'' that are
forever beyond our cognitive grasp?  What was wanted was an account of
``meaning'' analytically connected to truth, completely divorced from
human concerns.  But this is self-defeating, for such a conception
\emph{eo ipso} completely divorces, unbridgeably separates, semantics
from the \emph{fundamental sources} of scientific
knowledge---experimental knowledge---which in the end must ground the
empirical content and significance of our theoretical representations.
Relations of ``direct designation'' serve---can serve---no
philosophical or foundational purpose.  They can tell us nothing about
meaning.\footnote{\citeN[p.~50, emphases his]{stein-yes-but} puts
  the point forcefully, when he imagines Kant posing the following
  dilemma to Leibniz's ghost:
  \begin{quote}
    How can you \emph{know} that things are as you say they are?  If
    the claimed ``reference'' of the theory is something beyond its
    correctness and adequacy in representing phenomena -- if, that is,
    for a given theory, which (we may suppose) does represent
    phenomena correctly and adequately, there are still two
    possibilities: (a) that it is (moreover) \emph{true}, and (b) that
    it is (nonetheless) \emph{false} -- then how in the world could we
    ever tell what the actual case is?
  \end{quote}
  Stein talks here of representing phenomena ``correctly and
  adequately''.  His distinction, as I understand it, is related to my
  own ``appropriately and adequately'' in interesting ways.  For
  Stein, ``correctly'' is something like my ``appropriately and
  accurately'', and ``adequately'' is something like covering a large
  enough extent to think one has latched onto \emph{general} features
  of the world, not just something peculiar to the idiosyncracy of
  \emph{this} kind of system studied by \emph{this} kind of
  experiment.  This kind of comprehension in coverage is a fundamental
  part of the epistemic content of a good theory, but it would take us
  too far afield to discuss it here.}  They are vacuous chicanery,
nonsensical will-o-the-wisps leading us to a morass of philosophical
quicksand into which we hopelessly sink, suffocating on our own
confusion.

These criticisms assume a link between semantics and knowledge in all
its human forms, as explicated and discussed in
\S\ref{sec:forms-aspects-know}.  I think this must be right, that
there must be such a link between semantics and real human knowledge.
A semantics divorced from our actual state of knowledge (as achieved
state, as provider of evidentiary relations for epistemic warrant, as
guide to future investigation), and from the ways we have of improving
our epistemic state, is useless.

We must expel from philosophy the myth of a ``human-free'' semantics.
\emph{All} aspects of scientific knowledge, not just the
``ontological'', should be reflected to at least some degree in a
theory's semantic content, and most of all that knowledge ultimately
grounding the semantic content---and that is indubitably,
inextricably, inevitably \emph{experimental} in large part.  This view
may entail a blurring of the lines between the traditional conception
of semantics and pragmatics (in the sense of semiotic), but that, I
think, is all to the good, since the traditional notions are
(\emph{pace} Carnap, Suppes, \emph{et al}.\@) appropriate for
mathematics, not the empirical sciences.  In any event, I am not
convinced that one needs to lose a sharp distinction between semantics
in a formal sense and pragmatics in order to ground the kind of view I
advocate---one needs only to characterize semantics in a way that is
not wholly ``ontological'', based on a primitive relation of
designation.\footnote{I want to emphasize that I am not opposed to
  referential relations and the idea of designation itself in a
  semantics for theories.  I am opposed only to the idea that such
  relations be primitive.}

As I have argued, in order to know how to investigate whether or not a
theory provides an accurate representation of a system---whether that
system falls in the theory's regime of applicability---one must be
able to verify first whether or not the system satisfies the theory's
kinematical constraints, \emph{i}.\emph{e}., whether or not it falls
in the theory's regime of propriety.  Our problem, therefore,
% the analogue to Stalnaker's,
is to move in a principled way from an understanding of how to verify
whether or not a theory's kinematical constraints are satisfied (which
generically involves including schematized representations of the
observer in experimental models) to an understanding of how to use the
resources of the theory to represent a system once we have verified
the kinematical constraints are satisfied.
% As Stalnaker says, moreover, our account of this process should
% ``[explain] why we use the method we do use to [verify] them.''
Because, as I argued, the identification of (parts) of a concrete
model of a system with (parts) of the formal structures of a theory is
a pragmatic choice, the problem is to find justifications for those
choices in the success and fruitfulness of the subsequent use we make
of theory based on them.  Pragmatics, on this picture, still has to do
with the way that the contexts of individuals shape the expression of
the meaning of their utterances, and that context includes,
\emph{inter alia}, the individuals' beliefs; but what individuals
believe and express, in any given context, does not bear on the
objectivity of the knowledge embodied in the epistemic content of a
scientific theory, of which the semantics ought to ground the
articulation and representation.

Semantics, therefore, must ground analysis of epistemology, and be
grounded in turn by our grasp of it.

\noindent NOT:
\begin{equation*}
  \begin{split}
    \text{semantics} &\approx \text{ontology} \\
    \text{pragmatics} &\approx \text{use} \\       
  \end{split}
\end{equation*}
RATHER:
\begin{equation*}
  \begin{split}
    \text{semantics} &\approx \text{epistemology, methodology} \\
    \text{pragmatics} &\approx \text{acceptance, choice} \\       
  \end{split}
\end{equation*}
The Slogan:
\begin{center}
  Meaning comes before truth.
\end{center}

\section{Valediction}
\label{sec:valediction}

The miracle of science is that theory and experiment are consonant
with each other; the necessity of science is that they are
inextricably so---not, however, as equals.  Theory plays Boswell to
the subtle and tragic clown of experiment's Johnson.

\section[Appendix: Pr\'ecis]{Appendix:
  Pr\'ecis\symbolfootnote[2]{This appendix does not appear in the
    published version of the paper, due to length constraints.}}
\label{sec:precis}

This is a long and dense paper, and the overall argument has not leant
itself to a clean, simple, linear exposition.  I therefore sum up here
the main claims of the paper, along with my main criticisms of what I
call the standard view of theories.

First, I briefly rehearse the argument of \S\ref{sec:complex-simple}
for the inadequacy of the standard view. \skipline[-.25]
\begin{enumerate}
  \noitemsep
    \item The standard view is that a theory is characterized by a
  family of models, where ``characterized by'' is construed broadly.
    \item Correlatively, because those models are assumed to have
  intrinsic physical significance, it assumes a clean segregation of
  the theoretical from the practical forms and aspects of scientific
  knowledge (discussed in greater detail in
  \S\ref{sec:what-is-theory}).
    \item The ``models'' are stipulated to be:
  \begin{enumerate}
      \item\label{item:solns} either something like a family of
    (closed form) solutions to the theory's equations of motion or
    field equations, or a family of some mathematical structures with
    the right representational capacities;
      \item\label{item:syss} or else the family of physical systems
    that the equations' solutions or the theory's other mathematical
    structures represent.
  \end{enumerate}
    \item Both possible stipulations are, however, epistemologically
  speaking, acts of faith:
  \begin{enumerate}
      \item for almost all known theories we have very little
    knowledge of the family of solutions (or other relevant
    structures) because the general problems are mathematically
    intractable, which makes both stipulations~\ref{item:solns}\@ and
    \ref{item:syss}\@ untenable;
      \item for all real physical systems, their true complexity is
    such that we cannot possibly know to which solutions of which
    theories they correspond, which makes
    stipulation~\ref{item:syss}\@ untenable.
  \end{enumerate}
    \item To grasp either horn of the dilemma, in any event, is to be
  forced to conclude that we never have real knowledge of any
  sophisticated scientific theory and how it represents the world.
    \item It thus becomes utterly mysterious how the models are
  supposed to have intrinsic physical significance.
\end{enumerate}

\noindent Second, my discontents with the standard view, which are
part and parcel its inadequacy. \skipline[-.25]
\begin{enumerate}
  \noitemsep
    \item It assumes that only one form (theoretical), in one of its
  aspects (achieved state as represented purely by formalism),
  suffices for characterizing a theory
  (\S\S\ref{sec:forms-aspects-know}--\ref{sec:sems-epist-onto}).
    \item Thus, it cannot demarcate, much less identify, the family of
  physical systems the theory appropriately and adequately treats
  (\S\S\ref{sec:what-is-theory}--\ref{sec:what-theory-is},
  \S\ref{sec:breakdown-regimes}).
    \item Thus, it cannot explain the epistemic warrant we have for
  trusting and using the theory, and for believing that the
  understanding and comprehension it seems to give us of the world is
  indeed about the world (\S\ref{sec:theor-exper-contact},
  \S\ref{sec:sems-epist-onto}).
\end{enumerate}

\noindent Finally, the positive claims I defend. \skipline[-.25]
\begin{enumerate}
  \noitemsep
    \item A theory is characterized by its epistemic content, the sum
  total of all forms of knowledge it embodies, in all their aspects
  and relations to each other, as determined by our actual current
  state of knowledge about the world and about how to investigate it.
  (\S\ref{sec:what-is-theory}, \S\ref{sec:what-theory-is},
  \S\S\ref{sec:forms-aspects-know}--\ref{sec:theor-exper-contact})
    \item The total family of physical systems the theory
  appropriately and adequately treats, a proper subset of all those
  ``possibly representable'' by all its formal models, forms an
  essential part of that epistemic content.
  (\S\S\ref{sec:need-schem-obsr}--\ref{sec:what-theory-is},
  \S\ref{sec:breakdown-regimes})
    \item There are (at least) two forms of knowledge
  (\S\ref{sec:forms-aspects-know}): \skipline[-.3]
  \begin{enumerate}
      \item theoretical (what can be learned from books);
      \item practical (what can be fully understood only by doing).
  \end{enumerate}
    \item Those forms have (at least) four aspects
  (\S\S\ref{sec:forms-aspects-know}--\ref{sec:theor-exper-contact}):
  \skipline[-.3]
  \begin{enumerate}
      \item as achieved state;
      \item as susceptible of justification, and so involving a
    structure of evidential relations;
      \item as ground for epistemic warrant, and so involving a
    structure of evidential relations;
      \item as an enterprise, an activity aimed at increasing the
    first aspect as constrained by the second and third.
  \end{enumerate}
    \item Much of all of those forms and aspects is grounded on and
  embodied in our knowledge of how to schematically represent the
  observer in models of actual experiments and observations
  (\S\ref{sec:need-schem-obsr},
  \S\S\ref{sec:forms-aspects-know}--\ref{sec:breakdown-regimes}).
  Perhaps most importantly, schematization of the observer is needed
  to:
  \begin{enumerate}
      \item lay down adequate definitions for physical quantities
    (\S\ref{sec:breakdown-regimes});
      \item determine the theory's breakdown scales
    (\S\ref{sec:breakdown-regimes});
      \item demarcate the theory's regime of applicability, the total
    family of systems the theory appropriately and adequately treats,
    as delimited by the breakdown scales
    (\S\ref{sec:need-schem-obsr}, \S\ref{sec:breakdown-regimes});
      \item adjudicate whether and, if so, elucidate how a given
    experiment lends confirmatory support to (some part of) a theory
    (\S\ref{sec:theor-exper-contact}).
  \end{enumerate}
    \item A good semantics for theory should respect the fact that the
  epistemic content of a theory is not cleanly separable into formal
  and practical parts; in particular, semantics should be based on and
  reflect epistemology, not ontology (\S\ref{sec:sems-epist-onto}).
\end{enumerate}

\addcontentsline{toc}{section}{\hspace*{-1.3em}\numberline{}References}


\begin{thebibliography}{}

\bibitem[\protect\citeauthoryear{{Abbott, B. \emph{et al}.{\space}(LIGO
  Scientific Collaboration and Virgo Collaboration)}}{{Abbott, B. \emph{et
  al}.{\space}(LIGO Scientific Collaboration and Virgo
  Collaboration)}}{2016a}]{abbott-et-obs-gw-bin-bh}
{Abbott, B. \emph{et al}.{\space}(LIGO Scientific Collaboration and Virgo
  Collaboration)} (2016a, February).
\newblock Observation of gravitational waves from a binary black hole merger.
\newblock {\em Physical Review Letters\/}~{\em 116\/}(6), 061102.
\newblock doi:\href{http://dx.doi.org/10.1103/PhysRevLett.116.061102}
  {10.1103/PhysRevLett.116.061102}. Preprint:
  \href{https://arxiv.org/abs/1602.03837} {arXiv:1602.03837 [gr-qc]}.

\bibitem[\protect\citeauthoryear{{Abbott, B. \emph{et al}.{\space}(LIGO
  Scientific Collaboration and Virgo Collaboration)}}{{Abbott, B. \emph{et
  al}.{\space}(LIGO Scientific Collaboration and Virgo
  Collaboration)}}{2016b}]{abbott-et-tests-gr-gw150914}
{Abbott, B. \emph{et al}.{\space}(LIGO Scientific Collaboration and Virgo
  Collaboration)} (2016b, June).
\newblock Tests of general relativity with gw150914.
\newblock {\em Physical Review Letters\/}~{\em 116\/}(22), 221101.
\newblock doi:\href{http://dx.doi.org/10.1103/PhysRevLett.116.221101}
  {10.1103/PhysRevLett.116.221101}.

\bibitem[\protect\citeauthoryear{Benedict}{Benedict}{1969}]{benedict-funds-temp-press}
Benedict, R. (1969).
\newblock {\em Fundamentals of Temperature, Pressure and Flow Measurements}.
\newblock New York: John Wiley \& Sons, Inc.

\bibitem[\protect\citeauthoryear{Bogen and Woodward}{Bogen and
  Woodward}{1988}]{bogen-woodward-sav-pha}
Bogen, J. and J.~Woodward (1988, July).
\newblock Saving the phenomena.
\newblock {\em The Philosophical Review\/}~{\em \textsc{xcvii}\/}(3), 303--352.
\newblock doi:\href{http://dx.doi.org/10.2307/2185445} {10.2307/2185445}.

\bibitem[\protect\citeauthoryear{Buchwald}{Buchwald}{1994}]{buchwald-creat-sci-effs-hertz}
Buchwald, J. (1994).
\newblock {\em The Creation of Scientific Effects: Heinrich Hertz and Electric
  Waves}.
\newblock Chicago: University of Chicago Press.

\bibitem[\protect\citeauthoryear{Butterfield}{Butterfield}{2018}]{butterfield-dual-equiv-phys-thrs}
Butterfield, J. (2018).
\newblock On dualities and equivalences between physical theories.
\newblock Preprint: \href{https://arxiv.org/abs/1806.01505} {arXiv:1806.01505
  [physics.hist-ph]}.

\bibitem[\protect\citeauthoryear{Carnap}{Carnap}{1942}]{carnap-intro-semantics}
Carnap, R. (1942).
\newblock {\em Introduction to Semantics}.
\newblock Number \textsc{i} in Studies in Semantics. Cambridge, MA: Harvard
  University Press.

\bibitem[\protect\citeauthoryear{Carnap}{Carnap}{1956}]{carnap-eso}
Carnap, R. (1956).
\newblock Empiricism, semantics and ontology.
\newblock In {\em Meaning and Necessity: A Study in Semantics and Modal
  Logic\/} (Second ed.)., pp.\  205--221. Chicago: The University of Chicago
  Press.
\newblock An earlier version was published in \emph{Revue Internationale de
  Philosophie} 4(1950):20--40.

\bibitem[\protect\citeauthoryear{Cartwright}{Cartwright}{1999}]{cartwright99}
Cartwright, N. (1999).
\newblock {\em The Dappled World: A Study of the Boundaries of Science}.
\newblock Cambridge: Cambridge University Press.

\bibitem[\protect\citeauthoryear{Chandrasekhar}{Chandrasekhar}{1961}]{chandrasekhar-hydro}
Chandrasekhar, S. (1961).
\newblock {\em Hydrodynamic and Hydromagnetic Stability}.
\newblock Oxford: Oxford University Press.

\bibitem[\protect\citeauthoryear{Cohen and Callender}{Cohen and
  Callender}{2009}]{cohen-callender-better-best}
Cohen, J. and C.~Callender (2009).
\newblock A better best system account of lawhood.
\newblock {\em Philosophical Studies\/}~{\em 145\/}(1), 1--34.
\newblock doi:\href{http://dx.doi.org/10.1007/s11098-009-9389-3}
  {10.1007/s11098-009-9389-3}.

\bibitem[\protect\citeauthoryear{Collins}{Collins}{1992}]{collins-chng-ordr-replic-induc}
Collins, H. (1992).
\newblock {\em Changing Order: Replication and Induction in Scientific
  Practice}.
\newblock Chicago: University of Chicago Press.

\bibitem[\protect\citeauthoryear{Collmar, Straumann, Chakrabarti, 't~Hooft,
  Seidel, and Israel}{Collmar et~al.}{1998}]{collmar-et-panel-proof-exist-bhs}
Collmar, W., N.~Straumann, S.~Chakrabarti, G.~'t~Hooft, E.~Seidel, and
  W.~Israel (1998).
\newblock Panel discussion: {T}he definitive proofs of the existence of black
  holes.
\newblock In F.~Hehl, C.~Kiefer, and R.~Metzler (Eds.), {\em Black Holes:
  Theory and Observation}, Number 514 in Lecture Notes in Physics, Chapter~22,
  pp.\  481--489. Berlin: Springer-Verlag.
\newblock doi:\href{http://dx.doi.org/10.1007/978-3-540-49535-2_22}
  {10.1007/978-3-540-49535-2$\underscore$22}.

\bibitem[\protect\citeauthoryear{Curiel}{Curiel}{1999}]{curiel-sing}
Curiel, E. (1999, September).
\newblock The analysis of singular spacetimes.
\newblock {\em Philosophy of Science\/}~{\em 66}, S119--S145.
\newblock Supplement. Proceedings of the 1998 Biennial Meetings of the
  Philosophy of Science Association. Part I: Contributed Papers. Stable URL:
  $<$\url{http://www.jstor.org/stable/188766}$>$. A more recent, corrected,
  revised and extended version of the published paper is available at:
  $<$\url{http://strangebeautiful.com/phil-phys.html}$>$.

\bibitem[\protect\citeauthoryear{Curiel}{Curiel}{2014}]{curiel-cm-lag-not-ham}
Curiel, E. (2014).
\newblock Classical mechanics is {L}agrangian; it is not {H}amiltonian.
\newblock {\em British Journal for the Philosophy of Science\/}~{\em 65\/}(2),
  269--321.
\newblock doi:\href{http://dx.doi.org/10.1093/bjps/axs034}
  {10.1093/bjps/axs034}.

\bibitem[\protect\citeauthoryear{Curiel}{Curiel}{2017a}]{curiel-kins-dyns-struc-theors}
Curiel, E. (2017a).
\newblock Kinematics, dynamics, and the structure of physical theory.
\newblock Unpublished manuscript. Preprint:
  \href{https://arxiv.org/abs/1603.02999} {arXiv:1603.02999 [physics.hist-ph]}.
  Most recent draft available at
  $<$\url{http://strangebeautiful.com/phil-phys.html}$>$.

\bibitem[\protect\citeauthoryear{Curiel}{Curiel}{2017b}]{curiel-prpy-basis-sems}
Curiel, E. (2017b).
\newblock On the propriety of physical theories as a basis for their semantics.
\newblock Unpublished manuscript, draft available at
  $<$\url{http://strangebeautiful.com/phil-phys.html}$>$.

\bibitem[\protect\citeauthoryear{Curiel}{Curiel}{2019}]{curiel-many-defns-bh}
Curiel, E. (2019).
\newblock The many definitions of a black hole.
\newblock {\em Nature Astronomy\/}~{\em 3}, 27--34.
\newblock doi:\href{http://dx.doi.org/10.1038/s41550-018-0602-1}
  {10.1038/s41550-018-0602-1}. Free read-only SharedIt link:
  $<$\url{https://rdcu.be/bfNpM}$>$.

\bibitem[\protect\citeauthoryear{Curiel}{Curiel}{2020}]{curiel-fw-confirm-newt-abd}
Curiel, E. (2020).
\newblock Framework confirmation by {N}ewtonian abduction.
\newblock {\em Synthese\/}.
\newblock Part of the special issue ``Reasoning in Physics''. Published online.
  doi:\href{http://dx.doi.org/10.1007/s11229-019-02400-9}
  {10.1007/s11229-019-02400-9}. Preprint:
  \href{http://arxiv.org/abs/1804.07414} {arXiv:1804.07414 [physics.hist-ph]}.

\bibitem[\protect\citeauthoryear{da~Costa and French}{da~Costa and
  French}{2005}]{costa-french-models-sci-reason}
da~Costa, N. and S.~French (2005).
\newblock {\em Science and Partial Truth: A Unitary Approach to Models and
  Scientific Reasoning}.
\newblock Oxford: Oxford University Press.
\newblock doi:\href{http://dx.doi.org/10.1093/019515651X.001.0001}
  {10.1093/019515651X.001.0001}.

\bibitem[\protect\citeauthoryear{Davidson}{Davidson}{1969}]{davidson-true-facts}
Davidson, D. (1969, November).
\newblock True to the facts.
\newblock {\em Journal of Philosophy\/}~{\em 66\/}(21), 748--764.
\newblock doi:\href{http://dx.doi.org/10.2307/2023778} {10.2307/2023778}.

\bibitem[\protect\citeauthoryear{Demopoulos}{Demopoulos}{2013}]{demopoulos-log-phil-leg}
Demopoulos, W. (2013).
\newblock {\em Logicism and Its Philosophical Legacy}.
\newblock Cambridge: Cambridge University Press.

\bibitem[\protect\citeauthoryear{Earman}{Earman}{1978}]{earman-stat-thermo-rel}
Earman, J. (1978).
\newblock Combining statistical-thermodynamics and relativity theory:
  Methodological and foundations problems.
\newblock {\em PSA: Proceedings of the Biennial Meeting of the Philosophy of
  Science Association 1978\/}~{\em 2}, 157--85.
\newblock doi:\href{http://dx.doi.org/10.1086/psaprocbienmeetp.1978.2.192467}
  {10.1086/psaprocbienmeetp.1978.2.192467}.

\bibitem[\protect\citeauthoryear{Earman}{Earman}{1986}]{earman-primer-determ}
Earman, J. (1986).
\newblock {\em A Primer on Determinism}.
\newblock Dordrecht: D. Reidel Publishing Co.

\bibitem[\protect\citeauthoryear{Earman, Glymour, and Stachel}{Earman
  et~al.}{1977}]{earman-etal-found-st-theors}
Earman, J., C.~Glymour, and J.~Stachel (Eds.) (1977).
\newblock {\em Foundations of Space-Time Theories}.
\newblock Number \textsc{viii} in Minnesota Studies in Philosophy of Science.
  Minneapolis: University of Minnesota Press.
\newblock Freely available online:
  $<$\url{https://cla.umn.edu/mcps/publications/minnesota-studies-philosophy-science}$>$.

\bibitem[\protect\citeauthoryear{Eckart, H\"uttemann, Kiefer, Britzen,
  Zaja\v{c}ek, L\"ammerzahl, St\"ockler, Valencia-S., Karas, and
  Garc\'ia-Mar\'in}{Eckart et~al.}{2017}]{eckart-et-superm-bh-good-case}
Eckart, A., A.~H\"uttemann, C.~Kiefer, S.~Britzen, M.~Zaja\v{c}ek,
  C.~L\"ammerzahl, M.~St\"ockler, M.~Valencia-S., V.~Karas, and
  M.~Garc\'ia-Mar\'in (2017, May).
\newblock The {M}ilky {W}ay's supermassive black hole: {H}ow good a case is it?
\newblock {\em Foundations of Physics\/}~{\em 47\/}(5), 553--624.
\newblock doi:\href{http://dx.doi.org/10.1007/s10701-017-0079-2}
  {10.1007/s10701-017-0079-2}. Preprint:
  \href{https://arxiv.org/abs/1703.09118} {arXiv:1703.09118 [astro-ph.HE]}.

\bibitem[\protect\citeauthoryear{Elder}{Elder}{2020}]{elder-phd-epist-gw-astro}
Elder, J. (2020).
\newblock {\em The Epistemology of Gravitational Wave Astrophysics}.
\newblock Ph.\ D. thesis, University of Notre Dame, History and Philosophy of
  Science Program.
\newblock (expected).

\bibitem[\protect\citeauthoryear{Ellis}{Ellis}{1967}]{ellis-dyns-press-free-matt-gr}
Ellis, G. (1967, May).
\newblock Dynamics of pressure-free matter in general relativity.
\newblock {\em Journal of Mathematical Physics\/}~{\em 8\/}(5), 1171--1194.
\newblock doi:\href{http://dx.doi.org/10.1063/1.1705331} {10.1063/1.1705331}.

\bibitem[\protect\citeauthoryear{Ferrario and Schiaffonati}{Ferrario and
  Schiaffonati}{2012}]{ferrario-schiaffonati-form-meths-emp-pract-suppes}
Ferrario, R. and V.~Schiaffonati (2012).
\newblock {\em Formal Methods and Empirical Practices: Conversations with
  Patrick Suppes}.
\newblock Number 205 in CSLI Lecture Notes. Stanford, CA: CSLI Publications.

\bibitem[\protect\citeauthoryear{Fillion and Corless}{Fillion and
  Corless}{2014}]{fillion-corless-epist-mod-comput-err}
Fillion, N. and R.~Corless (2014).
\newblock On the epistemological analysis of modeling and computational error
  in the mathematical sciences.
\newblock {\em Synthese\/}~{\em 191\/}(7), 1451--1467.
\newblock doi:\href{http://dx.doi.org/10.1007/s11229-013-0339-4}
  {10.1007/s11229-013-0339-4}.

\bibitem[\protect\citeauthoryear{Fillion and Corless}{Fillion and
  Corless}{2019}]{fillion-corless-back-err-anal-pertbn}
Fillion, N. and R.~Corless (2019).
\newblock Backward error analysis for perturbation methods.
\newblock In N.~Fillion, R.~Corless, and I.~Kotsireas (Eds.), {\em Algorithms
  and Complexity in Mathematics, Epistemology, and Science: Proceedings of 2015
  and 2016 ACMES Conferences}, Number~82 in Fields Institute Communications,
  pp.\  35--80. New York: Springer.
\newblock doi:\href{http://dx.doi.org/10.1007/978-1-4939-9051-1_3}
  {10.1007/978-1-4939-9051-1$\underscore$3}.

\bibitem[\protect\citeauthoryear{Foias, Manley, Rosa, and Temam}{Foias
  et~al.}{2001}]{foias-et-nse-turb}
Foias, C., O.~Manley, R.~Rosa, and R.~Temam (2001).
\newblock {\em Navier-Stokes Equations and Turbulence}, Volume~83 of {\em
  Encyclopedia of Mathematics and Its Applications}.
\newblock Cambridge: Cambridge University Press.
\newblock doi:\href{http://dx.doi.org/10.1017/CBO9780511546754}
  {10.1017/CBO9780511546754}.

\bibitem[\protect\citeauthoryear{Franklin}{Franklin}{1986}]{franklin-neglect-exp}
Franklin, A. (1986).
\newblock {\em The Neglect of Experiment}.
\newblock Cambridge: Cambridge University Press.
\newblock doi:\href{http://dx.doi.org/10.1017/CBO9780511624896}
  {10.1017/CBO9780511624896}.

\bibitem[\protect\citeauthoryear{Friedman}{Friedman}{2001}]{friedman-dyns-reason}
Friedman, M. (2001).
\newblock {\em The Dynamics of Reason}.
\newblock Stanford, CA: CSLI Publications.
\newblock Delivered as the 1999 Kant Lectures at Stanford University.

\bibitem[\protect\citeauthoryear{Genzel, Eckart, Ott, and Eisenhauer}{Genzel
  et~al.}{1997}]{genzel-et-dark-mass-ctr-milky}
Genzel, R., A.~Eckart, T.~Ott, and F.~Eisenhauer (1997, October).
\newblock On the nature of the dark mass in the centre of the {M}ilky {W}ay.
\newblock {\em Monthly Notices of the Royal Astronomical Society\/}~{\em
  291\/}(1), 219--234.
\newblock doi:\href{http://dx.doi.org/10.1093/mnras/291.1.219}
  {10.1093/mnras/291.1.219}.

\bibitem[\protect\citeauthoryear{Geroch}{Geroch}{1969}]{geroch-lim-sts}
Geroch, R. (1969).
\newblock Limits of spacetimes.
\newblock {\em Communications in Mathematical Physics\/}~{\em 13\/}(3),
  180--193.
\newblock doi:\href{http://dx.doi.org/10.1007/BF01645486} {10.1007/BF01645486}.
  Open access at \url{http://projecteuclid.org/euclid.cmp/1103841574}.

\bibitem[\protect\citeauthoryear{Geroch}{Geroch}{1970}]{geroch-dom-dep}
Geroch, R. (1970).
\newblock Domain of dependence.
\newblock {\em Journal of Mathematical Physics\/}~{\em 11\/}(2), 437--449.
\newblock doi:\href{http://dx.doi.org/10.1063/1.1665157} {10.1063/1.1665157}.

\bibitem[\protect\citeauthoryear{Geroch}{Geroch}{1977}]{geroch-pred-gr}
Geroch, R. (1977).
\newblock Prediction in general relativity.
\newblock See \citeN{earman-etal-found-st-theors}, pp.\  81--93.
\newblock Freely available online:
  $<$\url{https://cla.umn.edu/mcps/publications/minnesota-studies-philosophy-science}$>$.

\bibitem[\protect\citeauthoryear{Geroch}{Geroch}{1996}]{geroch-pdes-phys}
Geroch, R. (1996).
\newblock Partial differential equations of physics.
\newblock In G.~Hall and J.~Pulham (Eds.), {\em General Relativity}, Aberdeen,
  pp.\  19--60. Scottish Universities Summer School in Physics.
\newblock Proceedings of the 46th Scottish Universities Summer School in
  Physics, Aberdeen, July 1995. Preprint:
  \href{http://arxiv.org/abs/gr-qc/9602055} {arXiv:gr-qc/9602055}.

\bibitem[\protect\citeauthoryear{Geroch}{Geroch}{2001}]{geroch-hyp-rel-diss-fl}
Geroch, R. (2001).
\newblock On hyperbolic ``theories'' of relativistic dissipative fluids.
\newblock \href{http://arxiv.org/abs/gr-qc/0103112} {arXiv:gr-qc/0103112v1}).

\bibitem[\protect\citeauthoryear{Ghez, Morris, Becklin, Tanner, and
  Kremenek}{Ghez et~al.}{2000}]{ghez-et-acc-stars-orbit-bh}
Ghez, A., M.~Morris, E.~Becklin, A.~Tanner, and T.~Kremenek (2000, September).
\newblock The accelerations of stars orbiting the {M}ilky {W}ay's central black
  hole.
\newblock {\em Nature\/}~{\em 407}, 349--351.
\newblock doi:\href{http://dx.doi.org/10.1038/35030032} {10.1038/35030032}.

\bibitem[\protect\citeauthoryear{Halvorson}{Halvorson}{2019}]{halvorson-real-qt}
Halvorson, H. (2019).
\newblock To be a realist about quantum theory.
\newblock In O.~Lombardi, S.~Fortin, C.~L\'opez, and F.~Holik (Eds.), {\em
  Quantum Worlds: Perspectives on the Ontology of Quantum Mechanics},
  Chapter~8, pp.\  133--163. Cambridge: Cambridge University Press.
\newblock doi:\href{http://dx.doi.org/10.1017/9781108562218.010}
  {10.1017/9781108562218.010}.

\bibitem[\protect\citeauthoryear{Halvorson and Tsementzis}{Halvorson and
  Tsementzis}{2017}]{halvorson-tsementzis-cats-sci-theor}
Halvorson, H. and D.~Tsementzis (2017).
\newblock Categories of scientific theories.
\newblock See \citeN{landry-cat-theor-work-philr}, Chapter~17, pp.\  402--429.
\newblock \href{http://dx.doi.org/10.1093/oso/9780198748991.003.0017}
  {doi:10.1093/oso/9780198748991.003.0017}.

\bibitem[\protect\citeauthoryear{Harper}{Harper}{2011}]{harper-newtons-sci-meth}
Harper, W. (2011).
\newblock {\em Isaac Newton's Scientific Method: Turning Data into Evidence
  about Gravity and Cosmology}.
\newblock Oxford: Oxford University Press.
\newblock
  doi:\href{http://dx.doi.org/10.1093/acprof:oso/9780199570409.001.0001}
  {10.1093/acprof:oso/9780199570409.001.0001}.

\bibitem[\protect\citeauthoryear{Haynes}{Haynes}{2014}]{crc-handbook-chem-phys-95}
Haynes, W. (Ed.) (2014).
\newblock {\em CRC Handbook of Chemistry and Physics\/} (95 ed.).
\newblock Boca Raton, FL: CRC Press.

\bibitem[\protect\citeauthoryear{Helmholtz}{Helmholtz}{1870}]{helmholtz-geom-axioms}
Helmholtz, H. (1870).
\newblock {\"U}ber den {U}rsprung und die {B}edeutung der geometrischen
  {A}xiome.
\newblock In {\em Popul\"are Wissenschaftliche Vortr\"age von H.~Helmholtz},
  Volume~3, pp.\  23--51. Braunschweig.
\newblock A lecture delivered in the Docentverein of Heidelberg, 1870. English
  translation by Howard Stein (unpublished), ``On the Origin and Significance
  of the Geometrical Axioms'', available at
  \url{http://strangebeautiful.com/other-texts/helmholtz-origin-signif-geom-axioms-stein.pdf.}

\bibitem[\protect\citeauthoryear{Hempel}{Hempel}{2001}]{hempel-stand-conc-sci-thrs}
Hempel, C. (2001).
\newblock On the `standard conception' of scientific theories.
\newblock In {\em The Philosophy of Carl G. Hempel: Studies in Science,
  Explanation and Rationality}, Chapter~11, pp.\  218--236. Oxford: Oxford
  University Press.
\newblock Originally published in M. Radner and S. Winokur (eds.),
  \emph{Theories and Methods of Physics and Psychology}, Minnesota Studies in
  Philosophy of Science Vol.~\textsc{iv}, Minneapolis: University of Minnesota
  Press, 1970, pp.~142--163.

\bibitem[\protect\citeauthoryear{Hertz}{Hertz}{1893}]{hertz-electric-waves}
Hertz, H. (1893).
\newblock {\em Electric Waves, Being Researches on the Propagation of Electric
  Action with Finite Velocity through Space}.
\newblock New York: Dover Press.
\newblock Trans.~D.~Jones. Originally published as \emph{ Untersuchungen \"uber
  die Ausbreitung der elektrischen Kraft}, Leipzig: J.~A.~Barth, 1892. This is
  a 1962 reprint of the English edition which first appeared in 1893, published
  by Macmillan and Sons.

\bibitem[\protect\citeauthoryear{Kovalevsky and Seidelmann}{Kovalevsky and
  Seidelmann}{2004}]{kovalevsky-seidelmann-astro}
Kovalevsky, J. and P.~Seidelmann (2004).
\newblock {\em Fundamentals of Astrometry}.
\newblock Cambridge: Cambridge University Press.
\newblock doi:\href{http://dx.doi.org/10.1017/CBO9781139106832}
  {10.1017/CBO9781139106832}.

\bibitem[\protect\citeauthoryear{Lakatos}{Lakatos}{1970}]{lakatos-fals-meth-sci-rsrch}
Lakatos, I. (1970).
\newblock Falsification and the methodology of scientific research programmes.
\newblock In I.~Lakatos and A.~Musgrave (Eds.), {\em Criticism and the Growth
  of Knowledge}, pp.\  91--196. Cambridge: Cambridge University Press.

\bibitem[\protect\citeauthoryear{Lamb}{Lamb}{1932}]{lamb-hydro}
Lamb, H. (1932).
\newblock {\em Hydrodynamics\/} (sixth ed.).
\newblock New York: Dover Publications.
\newblock The 1945 Dover reprint of the 1932 edition published by the Cambridge
  University Press.

\bibitem[\protect\citeauthoryear{Landau and Lifschitz}{Landau and
  Lifschitz}{1975}]{landau-lifschitz-fluid}
Landau, L. and E.~Lifschitz (1975).
\newblock {\em Fluid Mechanics\/} (Second ed.).
\newblock Oxford: Pergamon Press.
\newblock An expanded, revised edition of the original 1959 edition. Translated
  from the Russian by J. Sykes and W. Reid.

\bibitem[\protect\citeauthoryear{Landry}{Landry}{2017}]{landry-cat-theor-work-philr}
Landry, E. (Ed.) (2017).
\newblock {\em Categories for the Working Philosopher}.
\newblock Oxford: Oxford University Press.
\newblock doi:\href{http://dx.doi.org/10.1093/oso/9780198748991.001.0001}
  {10.1093/oso/9780198748991.001.0001}.

\bibitem[\protect\citeauthoryear{Laudan}{Laudan}{1977}]{laudan-prog-probs}
Laudan, L. (1977).
\newblock {\em Progress and Its Problems: Towards a Theory of Scientific
  Growth}.
\newblock Berkeley: University of California Press.

\bibitem[\protect\citeauthoryear{Lewis}{Lewis}{1970a}]{lewis-gen-sems}
Lewis, D. (1970a, December).
\newblock General semantics.
\newblock {\em Synthese\/}~{\em 22\/}(1--2), 18--67.
\newblock doi:\href{http://dx.doi.org/10.1007/BF00413598} {10.1007/BF00413598}.

\bibitem[\protect\citeauthoryear{Lewis}{Lewis}{1970b}]{lewis-def-theor-terms}
Lewis, D. (1970b, July).
\newblock How to define theoretical terms.
\newblock {\em The Journal of Philosophy\/}~{\em 67\/}(13), 427--446.
\newblock Stable URL: $<$\url{http://www.jstor.org/stable/2023861}$>$.

\bibitem[\protect\citeauthoryear{Lewis}{Lewis}{1975}]{lewis-langs-lang}
Lewis, D. (1975).
\newblock Languages and language.
\newblock In K.~Gunderson (Ed.), {\em Language, Mind and Knowledge}, Number
  \textsc{vii} in Minnesota Studies in Philosophy of Science, Chapter~1, pp.\
  3--35. Minneapolis: University of Minnesota Press.
\newblock Freely available online:
  $<$\url{https://cla.umn.edu/mcps/publications/minnesota-studies-philosophy-science}$>$.

\bibitem[\protect\citeauthoryear{Lewis}{Lewis}{1986}]{lewis-phil-papers-2}
Lewis, D. (1986).
\newblock {\em Philosophical Papers}, Volume~2.
\newblock Oxford: Oxford University Press, 1986.

\bibitem[\protect\citeauthoryear{Lockman}{Lockman}{2005}]{lockman-remote-obs-good-obs}
Lockman, F. (2005).
\newblock Can remote observing be good observing? {R}eflections on {P}rocrustes
  and {A}ntaeus.
\newblock \href{https://arxiv.org/abs/astro-ph/0507140}
  {arXiv:astro-ph/0507140}. A slightly edited version of a paper published in
  1993 in \emph{Observing at a Distance}, D. T. Emerson and R. G. Clowes
  (eds.), World Scientific, Singapore, pp.~325--340, with a new Afterword.

\bibitem[\protect\citeauthoryear{Lutz}{Lutz}{2014}]{lutz-right-synt-approach}
Lutz, S. (2014, August).
\newblock What's right with a syntactic approach to theories and models?
\newblock {\em Erkenntnis\/}~{\em 79\/}(8 Supplement), 1475--1492.
\newblock doi:\href{http://dx.doi.org/10.1007/s10670-013-9578-5}
  {10.1007/s10670-013-9578-5}.

\bibitem[\protect\citeauthoryear{M.}{M.}{2011}]{morrison-one-phn-many-mods}
M., M. (2011, June).
\newblock One phenomenon, many models: Inconsistency and complementarity.
\newblock {\em Studies in History and Philosophy of Science\/}~{\em 42\/}(2),
  342--351.
\newblock doi:\href{http://dx.doi.org/10.1016/j.shpsa.2010.11.042}
  {10.1016/j.shpsa.2010.11.042}.

\bibitem[\protect\citeauthoryear{Malament}{Malament}{1977}]{malament-obser-indist-sts}
Malament, D. (1977).
\newblock Observationally indistinguishable spacetimes: Comments on {G}lymour's
  paper.
\newblock See \citeN{earman-etal-found-st-theors}, pp.\  61--80.
\newblock Freely available online:
  $<$\url{https://cla.umn.edu/mcps/publications/minnesota-studies-philosophy-science}$>$.

\bibitem[\protect\citeauthoryear{Malament}{Malament}{2002}]{malament-rot-nogo}
Malament, D. (2002).
\newblock A no-go theorem about rotation in relativity theory.
\newblock In D.~Malament (Ed.), {\em Reading Natural Philosophy: Essays in the
  History and Philosophy of Science and Mathematics}. Chicago: Open Court
  Press.
\newblock Essays presented to Howard Stein in honor of his 70th birthday,
  delivered at a \emph{Festchrift} in Stein's honor at the University of
  Chicago, May, 1999.

\bibitem[\protect\citeauthoryear{Malament}{Malament}{2003}]{malament-rel-orbit-rot}
Malament, D. (2003).
\newblock On relative orbital rotation in general relativity.
\newblock In A.~Ashtekar, R.~Cohen, D.~Howard, J.~Renn, S.~Sarkar, and
  A.~Shimony (Eds.), {\em Revisiting the Foundations of Relativistic Physics:
  Festschrift for John Stachel}, pp.\  175--190. Dordrecht: Kluwer.

\bibitem[\protect\citeauthoryear{Manchak}{Manchak}{2009}]{manchak-know-glob-struc-st}
Manchak, J. (2009, January).
\newblock Can we know the global structure of spacetime?
\newblock {\em Studies in History and Philosophy of Modern Physics\/}~{\em
  40\/}(1), 53--56.
\newblock doi:\href{http://dx.doi.org/10.1016/j.shpsb.2008.07.004}
  {10.1016/j.shpsb.2008.07.004}.

\bibitem[\protect\citeauthoryear{Maxwell}{Maxwell}{1871}]{maxwell-intro-lect-exp-phys}
Maxwell, J.~C. (1871).
\newblock Introductory lecture on experimental physics.
\newblock See \citeN{maxwell-coll-paps}, pp.\  240--255.

\bibitem[\protect\citeauthoryear{Maxwell}{Maxwell}{1876}]{maxwell-sci-apparat}
Maxwell, J.~C. (1876).
\newblock General considerations concerning scientific apparatus.
\newblock See \citeN{maxwell-coll-paps}, pp.\  505--522.
\newblock Originally published in the \emph{Handbook to the Special Loan
  Collection of Scientific Apparatus}, 1876, South Kensington Museum, London:
  Chapman and Hall, 1--21.

\bibitem[\protect\citeauthoryear{Maxwell}{Maxwell}{1965}]{maxwell-coll-paps}
Maxwell, J.~C. (1965).
\newblock {\em The Scientific Papers of {J.~C.~Maxwell}}.
\newblock New York: Dover Publications, Inc.
\newblock W.~Niven (Ed.). Two volumes, published as one.

\bibitem[\protect\citeauthoryear{Newton}{Newton}{1999}]{newton-princ-cohen-whitman}
Newton, I. (1999).
\newblock {\em The Principia: Mathematical Principles of Natural Philosophy}.
\newblock Berkeley, CA: University of California Press.
\newblock Translation of \emph{Philosophi{\ae} Naturalis Principia Mathematica}
  by I. Cohen and A. Whitman, based on the Third Edition text of 1726.

\bibitem[\protect\citeauthoryear{Patton}{Patton}{2011}]{patton-reconsid-exps}
Patton, L. (2011, Fall).
\newblock Reconsidering experiments.
\newblock {\em HOPOS: The Journal of the International Society for the History
  of Philosophy of Science\/}~{\em 1\/}(2), 209--226.
\newblock doi:\href{http://dx.doi.org/10.1086/660167} {10.1086/660167}.

\bibitem[\protect\citeauthoryear{Patton}{Patton}{2012}]{patton-exp-thry-build}
Patton, L. (2012, February).
\newblock Experiment and theory building.
\newblock {\em Synthese\/}~{\em 184\/}(3), 235--246.
\newblock doi:\href{http://dx.doi.org/10.1007/s11229-010-9772-9}
  {10.1007/s11229-010-9772-9}.

\bibitem[\protect\citeauthoryear{Patton}{Patton}{2020}]{patton-expand-thry-test-gr}
Patton, L. (2020, February).
\newblock Expanding theory testing in general relativity: {LIGO} and
  parametrized theories.
\newblock {\em Studies in History and Philosophy of Modern Physics\/}~{\em 69},
  142--153.
\newblock doi:\href{http://dx.doi.org/10.1016/j.shpsb.2020.01.001}
  {10.1016/j.shpsb.2020.01.001}.

\bibitem[\protect\citeauthoryear{Peirce}{Peirce}{1878}]{peirce-doct-chnc}
Peirce, C.~S. (1878, March).
\newblock The doctrine of chances.
\newblock {\em Popular Science Monthly\/}~{\em 12}, 604--615.

\bibitem[\protect\citeauthoryear{Poincar\'e}{Poincar\'e}{1897}]{poincare-stab-sys-solar}
Poincar\'e, H. (1897).
\newblock {\em \french{Sur la stabilit\'e du Syst\`eme Solaire}}, pp.\
  B1--B16.
\newblock Paris: Gauthier-Villars.

\bibitem[\protect\citeauthoryear{Pope}{Pope}{2000}]{pope-turb-flows}
Pope, S. (2000).
\newblock {\em Turbulent Flows}.
\newblock Cambridge: Cambridge University Press.
\newblock doi:\href{http://dx.doi.org/10.1017/CBO9780511840531}
  {10.1017/CBO9780511840531}.

\bibitem[\protect\citeauthoryear{Popper}{Popper}{1959}]{popper-log-sci-disc}
Popper, K. (1959).
\newblock {\em The Logic of Scientific Discovery}.
\newblock London: Hutchinson \& Co.

\bibitem[\protect\citeauthoryear{Putnam}{Putnam}{1977}]{putnam-real-reas-apa}
Putnam, H. (1977, August).
\newblock Realism and reason.
\newblock {\em Proceedings and Addresses of the American Philosophical
  Association\/}~{\em 50\/}(6), 483--498.
\newblock Presidential address delivered before the Seventy-Third Annual
  Eastern Meeting of the American Philosophical Association, Boston, MA, Dec
  29, 1976. doi:\href{http://dx.doi.org/10.2307/3129784} {10.2307/3129784}.

\bibitem[\protect\citeauthoryear{Quine}{Quine}{1980}]{quine-2dogmas}
Quine, W. (1980).
\newblock Two dogmas of empiricism.
\newblock In {\em From a Logical Point of View: Nine Logico-Philosophical
  Essays\/} (2nd, revised ed.)., Chapter \textsc{\textsc{II}}, pp.\  20--46.
  Cambridge, MA: Harvard University Press.
\newblock Emended version of the article originally published in the
  \emph{Philosophical Review}, 1951, 60(1, January):20--43.

\bibitem[\protect\citeauthoryear{Reichenbach}{Reichenbach}{1965}]{reichenbach-rel-apriori-know}
Reichenbach, H. (1965).
\newblock {\em The Theory of Relativity and A Priori Knowledge}.
\newblock Berkeley, CA: University of California Press.

\bibitem[\protect\citeauthoryear{Reintjes and Temple}{Reintjes and
  Temple}{2020}]{reintjes-temnple-shock-wv-gr-met-smooth}
Reintjes, M. and B.~Temple (2020, March).
\newblock Shock wave interactions in general relativity: {T}he geometry behind
  metric smoothing and the existence of locally inertial frames.
\newblock {\em Archive for Rational Mechanics and Analysis\/}~{\em 235},
  1873--1904.
\newblock doi:\href{http://dx.doi.org/10.1007/s00205-019-01456-8}
  {10.1007/s00205-019-01456-8}. Preprint:
  \href{https://arxiv.org/abs/1610.02390} {arXiv:1610.02390 [gr-qc]}.

\bibitem[\protect\citeauthoryear{Smith}{Smith}{1999a}]{smith-mot-lunar-apsis}
Smith, G. (1999a).
\newblock The motion of the lunar apsis.
\newblock See \citeN{newton-princ-cohen-whitman}, Chapter 8.16, pp.\  257--264.

\bibitem[\protect\citeauthoryear{Smith}{Smith}{1999b}]{smith-newt-prob-moon-mot}
Smith, G. (1999b).
\newblock Newton and the problem of the moon's motion.
\newblock See \citeN{newton-princ-cohen-whitman}, Chapter 8.15, pp.\  252--257.

\bibitem[\protect\citeauthoryear{Stegm\"uller}{Stegm\"uller}{1979}]{stegmuller-struc-view-theors}
Stegm\"uller, W. (1979).
\newblock {\em The Structuralist View of Theories: {A} Possible Analogue of the
  Bourbaki Programme in Physical Science}.
\newblock Berlin: Springer-Verlag.
\newblock doi:\href{http://dx.doi.org/10.1007/978-3-642-95360-6}
  {10.1007/978-3-642-95360-6}.

\bibitem[\protect\citeauthoryear{Stein}{Stein}{1989}]{stein-yes-but}
Stein, H. (1989, June).
\newblock Yes, but\ldots: Some skeptical remarks on realism and anti-realism.
\newblock {\em Dialectica\/}~{\em 43\/}(1-2), 47--65.
\newblock doi:\href{http://dx.doi.org/10.1111/j.1746-8361.1989.tb00930.x}
  {10.1111/j.1746-8361.1989.tb00930.x}.

\bibitem[\protect\citeauthoryear{Stein}{Stein}{1992}]{stein-carnap-not-wrong}
Stein, H. (1992, November).
\newblock Was {C}arnap entirely wrong, after all?
\newblock {\em Synthese\/}~{\em 93}, 275--295.
\newblock doi:\href{http://dx.doi.org/10.1007/BF00869429} {10.1007/BF00869429}.

\bibitem[\protect\citeauthoryear{Stein}{Stein}{1994}]{stein-struct-know}
Stein, H. (1994).
\newblock Some reflections on the structure of our knowledge in physics.
\newblock In D.~Prawitz, B.~Skyrms, and D.~Westerst{\aa}hl (Eds.), {\em Logic,
  Metholodogy and Philosophy of Science}, pp.\  633--655. New York: Elsevier
  Science B.V.

\bibitem[\protect\citeauthoryear{Stein}{Stein}{2004}]{stein-enterprise}
Stein, H. (2004, May).
\newblock The enterprise of understanding and the enterprise of
  knowledge---{F}or {I}saac {L}evi's seventieth birthday.
\newblock {\em Synthese\/}~{\em 140\/}(1--2), 135--176.
\newblock doi:\href{http://doi.org/10.1023/B:SYNT.0000029946.38831.c9}
  {10.1023/B:SYNT.0000029946.38831.c9}.

\bibitem[\protect\citeauthoryear{Stein}{Stein}{sheda}]{stein-furth-consid-newt-meth}
Stein, H. (unpublisheda).
\newblock Further considerations on {N}ewton's method.
\newblock Unpublished manuscript. Available at
  $<$\url{http://strangebeautiful.com/other-minds.html}$>$.

\bibitem[\protect\citeauthoryear{Stein}{Stein}{shedb}]{stein-meta-meth-newton}
Stein, H. (unpublishedb).
\newblock On metaphysics and method in {N}ewton.
\newblock Unpublished manuscript. Available at
  $<$\url{http://strangebeautiful.com/other-minds.html}$>$.

\bibitem[\protect\citeauthoryear{Suppe}{Suppe}{1974}]{suppe-srch-underst-sci-thry}
Suppe, F. (1974).
\newblock The search for philosophic understanding of scientific theories.
\newblock In F.~Suppe (Ed.), {\em The Structure of Scientific Theories}, pp.\
  3--254. Chicago: University of Illinois Press.

\bibitem[\protect\citeauthoryear{Suppes}{Suppes}{1960}]{suppes-mngs-uses-mods}
Suppes, P. (1960, September).
\newblock A comparison of the meaning and uses of models in mathematics and the
  empirical sciences.
\newblock {\em Synthese\/}~{\em 12\/}(2--3), 287--301.
\newblock doi:\href{http://dx.doi.org/10.1007/BF00485107} {10.1007/BF00485107}.

\bibitem[\protect\citeauthoryear{Suppes}{Suppes}{1962}]{suppes-mods-data}
Suppes, P. (1962).
\newblock Models of data.
\newblock In E.~Nagel, P.~Suppes, and A.~Tarski (Eds.), {\em Logic, Methodology
  and Philosophy of Science}, pp.\  252--261. Palo Alto, CA: Stanford
  University Press.

\bibitem[\protect\citeauthoryear{Suppes}{Suppes}{1993}]{suppes-role-form-meths}
Suppes, P. (1993).
\newblock The role of formal methods in the philosophy of science.
\newblock In {\em Models and Methods in the Philosophy of Science: Selected
  Essays}, Chapter~1, pp.\  3--14. Berlin: Springer.
\newblock doi:\href{http://dx.doi.org/10.1007/978-94-017-2300-8_1}
  {10.1007/978-94-017-2300-8$\underscore$1}.

\bibitem[\protect\citeauthoryear{Suppes}{Suppes}{1998}]{suppes-pragm-phys}
Suppes, P. (1998).
\newblock Pragmatism in physics.
\newblock In P.~Weingartner, G.~Schurz, and G.~Dorn (Eds.), {\em The Role of
  Pragmatics in Contemporary Philosophy}, pp.\  236--253. Vienna:
  H\"older-Pichler-Tempsky.
\newblock Proceedings of the 20th International Wittgenstein Symposium, 10--16
  August 1997.

\bibitem[\protect\citeauthoryear{Thurston}{Thurston}{1994}]{thurston-proof-prog-math}
Thurston, W. (1994, April).
\newblock On proof and progress in mathematics.
\newblock {\em Bulletin of the American Mathematical Society\/}~{\em 30\/}(2),
  161--177.

\bibitem[\protect\citeauthoryear{{}van Fraassen}{{}van
  Fraassen}{1980}]{fraassen-sci-image}
{}van Fraassen, B. (1980).
\newblock {\em The Scientific Image}.
\newblock Oxford: Oxford University Press.
\newblock doi:\href{http://dx.doi.org/10.1093/0198244274.001.0001}
  {10.1093/0198244274.001.0001}.

\bibitem[\protect\citeauthoryear{{}van Fraassen}{{}van
  Fraassen}{2008}]{fraassen-sci-rep}
{}van Fraassen, B. (2008).
\newblock {\em Scientific Representation: Paradoxes of Perspective}.
\newblock Oxford: Oxford University Press.
\newblock
  doi:\href{http://dx.doi.org/10.1093/acprof:oso/9780199278220.001.0001}
  {10.1093/acprof:oso/9780199278220.001.0001}.

\bibitem[\protect\citeauthoryear{{}van Fraassen}{{}van
  Fraassen}{2012}]{fraassen-mod-meas-emp-grnd}
{}van Fraassen, B. (2012, December).
\newblock Modeling and measurement: {T}he criterion of empirical grounding.
\newblock {\em Philosophy of Science\/}~{\em 79\/}(5), 773--784.
\newblock doi:\href{http://dx.doi.org/10.1086/667847} {10.1086/667847}.

\bibitem[\protect\citeauthoryear{Weatherall}{Weatherall}{2017}]{weatherall-cats-class-st-thrs}
Weatherall, J. (2017).
\newblock Categories and the foundations of classical space-time theories.
\newblock See \citeN{landry-cat-theor-work-philr}, Chapter~13, pp.\  329--348.
\newblock doi:\href{http://dx.doi.org/10.1093/oso/9780198748991.003.0013}
  {10.1093/oso/9780198748991.003.0013}.

\end{thebibliography}
\end{document}